\documentclass[twocolumn,aps,prevon]{revtex4}
\usepackage{amsmath}
\usepackage{picins}
\usepackage{graphicx}
\usepackage{dcolumn}
\usepackage{bm}
\usepackage{multirow}
\usepackage[nice]{nicefrac}
\usepackage{latexsym}
\usepackage{relsize}
\usepackage{verbatim}
\newcommand{\la}{\langle}
\newcommand{\ra}{\rangle}

\begin{document}

\title{Equilibrium Shape of Crystals}
\author{T. L. Einstein}
\affiliation{Department of Physics and Condensed Matter Theory Center, University of Maryland, College Park, Maryland 20742-4111 USA}
\email{einstein@umd.edu             }
\homepage{www.physics.umd.edu/~einstein/}
\date{\today}
\begin{abstract}
This chapter discusses the equilibrium crystal shape (ECS) from a physical perspective, beginning with a historical introduction to the Wulff theorem. It takes advantage of excellent prior reviews, particularly in the late 1980’s, recapping highlights from them. It contains many ideas and experiments subsequent to those reviews. Alternatives to Wulff constructions are presented. Controversies about the critical behavior near smooth edges on the ECS are recounted, including the eventual resolution. Particular attention is devoted to the origin of sharp edges on the ECS, to the impact of reconstructed or adsorbed surface phases coexisting with unadorned phases, and to the role and nature of possible attractive step-step interactions.


\end{abstract}

\maketitle
\renewcommand{\thesection}{\arabic{section}}
\renewcommand{\thesubsection}{\thesection.\arabic{subsection}}
\vspace{8mm}
\hspace{-1mm}\fbox{\noindent
   \parbox{0.47\textwidth}{\textit{Reformatted version (with slightly modified bibliography, also alphabetized and including article titles) of a review appearing in} \textbf{Handbook of Crystal Growth, Fundamentals, 2nd ed}., edited by T. Nishinaga (Elsevier, Amsterdam, 2015--ISBN 9780444563699/eBook:9780444593764), vol.\ 1A (Thermodynamics and Kinetics), chap.\ 5, pp.\ 215--264; \texttt{http://www.sciencedirect.com/science/article/}\\ \texttt{pii/B9780444563699000058.}\hfill}}

\section{Introduction}\label{sec:level1}

The notion of equilibrium crystal shape is arguably the platonic ideal of crystal growth.  It underpins much of our thinking about crystals and, accordingly, has been the subject of several special reviews and tutorials [Rottman and Wortis 1984a, Wortis 1988, Zia 1988, Williams and Bartelt 1989, Bonzel 2003] and is a prominent section of most volumes and extended review articles and texts about crystals and their growth [Landau and Lifshitz 1980, van Beijeren and Nolden 1987, Nozi\`eres 1992, Pimpinelli and Villain 1998, Sekerka 2004].  In actual situations, there are many complications that thwart observation of such behavior, including kinetic barriers, impurities, and other bulk defects like dislocations.  Furthermore, the notion of a well-defined equilibrium shape requires that the crystal not make contact with a wall or surface, since that would alter its shape.  By the same token, the crystal cannot then be supported, so gravity is neglected.  For discussions of the effect of gravity or contact with walls, see, e.g., Nozi\`eres [1992].

Gibbs [1874-8] is generally credited with being the first to recognize that the equilibrium shape of a substance is that which, for a fixed volume, minimizes the (orientation-dependent) surface free energy integrated over the entire surface: the bulk free energy is irrelevant since the volume is conserved, while edge or corner energies are ignored as being higher-order effects that play no role in the thermodynamic limit. Herring [1951, 1953] surveys the early history in detail:  The formulation of the problem was also carried out independently by Curie [1885].  The solution of this ECS problem, the celebrated Wulff construction, was stated by by Wulff (1901), but his proof was incorrect.  Correct proofs were subsequently given by Hilton [1903], by Liebmann [1914], and by von Laue [1943], who presented a critical review.  However, these proofs, while convincing of the theorem, were not general (and evidently applied only to $T$=0, since they assumed the ECS to be a polyhedron and compared the sum over the facets of the surface free energy of each facet times its area with a similar sum over a similar polyhedron with the same facet planes but slightly different areas (and the same volume). Dinghas [1944] showed that the Brunn-Minkowski inequality could be used to prove directly that
any shape differing from that resulting from the Wulff construction has a higher surface free energy. Although Dinghas again considered only a special class of polyhedral shapes, Herring [1951, 1953] completed the proof by noting that Dinghas's method is easily extended to arbitrary shapes, since the inequality is true for convex bodies in general.  In their seminal paper on crystal growth, Burton, Cabrera, and Frank [1951] present a novel proof of the theorem in two dimensions (2D).

Since equilibrium implies minimum Helmholtz free energy for a given volume and number, and since the bulk free energy is ipso facto independent of shape, the goal is to determine the shape that minimizes the integrated surface free energy of the crystal.  The prescription takes the following form: One begins by creating a polar plot of the surface free energy as a function of orientation angle (of the surface normal) and draws a perpendicular plane (or line in 2D) through the tip of each ray. (There are many fine reviews of this subject by Chernov [1960], Frank [1962], Mullins [1962], and Jackson [1975].) Since the surface free energy in 3D (three dimensions) is frequently denoted by $\gamma$, this is often called a $\gamma$ plot.  The shape is then the formed by the interior envelope of these planes or lines, often referred to as a pedal.  At zero temperature, when the free energy is just the energy, this shape is a polyhedron in 3D and a polygon in 2D, each reflecting the symmetry of the underlying lattice.
At finite temperature the shapes become more complex.  In 2D the sharp corners are rounded.  In 3D, the behavior is richer, with two possible modes of evolution with rising temperature.  For what Wortis terms type-A crystals, all sharp boundaries smoothen together, while in type-B, first the corners smooth, then above a temperature denoted $T_0$ the edges also smooth. The smooth regions correspond to thermodynamic rough phases, with height-height correlation functions that diverge for large lateral separation $l$---like $l^{\alpha}$, with $\alpha$ (typically $0 < \alpha < 1$) called the roughening exponent---in contrast to facets, where they attain some finite value as $l \rightarrow \infty$ [Pimpinelli and Villain 1998].  The faceted regions in turn correspond to ``frozen" regions.  Pursuing the correspondence, sharp and smooth edges correspond to first-order and second-order phase transitions, respectively.

The aim of this chapter is primarily to explore physical ideas regarding ECS and the underlying Wulff constructions.  This topic has also attracted considerable interest in the mathematics community.  Readers interested in more formal and sophisticated approaches are referred to two books, Dobrushin et al.\ [1992] and Cerf and Picard [2006] and to many articles, including De Coninck et al.\ [1989], Fonesca [1991], Pfister [1991], Fonesca and M\"uller [1992], Dacorogna and Pfister [1992], Dobrushin et al.\ [1993], Miracle-Sole and Ruiz [1994], Almgren and Taylor [1996], Peng et al.\ [1999], Miracle-Sole [1999, 2013].

\section{From Surface Free Energies to Equilibrium Crystal Shape}
\subsection{General Considerations}\label{s:gen}

\begin{figure}
\label{f:nelson}
\centering
  \includegraphics[trim=60 180 40 50,clip,width=8cm,height=5cm]{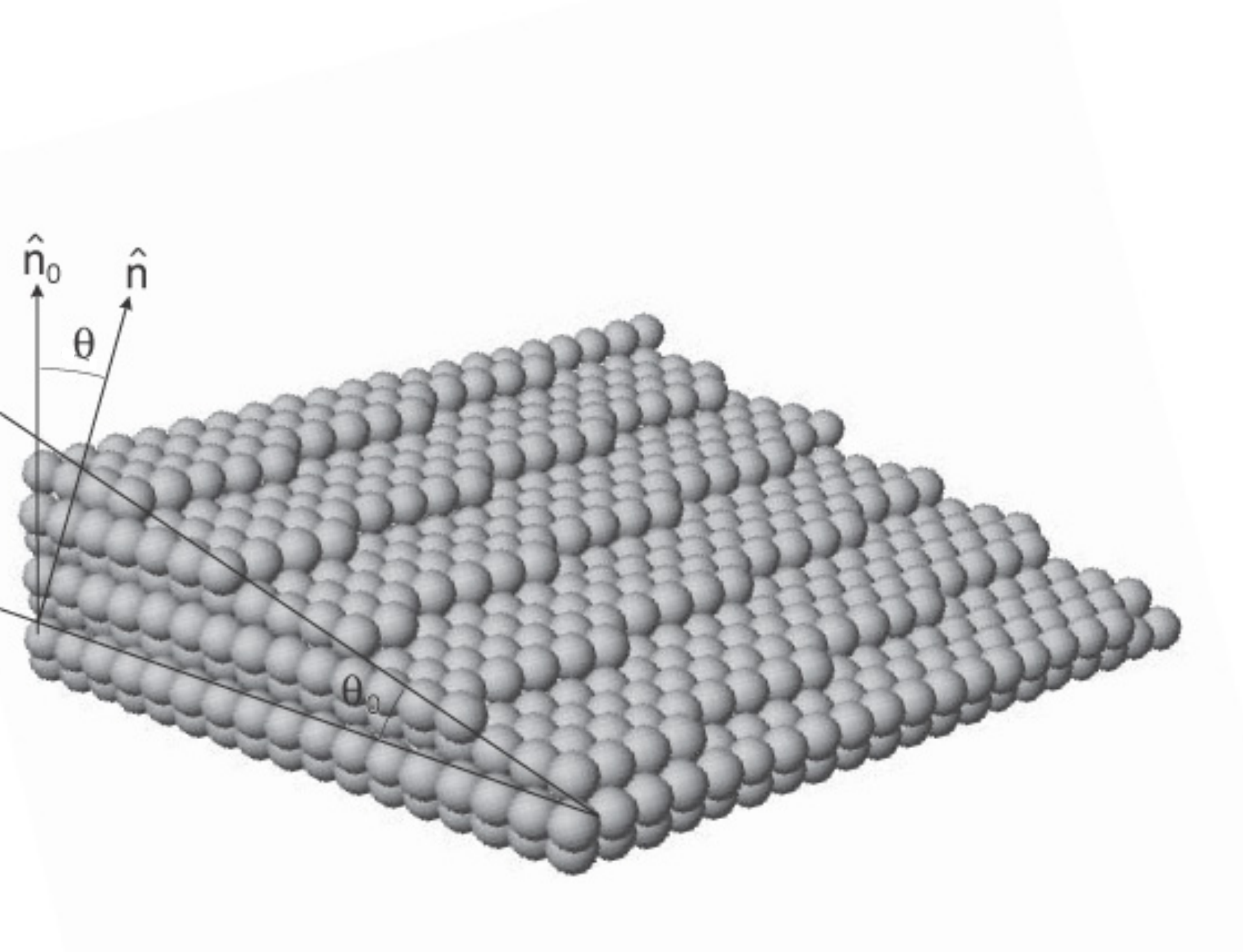}
  \caption{Portion of a (3,$\bar{2}$,16) surface, vicinal to an fcc (001), to illustrate a misoriented, vicinal surface.  The vicinal-surface and terrace normals are $\hat{n}$ = (3, -2, 16)/$\sqrt{269}$ and $\hat{n}_0$ = (0, 0, 1), respectively. The polar angle $\theta$ (with respect to the (001) direction),  denoted $\phi$ in the original figure (consistent with most of the literature on vicinal surfaces), is $\arccos(16/\sqrt{269})$, while azimuthal angle $\varphi$ (denoted $\theta$ in most of the literature on vicinal surfaces), indicating how much $\hat{n}$ is rotated around $\hat{n}_0$ away from the
vertical border on which $\theta_0$ is marked, is clearly $\arctan(1/5)$; $\tan \theta_0 = \tan \theta_0 \cos\varphi$. Since $h$ is $a_1/\sqrt{2}$, where $a_1$ is the nearest-neighbor spacing, the mean distance $\ell$ (in a terrace plane) between steps is $a_1/(\sqrt{2} \tan \theta) = 8\sqrt{2/13} a_1 = 3.138a_1$.  While
the average distance from one step to the next along a principal, (110) direction looks like 3.5$a_1$, it is in fact
$a_1 /(\sqrt{2} \tan \theta_0) = 3.2a_1$.
The ``projected area" of this surface segment, used to compute the surface free energy $f_p$, is the size of a (001) layer: $20a_1 \times 17a_1 = 340a_1^2$; the width is 20$a_1$. In ``Maryland notation" [see text] $z$ is in the $\hat{n}_0$ direction, while the normal to the vicinal, $\hat{n}$, lies in the $x-z$ plane and $y$ runs along the mean direction of the edges of the steps. In most discussions, $\varphi =0$, so that this direction would be that of the upper and lower edges of the depicted surface. Adapted from Nelson et al.\ [1993]
}
\end{figure}

To examine this process more closely, we examine the free energy expansion for a vicinal surface, that is a surface misoriented by some angle $\theta$ from a facet direction.  Cf.\ Fig.~1. Unfortunately, this polar angle is denoted by $\phi$ in much of the literature on vicinal surfaces, with $\theta$ used for in-plane misorientation; most reviews of ECS use $\theta$ for the polar angle, as we shall here.  The term ``vicinal" implies that the surface is in the vicinity of the orientation.  It is generally assumed that the surface orientation itself is rough (while the facet direction is below its roughening temperature and so is smooth).  We consider the projected surface free energy $f_p(\theta,T)$ [Jayaprakash et al.\ 1984] (with the projection being onto the low-index reference, facet direction of terraces) :

\begin{equation}
\label{e:fp-th}
f_p(\theta,T) = f_0(T) + \beta(T) \frac{|\tan \theta|}{h} + g(T) |\tan \theta|^3 + c (\tan \theta)^4.
\end{equation}

\noindent The first term is the surface free energy per area of the terrace (facet) orientation; it is often denoted $\sigma$.  The average density of steps (the inverse of their mean separation $\la \ell \ra$) is $\tan \theta /h$, where $h$ is the step height.  In the second term $\beta(T)$ is the line tension or free energy per length of step formation. (Since 2D is a dimension smaller than 3D, one uses $\beta$ rather than $\gamma$. Skirting over the difference in units resulting from the dimensional difference, many use $\gamma$ in both cases.  While step free energy per length and line tension are equivalent for these systems, where the surface is at constant (zero) charge, they are inequivalent in electrochemical systems, where it is the electrode potential conjugate to the surface charge that is held fixed [Ibach and Schmickler, 2003])  The third term is associated with interactions between steps, in this case assumed to be proportional to $\ell^{-2}$ (so that this term, which also includes the $\ell^{-1}$ density of steps, goes like $\ell^{-3}$). The final term is the leading correction.

The $\ell^{-2}$ interaction is due to a combination (not a simple sum) of two repulsive potential energies: the entropic repulsion due to the forbidden crossing of steps and the elastic repulsions due to dipolar strains near each step.  An explicit form for $g(T)$ is given in Eq.~(\ref{e:gA}) below.  The $\ell^{-2}$ of the entropic interaction can be understood from viewing the step as performing a random walk in the direction between steps (the $x$ direction in ``Maryland notation"\footnote{This term was coined by a speaker at a workshop in Traverse City in August 1996---see Duxbury \& Pence [1997] for the proceedings---and then used by several other speakers.} as a function of the $y$ direction (which is time-like in the fermion transcription to be discussed later)---cf.\ Fig.~1, so the distance ($y$) it must go till it touches a neighboring step satisfies $\ell^2 \propto y$.   To get a crude understanding of the origin of the elastic repulsion, one can imagine that since a step is unstable relative to a flat surface, it will try to relax toward a flatter shape, pushing atoms away from the location of the step by a distance decaying with distance from the step.  When two steps are close to each other, such relaxation will be frustrated since atoms on the terrace this pair of steps are pushed in opposite directions, so relax less than if the steps are widely separated, leading to a repulsive interaction.  Analyzed in more detail [Marchenko and Parshin 1980, Nozi\`eres 1992, Stewart et al.\ 1994], this repulsion is dipolar and so proportional to $\ell^{-2}$.  However, attempts to reconcile the prefactor with the elastic constants of the surface have met with limited success.  The quartic term in Eq.~(\ref{e:fp-th}) is due to the leading ($\ell^{-3}$) correction to the elastic repulsion [Najafabadi and Srolovitz 1994], a dipole-quadrupole repulsion.  It generally has no significant consequences but is included to show the leading correction to the critical behavior near a smooth edge on the ECS, to be discussed below.

The absence of a quadratic term in Eq.~(\ref{e:fp-th}) reflects that there is no $\ell^{-1}$ interaction between steps.  In fact, there are some rare geometries, notably vicinals to (110) surfaces of fcc crystals (Au in particular) that exhibit what amounts to $\ell^{-1}$ repulsions that lead to more subtle behavior [Carlon and van Beijeren 2000].  Details about this faxcinating idiosyncratic surface are beyond the scope of this chapter; readers should see the thorough, readable discussion by van Albada et al.\ [2002].

As temperature increases, $\beta(T)$ decreases due to increasing entropy associated with step-edge excitations (via the formation of kinks).  Eventually $\beta(T)$ vanishes at a temperature $T_R$ associated with the roughening transition.  At and above this $T_R$ of the facet orientation, there is a profusion of steps, and the idea of a vicinal surface becomes meaningless. For rough surfaces the projected surface free energy $f_p(\theta,T)$ is quadratic in $\tan \theta$.  To avoid the singularity at $\theta = 0$ in the free energy expansion that thwarts attempts to proceed analytically, some treatments, notably Bonzel and Preuss [1995], approximate $f_p(\theta,T)$ as quadratic in a small region near $\theta = 0$.   It is important to recognize that the vicinal orientation is thermodynamically rough, even though the underlying facet orientation is smooth.  The two regions correspond to incommensurate and commensurate phases, respectively.  Thus, in a rough region the mean spacing $\langle \ell \rangle$ between steps is not in general simply related to (i.e. an integer multiple plus some simple fraction) the atomic spacing.

Details of the roughening process have been reviewed by Weeks [1980] and van Beijeren and Nolden [1987]; the chapter by Akutsu in this Handbook provides an up-to-date account.  However, for use later, we note that much of our understanding of this process is rooted in the mapping between the restricted BCSOS (body-centered [cubic] solid-on-solid) model and the exactly-solvable [Lieb 1967, Lieb and Wu 1972] symmetric 6-vertex model [van Beijeren 1977], which has a transition in the same universality class as roughening.  (This BCSOS model is based on the BCC crystal structure, involving square net layers with ABAB stacking, so that sites in each layer are lateral displaced to lie over the centers of squares in the preceding [or following] layer.  Being an SOS model means that for each column of sites along the vertical direction there is a unique upper occupied site, with no vacancies below it nor floating atoms above it.  Viewed from above, the surface is a square network with one pair of diagonally opposed corners on A layers and the other pair on B layers.  The restriction is that neighboring sites must be on adjacent layers (so that their separation is the distance from a corner to the center of the BCC lattice).  There are then 6 possible configurations: two in which the two B corners are both either above or below the A corners and four in which one pair of catercorners are on the same layer and the other pair are on different layers (one above and one below the first pair).  In the symmetric model, there are three energies, $-\epsilon$ for the first pair, and $\pm \delta/2$ for the others, the sign depending on whether the catercorner pair on the same lattice are on A or B [Nolden and van Beijeren 1994].  The case $\delta$=0 corresponds to the $F$-model, which has an infinite-order phase transition and an essential singularity at the critical point, in the class of the Kosterlitz-Thouless [1973] transition [Kosterlitz 1974].  (In the ``ice" model, $\epsilon$ also is 0.) For the asymmetric 6-vertex model, each of the 6 configurations can have a different energy; this model can also be solved exactly [Yang 1967, Sutherland et al.\ 1967].

\subsection{More Formal Treatment}

To proceed more formally, we largely follow Wortis [1988].  The shape of a crystal is given by the length $R(\mathbf{\hat{h}})$ of a radial vector to the crystal surface for any direction $\mathbf{\hat{h}}$.  The shape of the crystal is defined as the thermodynamic limit of this crystal for increasing volume $V$, specifically,
\begin{equation}
\label{e:rh1}
 r(\mathbf{\hat{h}},T) \equiv \lim_{V \rightarrow \infty}[R(\mathbf{\hat{h}})/\alpha V^{1/3}],
\end{equation}
where $\alpha$ is an arbitrary dimensionless variable.  This function $r(\mathbf{\hat{h}},T)$ corresponds to a free energy.  In particular, since both independent variables are field-like (and so intrinsically intensive), this is a Gibbs-like free energy.  Like the Gibbs free energy, $r(\mathbf{\hat{h}},T)$ is continuous and convex in $\mathbf{\hat{h}}$.

\begin{figure}
  \centering
    \includegraphics[width=8cm]{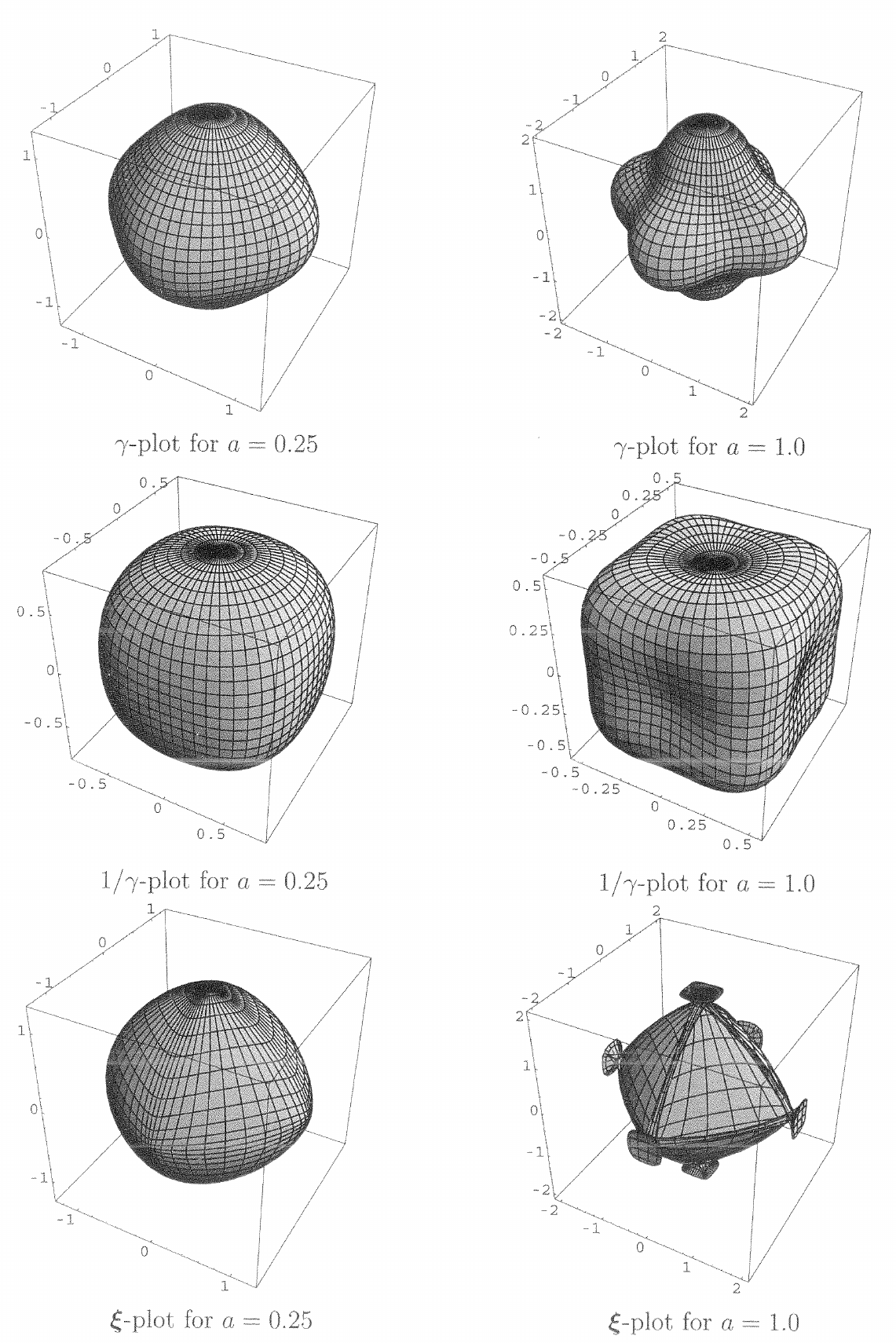}
  \caption{$\gamma$-plots (plots of $f_i(\mathbf{\hat{m}})$, $1/\gamma$-plots and $\mathbf{\xi}$-plots for Eq.~(\ref{e:Q4}) for positive values of $a$. For  $a$ = 1/4, all orientations appear on the ECS. For $a$ = 1.0, the $1/\gamma$-plot has
concave regions, and the $\mathbf{\xi}$-plot has ears and flaps that must be truncated to give the ECS, essentially an octahedron with curved faces.   From Sekerka [2004], which shows in a subsequent figure that the $\gamma$- and $1/\gamma$-plots for  $a$ = -0.2 and -0.5 resemble the $1/\gamma$- and $\gamma$-plots, respectively, for $a$ = 1/4 and 1.}\label{f:Sekerka4}
\end{figure}

The Wulff construction then amounts to a Legendre transformation \footnote{As exposited clearly in Callen [1985], one considers a [convex] function $y = y(x)$ and denotes its derivative as $p = \partial y/\partial x$.  If one then tries to consider $p$ instead of $x$ as the independent variable, there is information lost: one cannot reconstruct $y(x)$ uniquely from $y(p)$.  Indeed, $y = y(p)$ is a first-order differential equation, whose integration gives $y = y(x)$ only to within an undetermined integration constant.  Thus, $y = y(p)$ corresponds to a family of displaced curves, only one of which is the original $y = y(x)$.  The key concept is that the locus of points satisfying $y = y(x)$ can be equally well represented by a family of lines tangent to $y(x)$ at all $x$, each with a $y$-intercept $\psi$ determined by the slope $p$ at $(x,y(x))$.  That is, $\psi = \psi(p)$ contains all the information of $y = y(x)$. Recognizing that $p = (y - \psi)/(x - 0)$, one finds the transform $\psi = y - px$.  Readers should recall that this is the form of the relationship between thermodynamic functions, particularly the Helmholtz and the Gibbs free energies.} to $r(\mathbf{\hat{h}},T)$ from the orientation $\mathbf{\hat{m}}$-dependent interfacial free energy $f_i(\mathbf{\hat{m}},T)$ (or in perhaps more common but less explicit notation, $\gamma(\mathbf{\hat{m}},T)$), which is $f_p(\theta,T)/\cos(\theta)$.  For liquids, of course, $f_i(\mathbf{\hat{m}},T)$ is spherically symmetric, as is the equilibrium shape. (Herring [1953] mentions rigorous proofs of this problem by Schwarz in 1884 and by Minkowski in 1901.)  For crystals, $f_i(\mathbf{\hat{m}},T)$ is not spherically symmetric but does have the symmetry of the crystal lattice.   For a system with cubic symmetry, one can write
\begin{equation}\label{e:Q4}
f_i(\mathbf{\hat{m}},T) = \gamma_0(T)\, [1 + a(T)(m_x^4 + m_y^4 + m_z^4)],
\end{equation}
where $\gamma_0(T)$ and $a(T)$ are constants.  As illustrated in Fig.~\ref{f:Sekerka4}, for $a$=1/4 the asymmetry leads to minor distortions, which are rather inconsequential.  However, for $a$=1 the enclosed region is no longer convex, leading to an instability to be discussed shortly

 One considers the change in the interfacial free energy associated with changes in shape.  The constraint of constant volume is incorporated by subtracting from the change in the integral of $f_i(\mathbf{\hat{m}},T)$ the corresponding change in volume, multiplied by a Lagrange multiplier $\lambda$.  Herring [1951, 1953] showed that this constrained minimization problem has a unique and rather simple solution that is physically meaningful in the limit that it is satisfactory to neglect edge, corner, and kink energies in $f_i(\mathbf{\hat{m}},T)$, that is in the limit of large volume.  In this case $\lambda \propto V^{-1/3}$; by choosing the proportionality constant as essentially the inverse of $\alpha$, we can write the result as
\begin{equation}
\label{e:rh2}
  r(\mathbf{\hat{h}},T) = \min_{\mathbf{\hat{m}}} \left( \frac{f_i(\mathbf{\hat{m}},T)}{\mathbf{\hat{m}}\cdot \mathbf{\hat{h}}} \right)
\end{equation}

\begin{figure}
\centering
  \includegraphics[width=8cm]{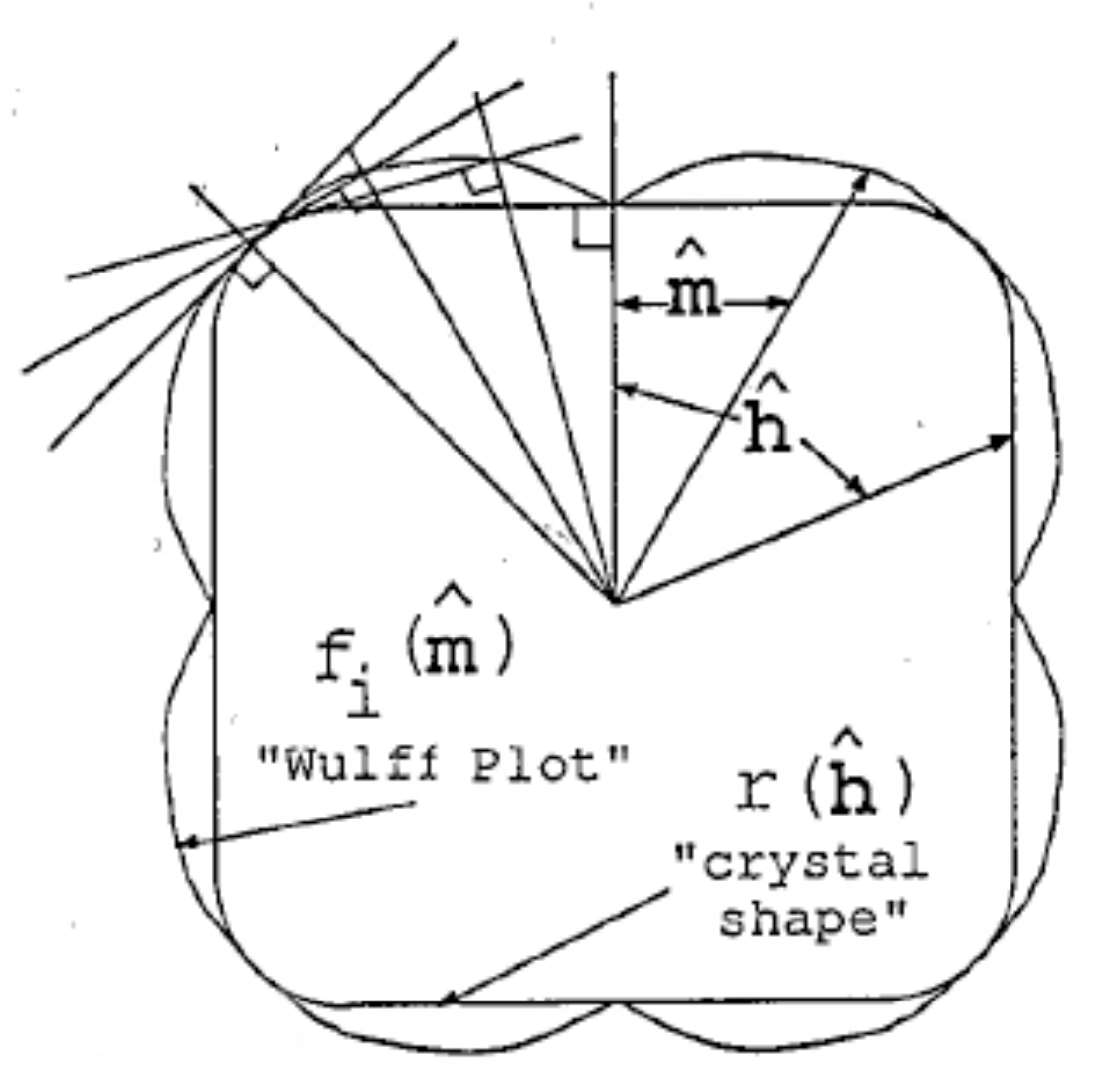}
  \caption{Schematic of the Wulff construction. The interfacial free energy per unit area f$_i\mathbf{\hat{m}}$
is plotted in polar form (the ``Wulff plot" or ``$\gamma$-plot"). One draws a radius vector in each direction $\mathbf{\hat{m}}$ and constructs a perpendicular plane where this vector hits the Wulff plot. The interior
envelope of the family of ``Wulff planes" thus formed, expressed algebraically in Eq.~(\ref{e:rh2}), is the crystal shape, up to an arbitrary overall scale factor which may be chosen as unity." From Wortis [1988] 
}
\label{f:Wortis17}
\end{figure}

The Wulff construction is illustrated in Fig.~\ref{f:Wortis17}.  The interfacial free energy $f_i(\mathbf{\hat{m}})$, at some assumed $T$, is displayed as a polar plot.  The crystal shape is then the interior envelope of the family of perpendicular planes (lines in 2D) passing through the ends of the radial vectors $\mathbf{\hat{m}} f_i(\mathbf{\hat{m}})$.  
Based on Eq.~(\ref{e:rh2}) 
one can, at least in principle, determine $\mathbf{\hat{m}(\hat{h})}$ or $\mathbf{\hat{h}(\hat{m})}$, which thus amounts to the equation of state of the equilibrium crystal shape.  One can also write the inverse of Eq.~(\ref{e:rh2}):
\begin{equation}
\label{e:rhinv}
  \frac{1}{f_i(\mathbf{\hat{m}},T)} = \min_{\mathbf{\hat{m}}} \left( \frac{1/f_i(\mathbf{\hat{h}},T)}{\mathbf{\hat{m}}\cdot \mathbf{\hat{h}}} \right)
\end{equation}
Thus, a Wulff construction using the inverse of the crystal shape function yields the inverse free energy.

To be more explicit, consider consider the ECS in Cartesian coordinates $z(x,y)$, i.e. $\mathbf{\hat{h}} \propto (x, y, z(x,y))$, assuming (without dire consequences [Wortis 1988]) that $z(x,y)$ is single-valued.  Then in order for any displacement to be tangent to $\mathbf{\hat{h}}$, $dz -p_x\, dx -p_y\, dy = 0$
\begin{equation}
\label{e:mLandau}
 \mathbf{\hat{h}} = \frac{1}{\sqrt{1 + p_x^2 + p_y^2}}(-p_x z,-p_y z,1),
\end{equation}
\noindent where $p_x$ is shorthand for $\partial z/ \partial x$.

Then the total free energy and volume are
\begin{eqnarray}\label{e:fv}
F_i(T) &=& \iint \, f_p(p_x,p_y) \, dx\, dy \nonumber \\
 V &=& \iint z(x,y)\, dx\, dy
\end{eqnarray}
\noindent where $f_p$ , which incorporates the line-segment length, is: $f_p \equiv [1 + p_x^2 + p_y^2]^{1/2}\, f_i$.  Minimizing $F_i$ subject to the constraint of fixed $V$ leads to the Euler-Lagrange equation
\begin{equation}\label{e:EulLag}
 \frac{\partial}{\partial x}\, \frac{f_p(\partial_x z,p_y)}{p_x} + \frac{\partial}{\partial y}\, \frac{f_p(p_x,p_y)}{p_y} = -2 \lambda
\end{equation}
(Actually one should work with macroscopic lengths, then divide by the $V^{1/3}$ times the proportionality constant. Note that this leaves $p_x$ and $p_y$ unchanged.  [Wortis 1988])  On the right-hand side $2\lambda$ can be identified as the chemical potential $\mu$, so that the constancy of the left hand side is a reflection of equilibrium.  Eq.~(\ref{e:EulLag}) is strictly valid only if the derivatives of $f_p$ exist, so one must be careful near high-symmetry orientations below their roughening temperature, for which facets occur.  To show that this highly nonlinear second-order partial differential equation with unspecified boundary conditions is equivalent to Eq.~(\ref{e:rh2}), we first note that the first integral of  Eq.~(\ref{e:EulLag}) is simply

\begin{figure}[t]
\centering
  \includegraphics[width=6cm]{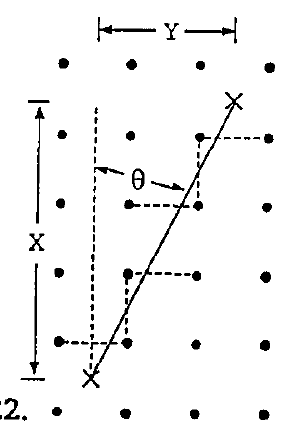}
  \caption{Kossel crystal at T = 0. The energy to cleave the crystal along the depicted
slanted interface  ($\tan \theta$ = Y/Z) is $\epsilon_1\, (|X| + |Y|)$. From Wortis [1988] 
}
\label{f:Wortis22}
\end{figure}

\begin{equation}\label{e:1st}
  z -xp_x  -yp_y = f_p(p_x,p_y)
\end{equation}
\noindent The right-hand side is just a function of derivatives, consistent with this being a Legendre transformation.  Then differentiating yields
\begin{equation}\label{e:xy}
x = - \partial f_p/\partial(p_x), \quad y = - \partial f_p/\partial(p_y)
\end{equation}
Hence, one can show that
\begin{equation}\label{e:xy2}
  z(x,y) = \min_{p_x,p_y} (f_p(p_x,p_y) + xp_x  + yp_y)
\end{equation}

\section{Applications of Formal Results}

\subsection{Cusps and Facets}

The distinguishing feature of Wulff plots of faceted crystals compared to liquids is the existence of [pointed] cusps in  $f_i(\mathbf{\hat{m}},T)$, which underpin these facets.
The simplest way to see why the cusp arises is to examine a square lattice with nearest-neighbor bonds having bond energy $\epsilon_1$, often called a 2D Kossel [1927, 1934] crystal; note also Stranski [1928].  In this model, the energy to cleave the crystal is the Manhattan distance between the ends of the cut; i.e., as illustrated in Fig.~\ref{f:Wortis22}, the energy of severing the bonds
between (0,0) and ($X,Y$) is just $+\epsilon_1\, (|X| + |Y|)$.  The interfacial area, i.e. length, is $2(X^2 + Y^2)$ since the cleavage creates \emph{two} surfaces.  At $T$ = 0, entropy plays no role so that
\begin{equation}\label{e:cuspKossel}
  f_i(\theta) = \frac{\epsilon_1}{2}\left(|\sin \theta| + |\cos \theta|\right) \sim \frac{\epsilon_1}{2}\left(1 + |\theta| + \ldots \right)
\end{equation}
At finite $T$ fluctuations and attendant entropy do contribute, and the argument needs more care.

\begin{figure}[t]
\centering
  \includegraphics[width=7cm]{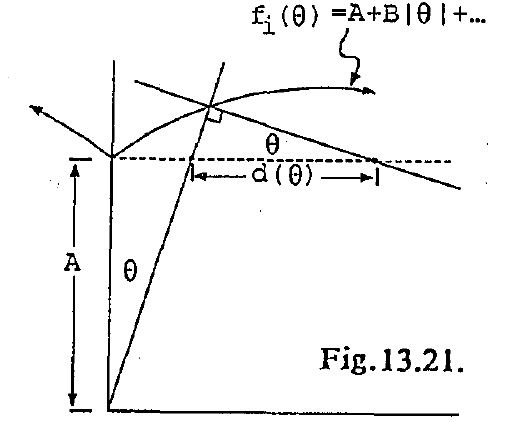}
  \caption{Wulff plot with a linear cusp at $\theta$ = 0. If $d(\theta) \rightarrow 0$ as• $\theta \rightarrow 0$, then there
is no facet corresponding to $\theta$ = 0, and the $\theta$ = 0 Wulff plane (dotted line) is tangent
to the crystal shape at just a single point. Since $d(\theta)$ = B, a cusp in the Wulff plot leads to a facet of the corresponding
orientat1on on the ECS. From Wortis [1988] 
}
\label{f:Wortis21}
\end{figure}

Recalling Eq.~(\ref{e:fp-th}) we see that if there is a linear cusp at $\theta = 0$, then
\begin{equation}\label{e:cusp}
  f_i(\theta,T) = f_i(0,T) + B(T) |\theta|,
\end{equation}
\noindent where $B = \beta(T)/h$, since the difference between $f_i(\theta)$ and $f_p(\theta)$ only appears at order $\theta^2$.  Comparing Eqs.~(\ref{e:cuspKossel}) and (\ref{e:cusp}), we see that for the Kossel square $f_i(0,0) = \epsilon_1/2$ and $B(0) = \epsilon_1/2$.  Further discussion of the 2D $f_i(\theta)$ is deferred to Section~4.3 below.

To see how a cusp in $f_i(\mathbf{\hat{m}},T)$ leads to a facet in the ECS, consider Fig.~\ref{f:Wortis21}: the Wulff plane for $\theta \gtrsim 0$ intersects the horizontal $\theta = 0$ plane at a distance $f_i(0) + d(\theta)$ from the vertical axis.  The crystal will have a horizontal axis if and only if $d(\theta)$ does not vanish as $\theta \rightarrow 0$.  From Fig.~\ref{f:Wortis21}, it is clear that $\theta \approx \sin \theta \approx B \theta/d(\theta)$ for $\theta$ near 0, so that $d(0) = B > 0$.  For a weaker dependence on $\theta$, e.g. $B |\theta|^\zeta$ with $\zeta > 1$, $d(0) = 0$, and there is no facet.  Likewise, at the roughening temperature $\beta$ vanishes and the facet disappears.

\begin{figure}[t]
  \centering
    \includegraphics[width=8cm]{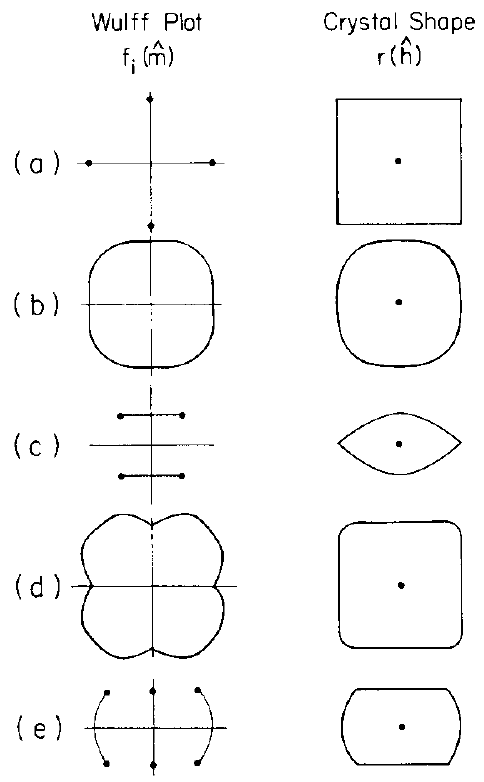}
  \caption{Some possible Wulff plots and corresponding equilibrium crystal shapes. Faceted and curved surfaces may appear, joined at sharp or
smooth edges in a variety of combinations. From Rottman and Wortis [1984a]; the ECS are also in Herring [1953]}
\label{f:Rott4}
\end{figure}

\subsection{Sharp Edges and Forbidden Regions}

When there is a sharp edge (or corner) on the ECS $r(\mathbf{\hat{h}},T)$, Wulff planes with a range of orientations $\mathbf{\hat{m}}$ will not be part of the inner envelope determining this ECS; they will lie completely outside it.  There is no portion of the ECS whose surface tangent has these orientations.  As in the analogous problems with forbidden values of the ``density" variable, the free energy $f_i(\mathbf{\hat{m}},T)$ is actually not properly defined for forbidden values of $\mathbf{\hat{m}}$; those unphysical values should actually be removed from the Wulff plot.
Fig.~\ref{f:Rott4} depicts several possible ECSs and their associated Wulff plots.
It is worth emphasizing that, in the extreme case of the fully faceted ECS at $T=0$, the Wulff plot is simply a set of discrete points in the facet directions.

Now if we denote by $\mathbf{\hat{m}_+}$  and $\mathbf{\hat{m}_-}$ the limiting orientations of the tangent planes approaching the edge from either side, then all intermediate values do not occur as stable orientations.  These missing, not stable, ``forbidden" orientations are just like the forbidden densities at liquid-gas transitions, forbidden magnetizations in ferromagnets at $T < T_c$ [Garc\'{\i}a 1984], and miscibility gaps in binary alloys.  Herring [1951, 1953] first presented an elegant way to determine these missing orientations using a spherical construction.  For any orientation $\mathbf{\hat{m}}$, this tangent sphere (often called a ``Herring sphere") passes through the origin and is tangent to the Wulff plot at $f_i(\mathbf{\hat{m}})$.  From geometry he invoked the theorem that an angle inscribed in a semicircle is a right angle.  Thence, if the orientation $\mathbf{\hat{m}}$ appears on the ECS, it appears at an orientation that points outward along the radius of that sphere.  Herring then observes that only if such a sphere lies completely inside the plot of $f_i(\mathbf{\hat{m}})$ does, that orientation appear on the ECS.  If some part were inside, its Wulff plane would clip off the orientation of the point of tangency, so that that orientation would be forbidden.

\begin{figure}[t]
\centering
  \includegraphics[width=8cm]{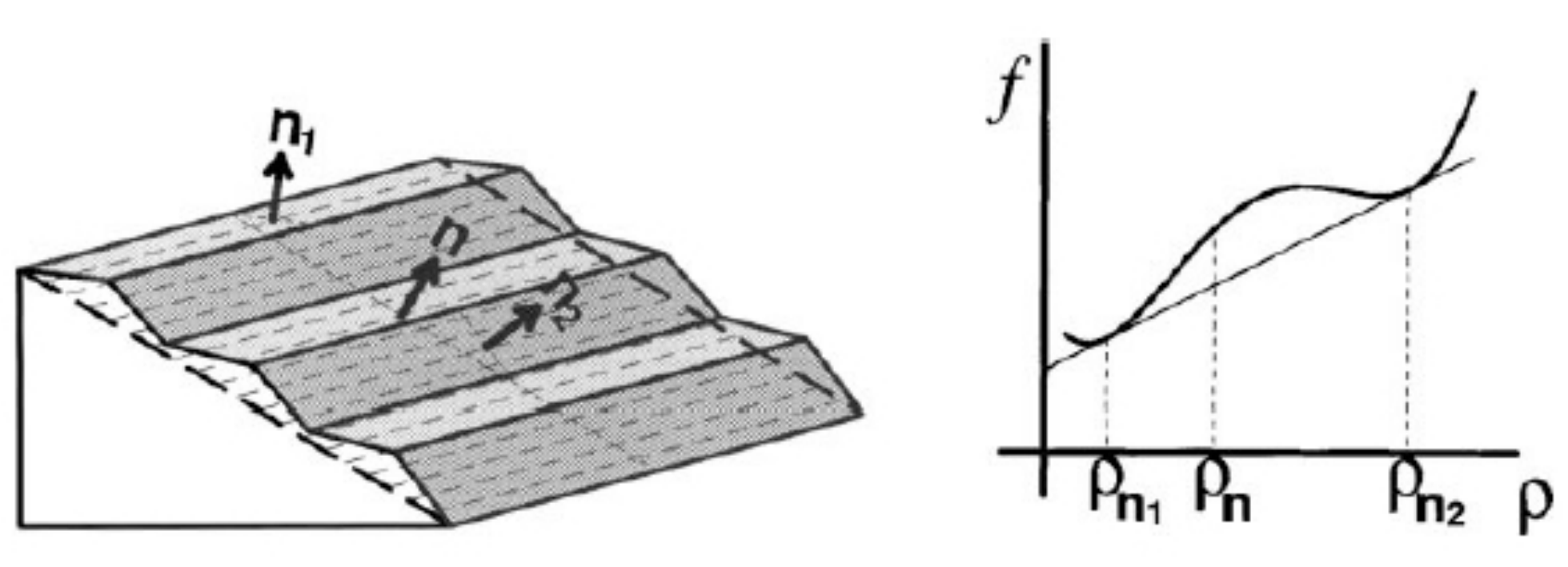}
  \caption{Illustration of how orientational phase separation occurs when a ``hill-and-valley'' structure has a lower total surface free energy per area than a flat surface as in Eq.~(\ref{e:HV}). The sketch of the free energy vs.\ $\rho \equiv \tan \theta$ shows that this situation reflects a region with negative convexity which is accordingly not stable.  The dashed line is the tie bar of a Maxwell or double-tangent construction. The misorientations are the coexisting slat-like planes, with orientations $\mathbf{n}_1$ and $\mathbf{n}_2$, in the hill-and-valley structure. From Jeong and Williams [1999]}
  \label{f:HV}
\end{figure}

\begin{figure}[b]
\centering
  \includegraphics[width=6cm]{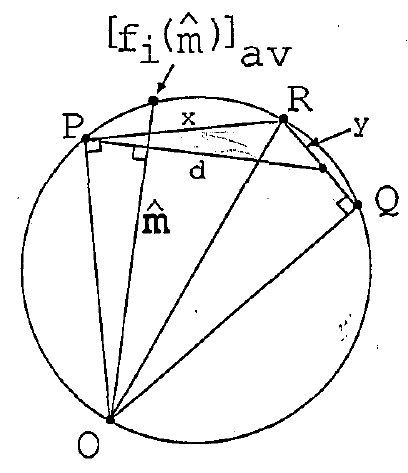}
  \caption{ECS analogue of the Maxwell double-tangent
construction. O is the center of the crystal.
Points P and Q are on the (stable) Wulff plot but the
region between them is unstable; hence, the ECS
follows PRQ and has an edge at R. An interface at
the intermediate orientation $\mathbf{\hat{m}}$ breaks up into the
orientations P and Q with relative proportions x:y; thus,
the average free energy per unit area
given by Eq.~(\ref{e:PQ}), which in turn shows that $f_i(\mathbf{\hat{m}})]_{\rm avr}$
lies on the circle.  From Wortis [1988]}
  \label{f:Wortis27}
\end{figure}

The origin of a hill-and-valley structure from the constituent free energies [Jeong and Williams 1999, Williams and Bartelt1996] is illustrated schematically in Fig.~\ref{f:HV}.  It arises when they satisfy the inequality
\begin{equation}\label{e:HV}
  f_i(\mathbf{\hat{m}}=\mathbf{n}_1) A_1 + f_i(\mathbf{n}_2) A_2 < f_i(\mathbf{n}) A,
\end{equation}
\noindent where $A_1$ and $A_2$ are the areas of strips of orientation $\mathbf{n}_1$ and $\mathbf{n}_2$, respectively, while $A$ is the area of the sum of these areas projected onto the plane bounded by the dashed lines in the figure.  This behavior, again, is consistent with the identification of the misorientation as a density (or magnetization)-like variable rather than a field-like one.

The details of the lever rule for coexistence regimes were elucidated by Wortis [1988]:  As depicted in Fig.~\ref{f:Wortis27}, which denotes as \textsf{P} and \textsf{Q} the two orientations bounding the region that is not stable, the lever-rule interpolations lie on segments of a spherical surface.  Let the edge on the ECS be at \textsf{R}.  Then an interfaced created at a forbidden $\mathbf{\hat{m}}$ will evolve toward a hill-and-valley structure with orientations  \textsf{P} and \textsf{Q} with a free energy per area of
\begin{equation}\label{e:PQ}
  [f_i(\mathbf{\hat{m}})]_{\rm avr} = \frac{\mathsf{x}f_i(\mathsf{P}) + \mathsf{y}f_i(\mathsf{Q})}{d}.
\end{equation}
It can then be shown that $\mathbf{\hat{m}}[f_i(\mathbf{\hat{m}})]_{\rm avr}$ lies on the depicted circle, so that the Wulff plane passes through the edge at \textsf{R}.

\subsection{Going Beyond Wulff Plots}

To determine the limits of forbidden regions, it is more direct and straightforward to carry out a polar plot of $1/f_i(\mathbf{\hat{m}})$ [Frank, 1963] rather than $f_i(\mathbf{\hat{m}})$, as discussed in Sekerka's [2004] review chapter.  Then a sphere passing through the origin becomes a corresponding plane; in particular, a Herring sphere for some point becomes a plane tangent to the plot of $1/f_i(\mathbf{\hat{m}})$.  If the Herring sphere is inside the Wulff plot, then its associated plane lies outside the plot of $1/f_i(\mathbf{\hat{m}})$. If, on the other hand, if some part of the Wulff plot is inside a Herring sphere, the corresponding part of the $1/f_i(\mathbf{\hat{m}})$ plot will be outside the plane.  Thus, if the plot of $1/f_i(\mathbf{\hat{m}})$ is convex, all its tangent planes will lie outside, and all orientations will appear on the ECS.  If it is not convex, it can be made so by adding tangent planes.  The orientations associated with such tangent planes are forbidden, so their contact curve with the $1/f_i(\mathbf{\hat{m}})$ plot gives the bounding stable orientations into which forbidden orientations phase separate.

\begin{figure}
  \centering
  \includegraphics[width=8cm]{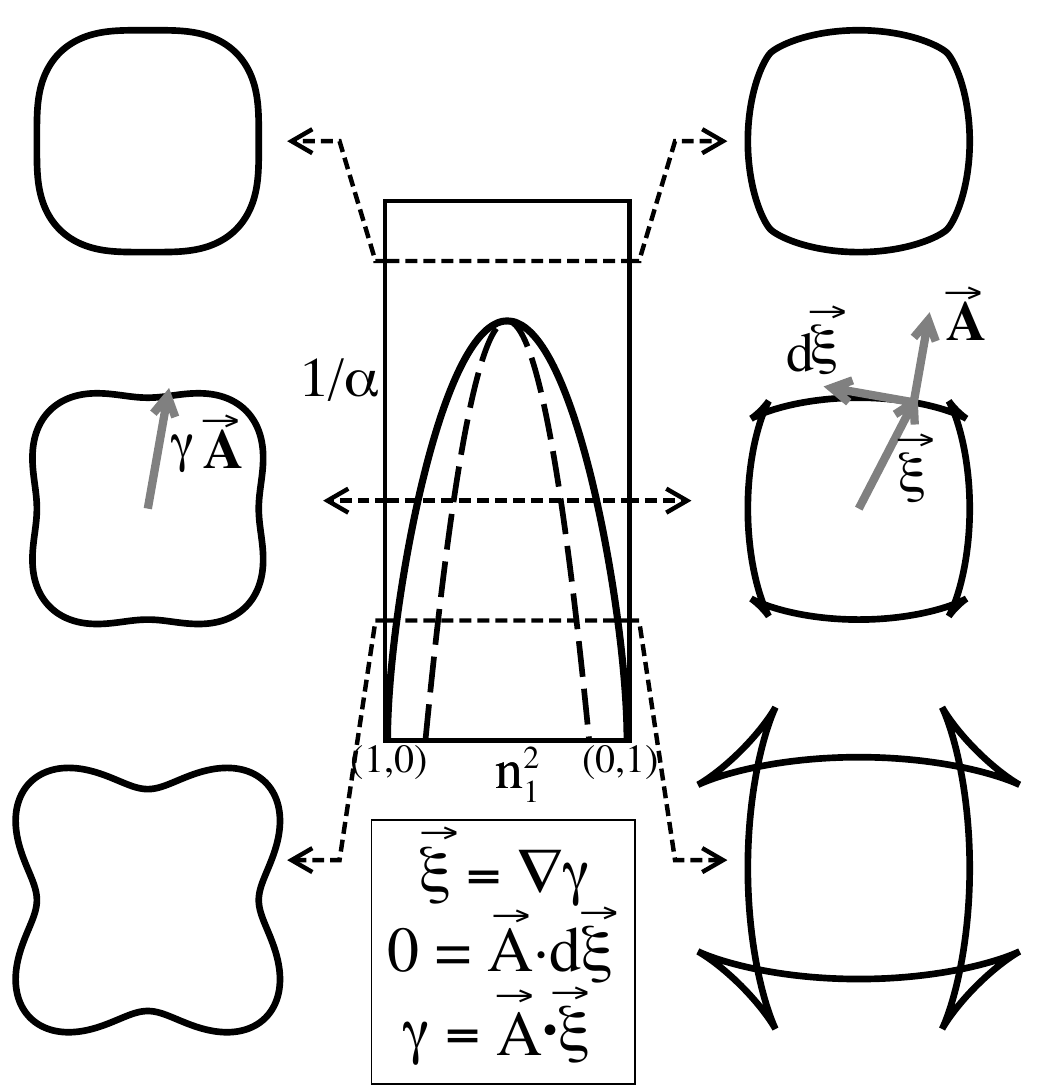}
  \caption{Graphical constructions for an anisotropic $f_i(\mathbf{\hat{m}})$
for various values of an anisotropy parameter $\alpha$, where $f_i \propto 1 + \alpha \cos^2 \theta \sin^2 \theta$. In the left column $f_i(\theta)$ is plotted from top to bottom for $\alpha$ = 1/2, 1, 2. Anisotropy increases with
positive $\alpha$, so 1/$\alpha$  corresponds in some sense to a temperature
in conventional plots. In the center panel, $n_1^2$ is $\cos^2 \theta$.  The shape resulting from the gradient construction with the ears removed is the Wulff ECS.  From Cahn and Carter [1996]}\label{f:CahnXi2D}
\end{figure}

Summarizing the discussion in Sekerka [2004], the convexity of $1/f_i(\mathbf{\hat{m}})$ can indeed be determined analytically since the curvature $1/f_i(\mathbf{\hat{m}})$ is proportional (with a positive-definite proportionality constant) to the stiffness, i.e.\ in 2D, $\gamma + \partial^2 \gamma/\partial \theta^2 = \tilde{\gamma}$, or preferably $\beta + \partial^2 \beta/\partial \theta^2 = \tilde{\beta}$ as in Eq.~(\ref{e:fp-th}) to emphasize that the stiffness and [step] free energy per length have different units in 2D from 3D.  Hence, $1/f_i(\mathbf{\hat{m}})$ is not convex where the stiffness is negative.  The very complicated generalization of this criterion to 3D is made tractable via the $\bm{\xi}$-vector formalism of Hoffman and Cahn [1972, 1974], where $\bm{\xi} = \mathbf{\nabla}(r \, f_i(\mathbf{\hat{m}}))$, where $r$ is the distance from the origin of the $\gamma$ plot.  Thus,

\begin{equation}\label{e:xi}
f_i(\mathbf{\hat{m}}) = \bm{\xi \cdot \hat{m}}, \quad  \mathbf{\hat{m}\,\cdot}\,d\bm{\xi} = 0,
\end{equation}

 \noindent which is discussed well by Wheeler [1999] and Sekerka [2004].  To elucidate the process, we consider just the 2D case [Cahn and Carter 1996]; cf.\ Fig.~\ref{f:CahnXi2D}.

\begin{figure}
  \centering
  \includegraphics[width=8cm]{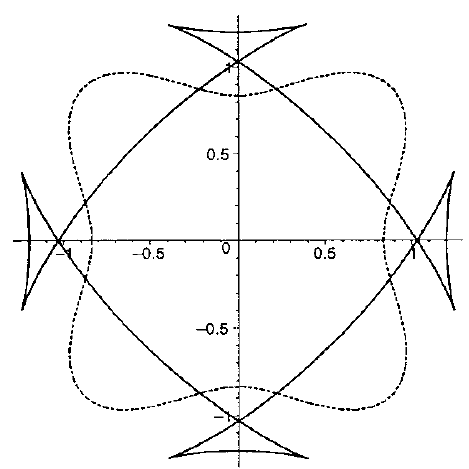}
  \caption{The solid curve is the $\bm{\xi}$ plot, while the dashed curve is the 1/$\gamma$-plot for
$f_i(\mathbf{\hat{m}}) \equiv \gamma \propto 1 + 0.2 \cos 4\theta$. For this case (but not for small values of $a$) the 1/$\gamma$-plot is not convex, and the $\bm{\xi}$ plot forms
``ears."  These ears are then removed, so that the equilibrium shape is given by the interior envelope of the $\bm{\xi}$ plot, in this case
having four corners. From Wheeler [1999]}\label{f:Wheeler2}
\end{figure}

The solid curve in Fig.~\ref{f:Wheeler2} is the $\bm{\xi}$ plot and the dashed curve is the 1/$\gamma$-plot for
$f_i(\mathbf{\hat{m}}) \equiv \gamma \propto 1 + 0.2 \cos 4\theta$. For this case the 1/$\gamma$-plot is not convex and the $\bm{\xi}$ plot forms
``ears.'' The equilibrium shape is given by the interior envelope of the $\bm{\xi}$ plot; in this case
it exhibits four corners.

Pursuing this analogy, we see that if one cleaves a crystal at some orientation $\mathbf{\hat{m}}$ that is not on the ECS, i.e. between $\mathbf{\hat{m}_+}$  and $\mathbf{\hat{m}_-}$, then this orientation will break up into segments with orientations $\mathbf{\hat{m}_+}$  and $\mathbf{\hat{m}_-}$ such that the net orientation is still $\mathbf{\hat{m}}$, providing another example of the lever rule associated with Maxwell double-tangent constructions for the analogous problems.  The time to evolve to this equilibrium state depends strongly on the size of energy barriers to mass transport in the crystalline material; it could be exceedingly long.   To achieve rapid equilibration, many nice experiments were performed on solid hcp $^4$He bathed in superfluid $^4$He, for which equilibration occurs in seconds (Balibar and Castaing [1980], Keshishev et al.\ [1981], Wolf et al.\ [1983, 1985], and many more; see Balibar et al.\ [2005] for a comprehensive recent review.  Longer but manageable equilibration times are found for Si and for Au, Pb, and other soft transition metals.

\section{Some Physical Implications of Wulff Constructions}

\subsection{Thermal Faceting and Reconstruction}

A particularly dramatic example is the case of surfaces vicinal to Si (111) by a few degrees.  In one misorientation direction the vicinal surface is stable above the reconstruction temperature of the (111) facet but below that temperature, $f_i(111)$ decreases significantly so that the original orientation is no longer stable and phase separates into reconstructed (111) terraces and more highly misoriented segments [Phaneuf, 1987; Bartelt, 1989].  The correspondence to other systems with phase separation at first-order transitions is even more robust.  Within the coexistence regime one can in mean field determine a spinodal curve.  Between it and the coexistence boundary one observes phase separation by nucleation and growth, as for metastable systems; inside the spinodal one observes much more rapid separation with a characteristic most-unstable length [Phaneuf, 1993].  This system is discussed further in Section~7.1 below.  Furthermore, there are remarkably many ordered phases at larger misorientations [Olshanetsky and Mashanov 1981, Baski and Whitman 1997].

Wortis [1988] describes  ``thermal faceting" experiments in which metal crystals, typically late transition or noble metal elements like Cu, Ag, and Fe, are cut at a high Miller-index direction and polished.  They are then annealed at high temperatures.  If the initial plane is in a forbidden direction, optical striations, due to hill and valley formation, appear once these structures have reached optical wavelengths.  While the characteristic size of this pattern continues to grow as in spinodal decomposition, the coarsening process is eventually slowed and halted by kinetic limitations.

\begin{figure}
  \centering
  \includegraphics[width=8cm]{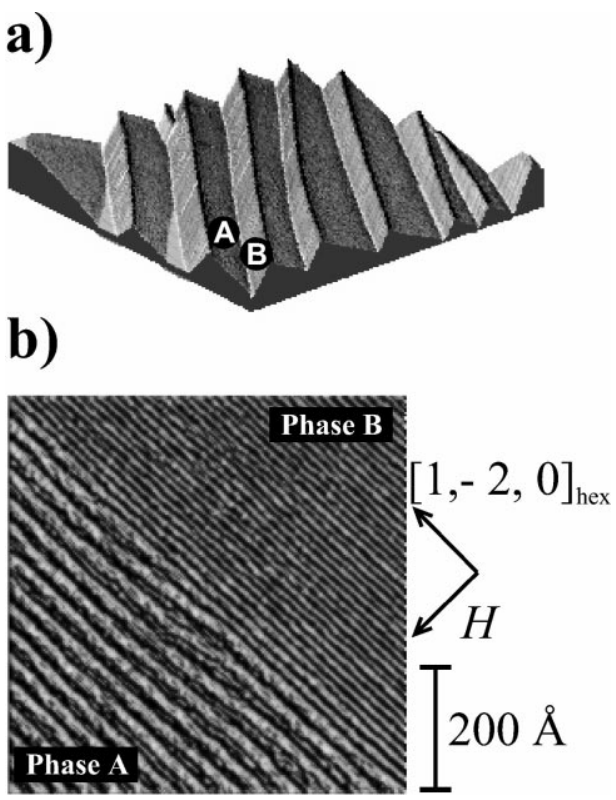}
  \caption{Morphology of the faceted Au(4,5,5) surface measured
at room temperature. (a) 3D plot of a large-scale (scan area: 1.4 mm $\times$ 1.4 mm)
STM image.  Phases A and B form the hill-and-valley morphology. (b) STM
image zoomed in on a boundary between the two phases. All steps
single-height, i.e.\ 2.35\AA\ high.
Phase B has smaller terraces, 13\AA\
wide, while phase A terraces are about 30\AA\ wide.  This particular surface has (2,3,3) orientation. From Rousset [1999]}\label{f:Rousset5}
\end{figure}

There are more recent examples of such phenomena. After sputtering and annealing above 800K, Au(4,5,5) at 300K forms a hill-and-valley structure of two Au(111) vicinal surfaces, one that is reconstructed and the other which is not, as seen in Fig.~\ref{f:Rousset5}.  This seems to be an equilibrium phenomenon: it is reversible and independent of cooling rate [Rousset 1999].  Furthermore, while it has been long known that adsorbed gases can induce faceting on bcc (111) metals [Bonczek et al.\ 1980], ultrathin metal films have also been found to produce faceting of W(111), W(211), and Mo(111) [Madey et al.\ 1996, 1999].

\subsection{Types A \& B}
The above analysis indicates that at $T=0$ the ECS of a crystal is a polyhedron having the point symmetry of the crystal lattice, a result believed to be general for finite-range interactions [Fisher 1983].  All boundaries between facets are sharp edges, with associated forbidden non-facet orientation; indeed, the Wulff plot is just a set of discrete points in the symmetry directions.  At finite temperature, two possibilities have been delineated (with cautions [Wortis 1988], labeled (nonmnemonically) A and B.  In type-A, there are smooth curves between facet planes rather than edges and corners.  Smooth here means, of course, that not only is the ECS continuous, but so is its slope, so that there are no forbidden orientations anywhere.  This situation corresponds to continuous phase transitions.  In type B, in contrast, corners round at finite $T$ but edges stay sharp until some temperature $T_0$.  For $T_0 < T < T_1$ there are some rounded edges and some sharp edges, while above $T_1$ all edges are rounded.

\begin{figure}
  \centering
    \includegraphics[width=8cm]{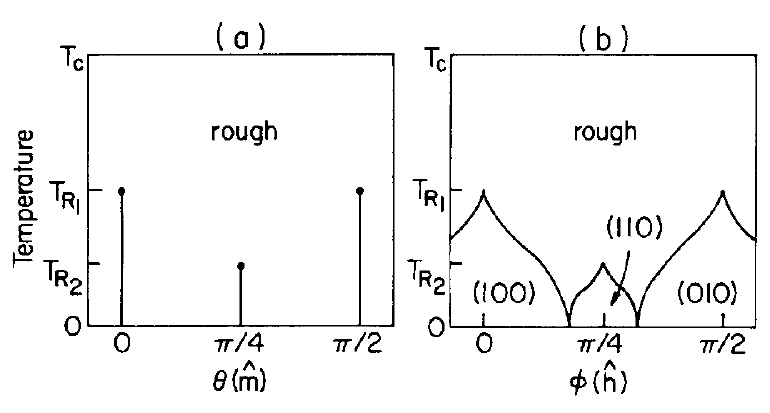}
  \caption{Interfacial phase diagrams for simple-cubic nearest-neighbor
 Kossel crystal with nearest-neighbor as well as (weak) next-nearest-neighbor (NNN)
attractions.  The angular variables $\theta$ and $\phi$ (not to be confused with $\varphi$, cf.\ Section \ref{s:gen}) interfacial orientation ($\mathbf{\hat{m}}$) and ECS ($\mathbf{\hat{h}}$), respectively, in an
equatorial section of the full 3D phase diagram. a) The $T-\theta$ phase diagram (a) shows the locus of cusps in the Wulff plot along
the symmetry directions below the respective roughening temperatures.  For no NNN interaction ($\epsilon_2 = 0$), there are only cusps at vertical lines at 0 and $\pi/2$.  b) The $T-\mathbf{\hat{h}}$ phase diagram gives the faceted areas of the crystal shape.
The NNN attraction leads to additional (111) [not seen in the equitorial plane] and (110) facets at low enough temperature. Thus, for $\epsilon_2 = 0$ the two bases of the (100) and (010) phases meet and touch each other at (and only at) $\phi = \pi/4$ (at $T=0$), with no intervening (110) phase.   Each
type of facet disappears at its own roughening temperature.  Above
the phase boundaries enclosing those regions, the crystal surfaces
are smoothly curved (i.e., thermodynamically ``rough").  This behavior is consistent with the observed phase diagram of hcp $^4$He. From Rottman and Wortis [1984a,b]}\label{f:Rott9}
\end{figure}

\begin{figure}
  \centering
    \includegraphics[width=8cm]{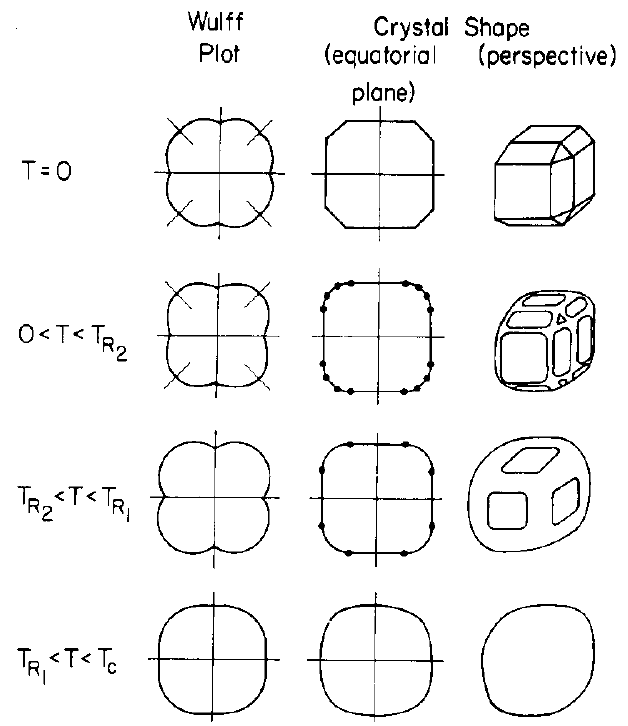}
  \caption{Representative Wulff plots and ECS's for the crystal with weak NNN attractions whose phase diagram is shown in Fig.\ref{f:Rott9}. At low enough
temperature there are (100), (110) and (111) facets. For weak attraction the (110) and (111)
facets roughen away below the (100) roughening temperature.  For $\epsilon_2 = 0$, $T_{R_2}=0$, so that the configurations in the second row do not occur; in the first row, the octagon becomes a square and the perspective shape is a cube. Facets are separated at $T > 0$ by curved surfaces, and all transitions
are second order. Spherical symmetry obtains as $T$ approaches melting at $T_c$. From Rottman and Wortis [1984]}\label{f:Rott10}
\end{figure}

\begin{figure}[t]
  \centering
    \includegraphics[width=8cm]{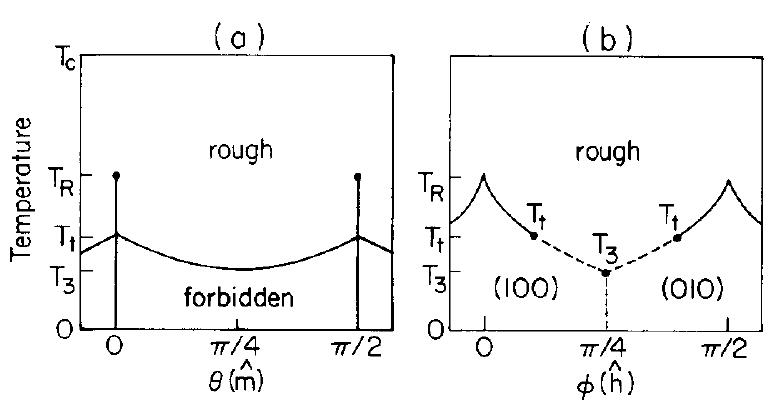}
  \caption{Interfacial phase diagram with (weak) next-nearest-neighbor (NNN) repulsion rather than attraction as in Fig.~\ref{f:Rott9}. The NNN repulsion stabilizes the (100) facets. Curved surfaces first appear at the cube corners and
then reach the equatorial plane at $T_3$. The transition at the equator
remains first order until a higher temperature $T_t$. The dotted boundaries are first order. A forbidden (coexistence)
region appears in the $T-\mathbf{\hat{h}}$ phase diagram. From Rottman and Wortis [1984a,b]}\label{f:Rott11}
\end{figure}

\begin{figure}
  \centering
    \includegraphics[width=8cm]{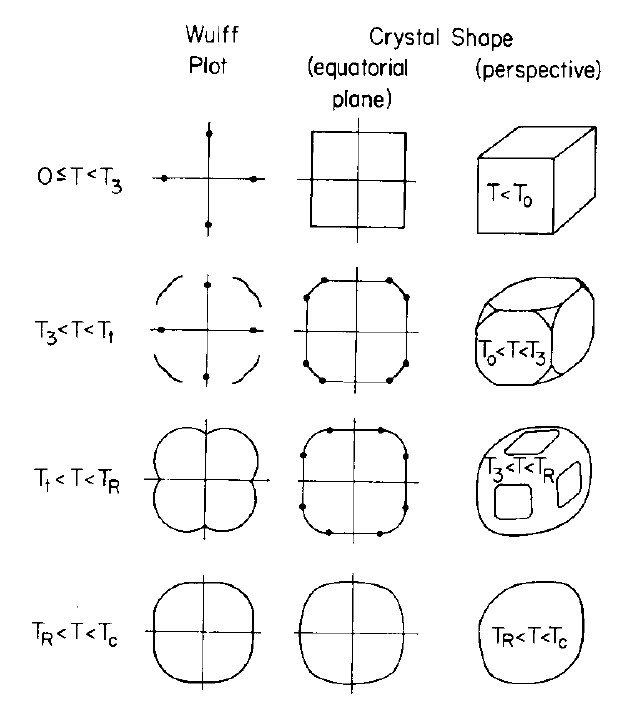}
  \caption{Representative Wulff plots and ECS's for the crystal with weak NNN repulsions whose phase diagram is shown in Fig.\ref{f:Rott11}.  Curved surfaces appear
first at the cube corners. Junctions between facets and curved
surfaces may be either first- or second-order  (sharp or smooth), depending on orientation
and temperature. From Rottman and Wortis [1984a,b]}\label{f:Rott12}
\end{figure}

Rottman and Wortis [1984a,b] presents a comprehensive catalogue of the orientation phase diagrams, Wulff plots, and ECSs for the cases of non-existent, weakly attractive, and weakly repulsive next-nearest-neighbor (NNN) bonds in 3D.  Figs.~\ref{f:Rott9} and \ref{f:Rott10} show the orientation phase diagrams and the Wulff plots with associated ECSs, respectively, for weakly attractive NNN bonds.  As indicated in the caption, it is easy to describe what then happens when $\epsilon_2 = 0$ and only \{100\} facets occur.  Likewise, Figs.~\ref{f:Rott11} and \ref{f:Rott12} show the orientation phase diagrams and the Wulff plots with associated ECSs, respectively, for weakly repulsive NNN bonds.

\subsection{2D Studies}
\label{s:2D}

Exploring the details is far more transparent in 2D than 3D.  The 2D case is physically relevant in that it describes the shape of islands of atoms of some species at low fractional coverage on an extended flat surface of the same or another material. An entire book is devoted to 2D crystals [Lyuksyutov et al 1992].  The 2D perspective can also be applied to cylindrical surfaces in 3D, as remarked by Nozi\`eres [1992].  Formal proof is also more feasible, if still arduous, in 2D; an entire book is devoted to this task [Dobrushin et al.\ 1992]; see also Pfister [1991] and Miracle-Sole and Ruiz [1994].

For the 2D nearest-neighbor Kossel crystal described above Wortis [1988] notes that at $T=0$ a whole class of Wulff planes pass through a corner.  At finite $T$ thermal fluctuations lift this degeneracy and the corner rounds, leading to type A behavior.  To gain further insight, we now include a next-nearest-neighbor (NNN) interaction $\epsilon_2$, so that
\begin{equation}\label{e:Kossel2}
f_i(\theta) = \frac{\epsilon_1 + \epsilon_2}{2}\left(|\cos \theta| + |\sin \theta|\right) + \frac{\epsilon_2}{2}\left(|\cos \theta| - |\sin \theta|\right)
\end{equation}
For favorable NNN bonds, i.e. $\epsilon_2 > 0$, one finds new \{11\} facets but still type-A behavior with sharp edges, while for unfavorable NNN bonds, i.e. $\epsilon_2 < 0$, there are no new facets but for finite $T$ the edges are no longer degenerate so that type-B behavior obtains.   Again recalling that $f_i(\theta) = f_p(\theta)\, |\cos \theta|$, we can identify $f_0 = \epsilon_2 + \epsilon_1/2$ and $\beta/h = \epsilon_1/2$, as noted in other treatments, e.g.\ Dieluweit et al. [2003]. That work, however, finds that such a model cannot adequately account for the orientation-dependent stiffness of islands on Cu(001).  Attempts to resolve this quandry using 3-site non-pairwise (trio) interactions [Stasevich et al.\ 2004, 2006] did not prove entirely satisfactory.   In contrast, on the hexagonal Cu(111) surface, only NN interactions are needed to account adequately for the experimental data [Stasevich et al.\ 2005, 2006]. In fact, for the NN model on a hexagonal grid, Zia [1986] found an exact and simple, albeit implicit, exact expression for the ECS.  However, on such (111) surfaces (and basal planes of hcp crystals) lateral pair interactions alone cannot break the symmetry to produce a difference in energies between the two kinds of step edges, viz.\ \{100\} and \{111\}) microfacets (A and B steps, respectively, with no relation to types A and B!).  The simplest viable explanation is an orientation-dependent trio interaction; calculations of such energies support this idea [Stasevich et al.\ 2005, 2006].

Strictly speaking, of course, there should be no 2D facet (straight edge) and accompanying sharp edges (corners) at $T > 0$  [Gallavotti 1972; Abraham and Reed 1974, 1976; Avron et al.\ 1982 and references therein] since that would imply 1D long-range order, which should not occur for short-range interactions.  Measurements of islands at low temperatures show edges that appear to be facets and satisfy Wulff corollaries such as that the ratio of the distances of two unlike facets from the center equals the ratio of their $f_i$ [Kodambaka et al.\ 2006].  Thus, this issue is often just mentioned in passing [Michely and Krug, 2004] or even ignored.  On the other hand, sophisticated approximations for $f_i(\theta)$ for the 2D Ising model, including NNN bonds have been developed, e.g. [Stasevich and Einstein 2007], allow numerical tests of the degree to which the ECS deviates from a polygon near corners of the latter.  One can also gauge the length scale at which deviations from a straight edge come into play by using that the probability per atom along the edge for a kink to occur is essentially the Boltzmann factor associated with the energy to create the kink [Weeks 2014].

Especially for heteroepitaxial island systems (when the island consists of a different species from the substrate), strain plays an important if not dominant role.  Such systems have been investigated, e.g., by F. Liu [2006], who points out that for such systems the shape does not simply scale with $\lambda$, presumably implying the involvement of some new length scale[s].  A dramatic manifestation of strain effects is the island shape transition of Cu on Ni(001), which changes from compact to ramified as island size increases [M\"uller et al.\ 1998].  For small islands, additional quantum-size and other effects lead to favored island sizes (magic numbers).

\section{Vicinal Surfaces--Entr\'ee to Rough Regions Near Facets}

In the rough regions the ECS is a vicinal surface of gradually evolving orientation.  To the extent that a local region has a particular orientation, it can be approximated as an infinite vicinal surface.  The direction perpendicular to the terraces (which are densely-packed facets) is typically called $\hat{z}$.  In ``Maryland notation" (cf.\ \S\ref{s:gen})the normal to the vicinal surface lies in the $x-z$ plane, and the distance $\ell$ between steps is measured along $\hat{x}$, while the steps run along the $\hat{y}$ direction.  In the simplest and usual approximation, the repulsions between adjacent steps arise from two sources: an entropic or steric interaction due to the physical condition that the steps cannot cross, since overhangs cannot occur in nature.  The second comes from elastic dipole moments due to local atomic relaxation around each step, leading to frustrated lateral relaxation of atoms on the terrace plane between two steps.  Both interactions are $\propto 1/\ell^2$.

The details of the distribution $\check{P}(\ell)$ of spacings between steps have been reviewed in many places [Jeong and Williams 1999, Einstein et al.\ 2001, Giesen 2001, Einstein 2007]   The average step separation $\la \ell \ra$ is the only characteristic length in the $\hat{x}$ direction.  N.B., $\la \ell \ra$ need not be a multiple of, or even simply related to, the substrate lattice spacing.  Therefore, we consider $P(s)= \la \ell \ra^{-1}\check{P}(\ell)$, where $s\equiv \ell/\la \ell \ra$, a dimensionless length.  For a ``perfect" cleaved crystal, $P(s)$ is just a spike $\delta (s-1)$.  For straight steps placed randomly at any position with probability $1/\la \ell \ra$, $P(s)$ is a Poisson distribution $\exp(-s)$.  Actual steps do meander, as one can study most simply in a terrace-step-kink (TSK) model.  In this model, the only excitations are kinks (with energy $\epsilon$) along the step.  (This is a good approximation at low temperature $T$ since adatoms or vacancies on the terrace cost several $\epsilon_1$ [4$\epsilon_1$ in the case of a simple cubic lattice].)  The entropic repulsion due to step meandering dramatically decreases the probability of finding adjacent steps at $\ell \ll \la \ell \ra$.  To preserve the mean of one, $P(s)$ must also be smaller than $\exp(-s)$ for large $s$.

\begin{figure}[t]
\includegraphics[width=8.5cm]{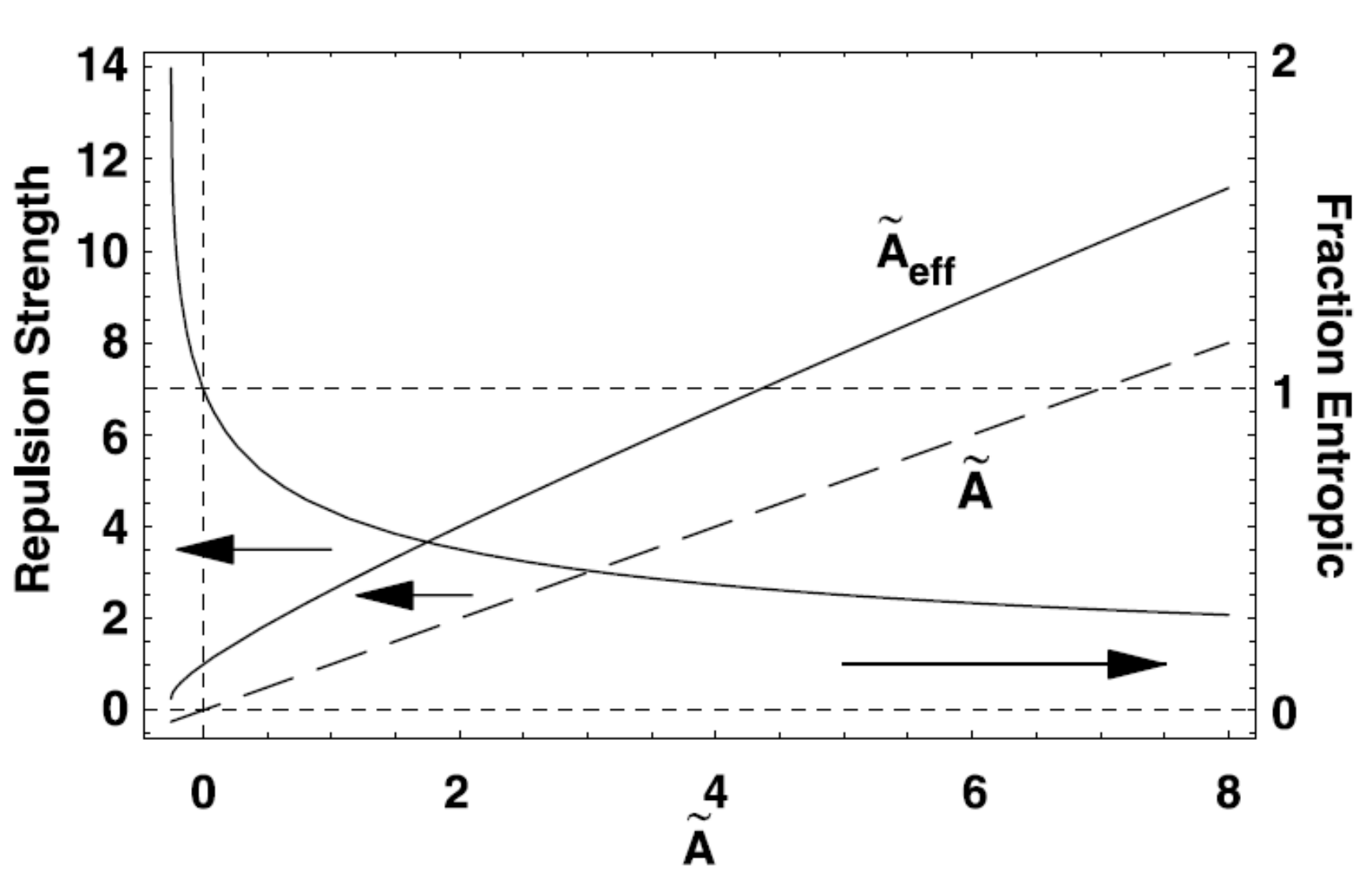}
\caption {Illustration of how entropic repulsion and energetic interactions combine, plotted vs.\ the dimensionless energetic interaction strength $\tilde A \equiv A\tilde \beta/(k_BT)^2$.  The dashed straight line is just $\tilde A$.  The solid curve above it is the combined entropic and energetic interactions, labeled $\tilde A_{\rm eff}$ for reasons explained below.  The difference between the two curves at any value of the abscissa is the dimensionless entropic repulsion for that $\tilde A$. The decreasing curve, scaled on the right ordinate, is the ratio of this entropic repulsion to the total dimensionless repulsion $\tilde A_{\rm eff}$.  It falls monotonically with $\tilde A$, passing through unity at $\tilde A = 0$.  See the discussion accompanying Eq.~(\ref{e:Aeff}) for more information and explicit expressions for the curves.  From Ei07.
}
{\label{f:enteng}}
\end{figure}

If there is an additional energetic repulsion $A/\ell^2$, the magnitude of the step meandering will decrease, narrowing $P(s)$.  As $A \rightarrow \infty$, the width approaches 0 ($P(s)\rightarrow \delta (s-1)$, the result for perfect crystals).  We emphasize that the energetic and entropic interactions do not simply add.  In particular, there is no negative (attractive) value of $A$ at which the two cancel each other.  (Cf.\ Eq.~(\ref{e:gA}) below.)  Thus, for strong repulsions, steps rarely come close, so the entropic interaction plays a smaller role, while for $A<0$, the entropic contribution increases, as illustrated in Fig.~\ref{f:enteng} and explicated below.  We emphasize that the potentials of both interactions decay as $\ell^{-2}$ (cf.\ Eq.~(\ref{e:gA} below), in contrast to some statements in the literature (in papers analyzing ECS exponents) that entropic interactions are strictly short range while energetic ones are long-range.

Investigation of the interaction between steps has been reviewed well in several places [Jeong and Williams 1999, Giesen 2001, Nozi\`eres 1999, Einstein 1996, Einstein et al.\ 2001].
The earliest studies seeking to extract $A$ from terrace-width distributions (TWDs) used the mean-field-like Gruber-Mullins [Gruber and Mullins 1967] approximation, in which a single active step fluctuates between two fixed straight steps $2\la \ell \ra$ apart.  Then the energy associating with the fluctuations $x(y,t)$ is
\begin{equation}
 \Delta {\cal E} = -\beta(0)L_y + \int_0^{L_y} \beta(\theta(y))\sqrt{1+\left(\frac{\partial x}{\partial y}\right)^2} dy ,
\label{e:fluc}
\end{equation}
\noindent where $L_y$ is the size of the system along the mean step direction (i.e. the step length with no kinks).  We expand $\beta(\theta)$ as the Taylor series $\beta(0) + \beta^\prime(0)\theta +
\nicefrac{1}{2}\,\beta^{\prime\prime}(0)\theta^2$ and recognize that the length of the line segment has increased from $dy$ to $dy/\cos \theta \approx dy(1 + \nicefrac{1}{2}\,\theta^2)$.  For close-packed steps, for which $\beta^\prime(0)=0$, it is well known that  (using $\theta \approx \tan \theta = \partial x/\partial y$)

\begin{equation}
 \Delta {\cal E} \approx \frac{\tilde\beta(0)}{2}\int_0^{L_y}\! \left(\frac{\partial x}{\partial y}\right)^2 \! dy, \quad  \tilde\beta(0) \equiv \beta(0)+\beta^{\prime\prime}(0),
\label{e:stiff}
\end{equation}
\noindent where $\tilde\beta$ is the step stiffness [Fisher et al. 1982]. N.B., the stiffness $\tilde\beta(\theta)$ has the same definition for steps with arbitrary in-plane orientation --- for which $\beta^{\prime}(\theta) \ne 0$ --- because to create such steps, one must apply a ``torque" [Leamy et al.\ 1975] which exactly cancels $\beta^{\prime}(\theta)$. (See Stasevich [2006, 2007] for a more formal proof.)

Since $x(y)$ is taken to be a single-valued function that is defined over the whole domain of $y$, the 2D configuration of the step can be viewed as the worldline of a particle in 1D by recognizing $y$ as a time-like variable.  Since the steps cannot cross, these particles can be described as spinless fermions in 1D, as pointed out first by de Gennes [1968] in a study of polymers in 2D. Thus, this problem can be mapped into the Schr\"odinger equation in 1D:  $\partial x/\partial y$ in Eq.~(\ref{e:stiff}) becomes $\partial x/\partial t$, with the form of a velocity, with the stiffness playing the role of an inertial mass.  This correspondence also applies to domain walls of adatoms on densely-covered crystal surfaces, since these walls have many of the same properties as steps.  Indeed, there is a close correspondence between the phase transition at smooth edges of the ECS and the commensurate-incommensurate phase transitions of such overlayer systems, with the rough region of the ECS corresponding to the incommensurate regions and the local slope related to the incommensurability [Pokrovsky and Talapov 1979, 1984, Villain 1980, Haldane and Villain 1981, Schulz et al.\ 1982].  Jayaprakash et al.\ [1984] provide the details of the mapping from a TSK model to the fermion picture, complete with fermion creation and annihilation operators.

In the Gruber-Mullins [1967] approximation, a step with no energetic interactions becomes a particle in a 1D infinite-barrier well of width $2\la \ell \ra$, with well-known ground-state properties

\begin{equation}
 \psi_0(\ell) \!  \propto \sin\left(\frac{\pi \ell}{2\la \ell \ra}\right); \; P(s)=\sin^2\left(\frac{\pi s}{2}\right); \; E_0= \frac{(\pi k_BT)^2}{8\tilde\beta\la \ell \ra^2}
\label{e:1Dfree}
\end{equation}
\noindent Thus, it is the kinetic energy of the ground state in the quantum model that corresponds to the entropic repulsion (per length) of the step.  In the exact solution for the free energy expansion of the equilibrium crystal shape [Akutsu et al.\ 1988], the numerical coefficient in the corresponding term is $\nicefrac{1}{6}$ rather than $\nicefrac{1}{8}$.  Note that $P(s)$ peaks at $s=1$ and vanishes for $s\ge 2$.

Suppose, next, that there is an energetic repulsion $U(\ell) = A/\ell^2$ between steps.  In the 1D Schr\"odinger equation, the prefactor of $-\partial^2 \psi(\ell)/\partial \ell^2$ is $(k_BT)^2/2\tilde\beta$, with the thermal energy $k_BT$ replacing $\hbar$.  (Like the repulsions, this term has units $\ell^{-2}$.) Hence, $A$ only enters the problem in the dimensionless combination $\tilde{A} \equiv A\tilde\beta/(k_BT)^2$ [Jeong and Weeks 1999].  In the Gruber-Mullins picture, the potential (per length) experienced by the single active particle is (with $\check\ell \equiv \ell - \la \ell \ra$):
\begin{equation}
\tilde{U}(\check\ell) = \frac{\tilde{A}}{(\check\ell\! -\! \la \ell \ra)^2} + \frac{\tilde{A}}{(\check\ell \! +\! \la \ell \ra)^2}
= \frac{2\tilde{A}}{\la \ell \ra^2} +\frac{6\tilde{A}\check\ell^2}{\la \ell \ra^4} +\mathcal{O}\left(\frac{\tilde{A}\check\ell^4}{\la \ell \ra^6}\right)
\label{e:sho}
\end{equation}

\noindent The first term is just a constant shift in the energy.  For $\tilde{A}$ sufficiently large, the particle is confined to a region $|\check\ell| \ll \la \ell \ra$, so that we can neglect the fixed walls and the quartic term, reducing the problem to the familiar simple harmonic oscillator, with the solution:

\begin{equation}
\! \psi_0(\ell) \!  \propto {\rm e}^{-\check\ell^2/4 w^2}; \;
 P_{\rm G}(s) \equiv \frac{1}{\sigma_G \sqrt{2\pi}}
              \exp\left[-\frac{(s-1)^2}{2\sigma_G^2}\right]
\label{e:1Dsho}
\end{equation}
\noindent where $\sigma_G = (48\tilde A)^{-1/4}$ and $w =\sigma_G \la \ell \ra$.

For $\tilde A$ of 0 or 2, the TWD can be computed exactly (See below).  For these cases, Eqs.~(\ref{e:1Dfree}) and (\ref{e:1Dsho}), respectively, provide serviceable approximations.  It is Eq.~(\ref{e:1Dsho}) that is prescribed for analyzing TWDs in the most-cited resource on vicinal surfaces [Jeong and Williams 1999].  Indeed, it formed the basis of initial successful analyses of experimental STM (scanning tunneling microscopy) data [Wang et al.\ 1990].  However, it has some notable shortcomings.  Perhaps most obviously, it is useless for small but not vanishing $\tilde A$, for which the TWD is highly skewed, not resembling a Gaussian, and the peak, correspondingly, is significantly below the mean spacing.  For large values of $\tilde A$, it significantly underestimates the variance or, equivalently, the value of $\tilde A$ one extracts from the experimental TWD width [Ihle et al.[1998]]: in the Gruber-Mullins approximation the TWD variance is the same as that of the active step, since the neighboring step is straight.  For large $\tilde A$ the fluctuations of the individual steps on an actual vicinal surface become relatively independent, so the variance of the TWD is the {\it sum} of the variance of each, i.e. twice the step variance.  Given the great (quartic) sensitivity of $\tilde A$ to the TWD width, this is problematic.  As experimentalists acquired more high-quality TWD data, other approximation schemes were proposed, all producing Gaussian distributions with widths $\propto \tilde A^{-1/4}$, but with proportionality constants notably larger than $48^{-1/4}= 0.38$.

For the ``free-fermion" ($\tilde A = 0$) case, Jo\'os et al.\ [1991] developed a sequence of analytic approximants to the exact but formidable expression [Dyson 1962, Mehta 2004] for $P(s)$.  They as well as a slightly earlier paper [Bartelt et al.\ 1990] draw the analogy between the TWD of vicinal surfaces and the distribution of spacings between interacting (spinless) fermions on a ring, the Calogero-Sutherland model Calogero 1969, Sutherland 1971, which in turn for three particular values of the interaction---in one case repulsive ($\tilde{A} = 2$), in another attractive ($\tilde{A} = -\nicefrac{1}{4}$), and lastly the free-fermion case ($\tilde{A} = 0$)---could be solved exactly by connecting to random matrix theory [Mehta 2004, Dyson 1970, Einstein 2007];   Fig.~5 of [Jo\'os 1991] depicts the three resulting TWDs.

These three cases can be well described by the Wigner surmise, for which there are many excellent reviews [Guhr et al.\ 1998, Mehta 2004, Haake 1991].  Explicitly, for $\varrho = 1$, 2, and 4,

\begin{equation}
  \label{e:Wigner}
   P_\varrho(s) =
    a_\varrho s^{\varrho} \exp \left(-b_\varrho s^2\right) \, ,
\end{equation}
\noindent where the subscript of $P$ refers to the exponent of $s$. In random matrix literature, the exponent of $s$, viz.\ 1, 2, or 4, is called $\beta$, due to an analogy with inverse temperature in one justification.  However, to avoid possible confusion with the step free energy per length $\beta$ or the stiffness $\tilde\beta$ for vicinal surfaces, I have sometimes called it instead by the Greek symbol that looked most similar, $\varrho$, and do so in this chapter.
The constants $b_\varrho$, which fixes its mean at
unity, and $a_\varrho$, which normalizes $P(s)$, are
\begin{equation}
    b_\varrho =
  \left[\frac{\Gamma \left(\frac{\varrho +2}{2}\right)}
             {\Gamma \left(\frac{\varrho +1}{2}\right)}\right]^2
\quad
    a_\varrho = \frac{2\left[\Gamma
\left(\frac{\varrho +2}{2}\right)\right]^{\varrho +1} }
                  { \left[\Gamma
\left(\frac{\varrho +1}{2}\right)\right]^{\varrho +2} } =
\frac{2b_\varrho^{(\varrho+1)/2}}{\Gamma \left(\frac{\varrho +1}{2}\right)}
\label{e:abr}
\end{equation}
\noindent Specifically, $b_\varrho = \pi/4$, $4/\pi$, and $64/9\pi$, respectively, while $a_\varrho = \pi/2$, $32/\pi^2$, and $(64/9\pi)^3$, respectively.

As seen most clearly by explicit plots, e.g.\ Fig.~4.2a of
Haake [1991], $P_1(s)$, $P_2(s)$, and $P_4(s)$ are
excellent approximations of the exact results for orthogonal,
unitary, and symplectic ensembles, respectively, and these simple
expressions are routinely used when confronting experimental data in a broad range of physical problems
[Guhr et al.\ 1998, Haake 1991].  (The agreement is particularly outstanding
for $P_2(s)$ and $P_4(s)$, which are the germane cases for vicinal
surfaces, significantly better than any other approximation [Gebremariam 2004].

Thus, the Calogero-Sutherland model provides a connection between random matrix theory, notably the Wigner surmise, and the distribution of spacings between fermions in 1D interacting with dimensionless strength $\tilde A$.  Specifically,
\begin{equation}
  \label{e:avrh}
  \tilde{A} = \frac{\varrho}{2}\left(\frac{\varrho}{2} -1\right)
     \quad \Leftrightarrow \quad \varrho = 1 +\sqrt{1 +4 \tilde{A}}.
\end{equation}
For an arbitrary system, there is no reason that $\tilde{A}$ should take on one of the three special values.  Therefore, we have used Eq.~(\ref{e:avrh}) for arbitrary $\varrho$ or $\tilde{A}$, even though there is no symmetry-based
justification of distribution based on the Wigner surmise of Eq.\ (\ref{e:Wigner}), and refer hereafter to this formula Eqs.\
(\ref{e:Wigner},\ref{e:abr}) as the GWD (generalized Wigner distribution).
Arguably the most convincing argument is a comparison of the predicted variance with numerical data generated from Monte Carlo simulations.  See Einstein [2007] for further discussion.

There are several alternative approximations that lead to a description of the TWD as a Gaussian.  Ihle, Pierre-Louis, and Misbah [1998], in particular, focus on the limit of large $\varrho$, neglecting the entropic interaction in that limit.  The variance $\sigma^2 \propto \tilde A^{-1/2}$, the proportionality constant is 1.8 times that in the Gruber-Mullins case.  This approximation is improved, especially for repulsions that are not extremely strong, by including the entropic interaction in an average way.  This is done by replacing $\tilde A$ by

\begin{equation}
  \label{e:Aeff}
  \tilde{A}_{\rm eff} = \left( \frac{\varrho}{2}\right)^2 = \tilde A +\frac{\varrho}{2}.
\end{equation}
\noindent
Physically,  $\tilde{A}_{\rm eff}$ gives the full strength of the inverse-square repulsion between steps, i.e. the modification due to the inclusion of entropic interactions.  Thus, in Eq.~(\ref{e:fp-th})

\begin{equation}
  \label{e:gA}
  g(T) = \frac{(\pi k_BT)^2}{6 h^3 \tilde{\beta}}\tilde{A}_{\rm eff}
  = \frac{(\pi k_BT)^2}{24 h^3 \tilde{\beta}}\left[ 1 +\sqrt{1\! +\! 4 \tilde{A}}   \right]^2.
\end{equation}

From Eq.~(\ref{e:Aeff}) it is obvious that the contribution of the entropic interaction, viz. the difference between the total and the energetic interaction, as discussed in conjunction with Fig.~\ref{f:enteng}, is $\varrho/2$.  Remarkably, the ratio of the entropic interaction to the total interaction is $(\varrho/2)/(\varrho/2)^2=2/\varrho$; this is the fractional contribution that is plotted in Fig.~\ref{f:enteng}.

\section{Critical Behavior of Rough Regions Near Facets}

\subsection{Theory}

\begin{figure}
\centering
  \includegraphics[width=7cm]{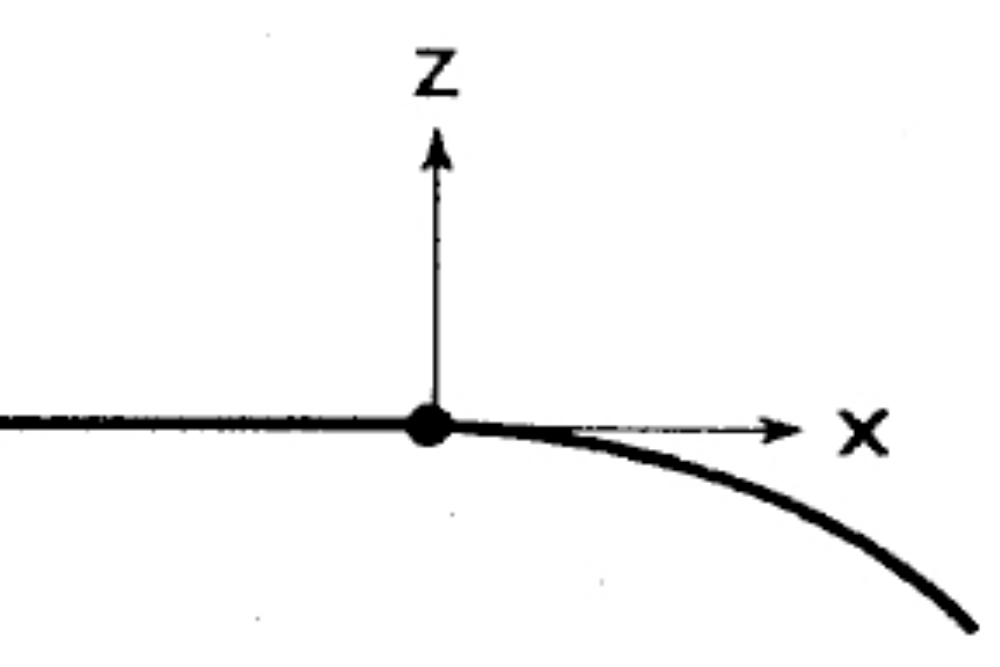}
  \caption{Critical behavior of the crystal shape near a smooth (second-order) edge, represented by the dot at $(x_0,z_0)$. The temperature is lower than the roughening temperature of the facet orientation, so that the region to the left of the dot is flat.  The curved region to the right of the dot correspond to a broad range of rough orientations.
In the thermodynamic limit, the shape of the smoothly curved region near the edge is described by the power law $z \sim z_0 -(x-x_0)^\vartheta$.
Away from the edge there are ``corrections to scaling", i.e. higher-order terms  (cf.\ Eq.~(\ref{e:zx}). For an actual crystal of any finite size, there is  ``finite-size rounding" near the edge, which smooths the singular behavior. Adapted from Jayaprakash and Saam [1984]}
  \label{f:JaySaamFE}
\end{figure}

Assuming (cf.\ Fig.~\ref{f:JaySaamFE}) $\mathbf{\hat{z}}$ the direction normal to the facet and $(x_0,z_0)$ denote the facet edge,  $z \sim z_0 -(x-x_0)^\vartheta$ for $x \ge x_0$. We show that the critical exponent $\vartheta$\footnote{The conventional designation of this exponent is $\lambda$ or $\theta$. However, these Greek letters are the Lagrange multiplier of the ECS and the polar angle, respectively.  Hence, we choose $\vartheta$ for this exponent.} has the value 3/2 for the generic smooth edge described by Eq.~(\ref{e:fp-th}) (with the notation of Eq.~(\ref{e:cusp})):.

\begin{equation}\label{e:fp-thB}
 f_p(p) = f_0 + Bp +gp^3 + cp^4.
\end{equation}
Then we perform a Legendre transformation [Callen, 1985] as in Andreev [1981] (foreshadowed in Landau and Lifshitz 1980) and Jayaprakash and Saam [1984]; explicitly
\begin{equation}\label{e:Legendre1}
  \frac{f_p(p) - \tilde{f}(\eta)}{p} = \left[ \frac{df_p}{dp} \equiv \eta \right] = B +3gp^2 + 4cp^3
  \end{equation}
Hence
\begin{equation}\label{e:Legendre2}
  \tilde{f}(\eta) = f_0 -2gp^3(\eta) -3cp^4(\eta)
\end{equation}
But from Eq.~(\ref{e:Legendre1})
\begin{equation}\label{e:Legendre3}
  p = \left(\frac{\eta \! -\! B}{3g} \right)^{1/2}\left[ 1 - \frac{2c}{3g}\left(\frac{\eta \! -\! B}{3g} \right)^{1/2} + \ldots \right]
\end{equation}
Inserting this into Eq.~(\ref{e:Legendre2}) gives
\begin{equation}\label{e:Legendre14}
 \tilde{f}(\eta) = f_0 -2g \left(\frac{\eta \! -\! B}{3g} \right)^{3/2} + c \left(\frac{\eta \! -\! B}{3g} \right)^2 + {\cal O} \left(\frac{\eta \! -\! B}{3g} \right)^{5/2}
\end{equation}
for $\eta \ge B$ and $\tilde{f}(\eta) = f_0$ for $\eta \le B$. (See Jayaprakash et al.\ [1983]; Jayaprakash and Saam [1984], van Beijeren and Nolden [1987].) Note that this result is true not just for the free fermion case but even when steps interact.  Jayaprakash et al.\ [1984] further show that the same $\vartheta$ obtains when the step-step interaction decreases with a power law in $\ell$ that is greater than 2.  We identify $\tilde{f}(\eta)$ with $r(\mathbf{\hat{h}})$, i.e.\  the magnetic-field-like variable discussed corresponds to the so-called Andreev field $\eta$. Writing $z_0 = f_0/\lambda$ and $x_0 = B/\lambda$, we find the shape profile 
\begin{eqnarray}\label{e:zx}
  \frac{z(x)}{z_0} &=& 1  -2\left(\frac{f_0}{g} \right)^{1/2} \left(\frac{x \! -\! x_0}{z_0} \right)^{3/2} \nonumber \\ && + \frac{c f_0}{g^2}\left(\frac{x \! -\! x_0}{z_0} \right)^2 + {\cal O} \left(\frac{x \! -\! x_0}{z_0} \right)^{5/2}
\end{eqnarray}
\noindent Note that the edge position depends only on the step free energy $B$, not on the step repulsion strength; conversely, the coefficient of the leading $(x-x_0)^{3/2}$ term is independent of the step free energy but varies as the inverse root of the total step repulsion strength, \i.e.\ as $g^{-1/2}$.

If instead of Eq.~(\ref{e:fp-thB}) one adopts the phenomenological Landau theory of continuous phase transitions [Andreev 1982] and performs an analytic expansion of $f_p(p)$ in $p$ [Cabrera 1964, Cabrera and Garcia 1982] (and truncate after a quadratic term $f_2 p^2$), then a similar procedure leads $\vartheta = 2$, which is often referred to as the ``mean-field" value.  This same value can be produced by quenched impurities, as shown explicitly for the equivalent commensurate-incommensurate transition by Kardar and Nelson [1985].

There are some other noteworthy results for the smooth edge.  As the facet roughening temperature is approached from below, the facet radius shrinks like $\exp[-\pi^2 T_R/4\{2\ln 2(T_R-T)\}^{1/2}]$ [Jayaprakash et al.\ 1983], in striking contrast to predictions by mean-field theory. The previous discussion implicitly assumes that the path along $x$ for which $\vartheta$ = 3/2 in Eq.~(\ref{e:zx}) is normal to the facet edge.  By mapping the crystal surface onto the asymmetric 6-vertex model, using its exact solution [Yang 1967, Sutherland et al.\ 1967], and employing the Bethe Ansatz to expand the free energy close to the facet edge, Dahmen et al.\ [1998] find that $\vartheta$ = 3/2 holds for any direction of approach along the rounded surface toward the edge, {\it except} along the tangential direction (the contour that is tangent to the facet edge at the point of contact $x_0$.  In that special direction, they find the new critical exponent $\vartheta_y$ = 3 (where the subscript $y$ indicates the direction perpendicular to the edge normal, $x$.  Akutsu et al.\ [1998] (also Akutsu and Akutsu [2006]) confirmed that this exact result was universally true for the Gruber-Mullins-Prokrovsky-Talapov free-energy expansion.  (The Pokrovsky-Talapov argument was for the equivalent commensurate-incommensurate transition.)  They also present numerical confirmation using their transfer-matrix method based othe product-wave-function renormalization group (PWFRG) [Nishino and Okunishi 1995; Okunishi et al.\ 1999]. Observing $\vartheta_y$ experimentally will clearly be difficult, perhaps impossible; the nature and breadth of crossover to this unique behavior has not, to the best of my knowledge, been published.  A third result is that there is a jump (for $T < T_R$) in the curvature of the rounded part near the facet edge that has a universal value [Akutsu et al.\ 1988; Sato and Akutsu 1995], distinct from the universal curvature jump of the ECS at $T_R$  [Jayaprakash et al.\ 1983]]

\subsection{Experiments on Lead}

Noteworthy initial experimental tests of $\vartheta$ = 3/2 include direct measurements of the shape of equilibrated crystals of $^4$He [Carmi et al.\ 1987] and Pb [Rottman...M\'etois 1984].   As in most measurements of critical phenomena, but even harder here, is the identification of the critical point, in this case the value of $x_0$ at which rounding begins.  Furthermore, as is evident from Eq.~(\ref{e:zx}), there are corrections to scaling, so that the ``pure" exponent 3/2 is seen only near the edge and a larger effective exponent will be found farther from the edge.  For crystals as large as a few mm at temperatures in the range 0.7K--1.1K, $^4$He $\vartheta = 1.55 \pm 0.06$ was found, agreeing excellently with the Pokrovsky-Talapov exponent. The early measurements near the close-packed (111) facets of Pb crystallites, at least two orders of magnitude smaller, were at least consistent with 3/2, stated conservatively as $\vartheta =1.60 \pm 0.15$ after extensive analysis.  S\'aenz and Garc\'{\i}a [1985] proposed that in Eq.~(\ref{e:fp-thB}) there can be a quadratic term, say $f_2 p^2$ (but neglect the possibility of a quartic term).  Carrying out the Legendre transformation then yields an expression with both $x-B$ and $(x -B +f_2^2/3g)^{3/2}$ terms which they claim will lead to effective values of $\vartheta$ between 3/2 and 2. This approach provided a competing model for experimentalists to consider but in the end seems to have produced little fruit.

\begin{figure}[t]
\includegraphics[width=8 cm]{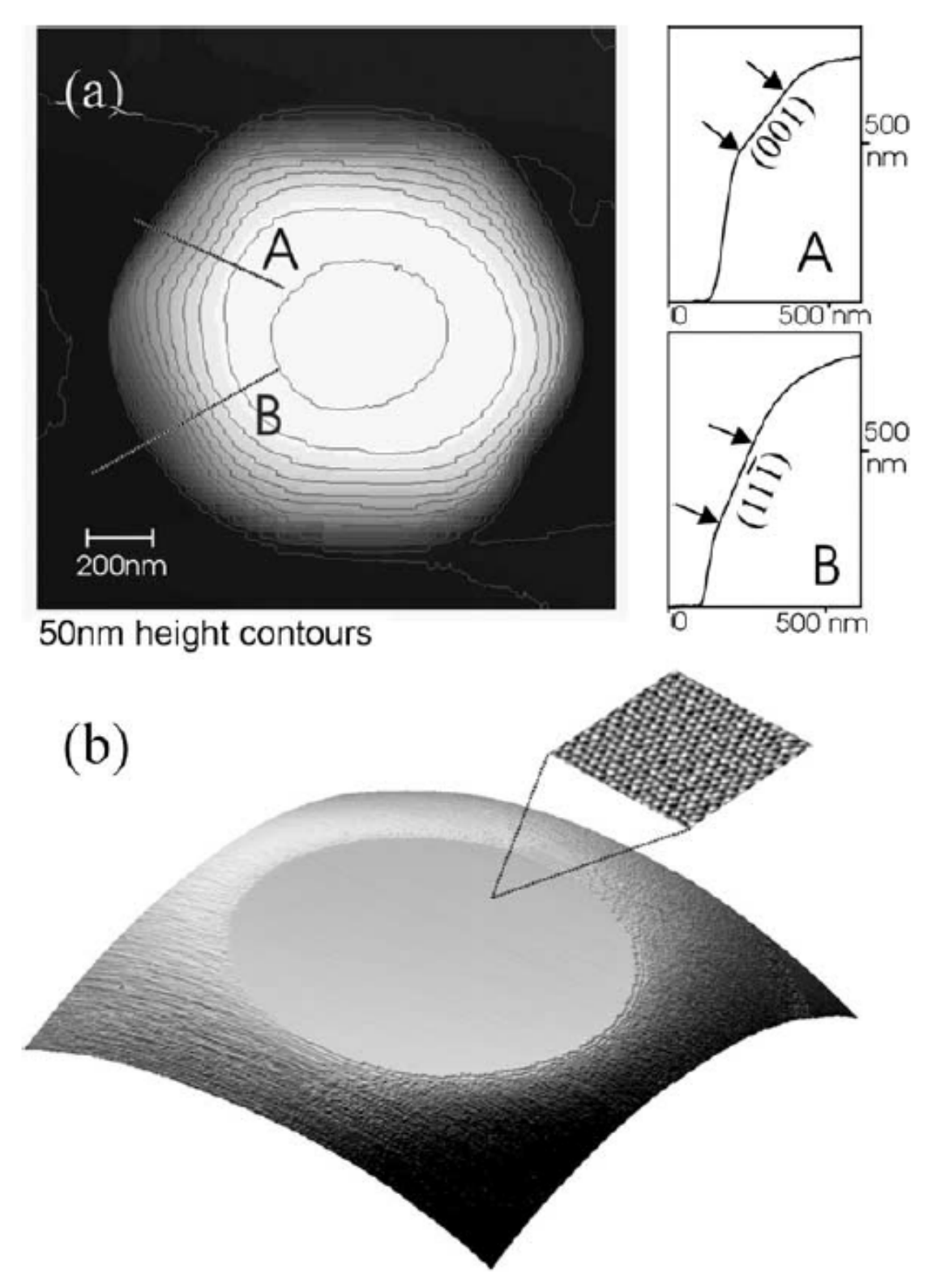}
\caption {(a) Micron-size lead crystal (supported on Ru) imaged with a variable-temperature
STM at $T = 95^{\circ}$C. Annealing at $T = 95^{\circ}$C for 20 hours allowed it to obtain its stable, regular shape. Lines marked A and
B indicate location of profiles. Profile A crosses a (0 0 1)-side
facet, while profile B a (1 1 1)-side facet. (b) 770 nm$\times$770 nm
section of the top part of a Pb-crystal. The insert shows a 5.3
nm$\times$5.3 nm area of the top facet, confirming its (1 1 1)-orientation.
Both the main image and the insert were obtained at $T = 110^{\circ}$C. From Th\"urmer et al.\ [2003].
}
{\label{f:ThurmerPb}}
\end{figure}

As seen in Fig.~\ref{f:ThurmerPb}, STM allows detailed measurement of micron-size crystal height contours and profiles at fixed azimuthal angles.
By using STM to locate the initial step down from the facet, first done by Surnev et al.\ [1998] for supported Pb crystallites, $x_0$ can be located independently and precisely.   However, from the 1984 Heyraud-M\'etois experiment [Rottman...M\'etois 1984] it took almost two decades until the Bonzel group could be fully confirm the $\vartheta$ = 3/2 behavior for the smooth edges of Pb(111) in a painstaking study [Nowicki et al.\ 2002b].  There were a number of noteworthy challenges.  While the close-packed 2D network of spheres has 6-fold symmetry, the top layer of a (111) facet of an fcc crystal (or of an (0001) facet of an hcp crystal) has only 3-fold symmetry due to the symmetry-breaking role of the second-layer.  There are two dense straight step edges, called A and B, with \{100\} and \{111\} microfacets, respectively. In contrast to noble metals, for Pb there is a sizeable (of order 10\%) difference between their energies.  Even more significant, when a large range of polar angles is used in the fitting, is the presence of small (compared to (111)) \{112\} facets for equilibration below 325K.  Due to the high atomic mobility of Pb that can lead to the formation of surface irregularities, Surnev et al.\ [1998] worked close to room temperature.  One then finds strong (3-fold) variation of $\vartheta$ with azimuthal angle, with $\vartheta$ oscillating between 1.4 and 1.7.  With a higher annealing temperature of 383K Nowicki et al.\ [2002b] report the azimuthal averaged value $\vartheta$ = 1.487 (but still with sizeable oscillations of about $\pm 0.1$); in a slightly early short report, they [Nowicki et al.\ 2002a] give a value $\vartheta$ = 1.47 for annealing at room temperature.  Their attention shifted to deducing the strength of step-step repulsions by measuring $g$ [Nowicki et al.\ 2003, Bonzel and Nowicki 2004].  In the most recent review of the ECS of Pb, Bonzel et al.\ [2007] rather tersely report that the Pokrovsky-Talapov value of 3/2 for $\vartheta$ characterizes the shape near the (111) facet and that imaging at elevated temperature is essential to get this result; most of their article relates to comparison of measured and theoretically calculated strengths of the step-step interactions.

Few other systems have been investigated in such detail.  Using scanning electron microscopy (SEM) M\'etois and Heyraud [1987] considered In, which has a tetragonal structure, near a (111) facet.  They analyzed the resulting photographs from two different crystals, viewed along two directions.  For polar angles $0^\circ \le \theta < \: \sim \! 5^\circ$ they find $\vartheta \approx 2$ while for $5^\circ \le \theta \le 15^\circ$ determine $\vartheta \approx 1.61$, concluding that in this window $\vartheta = 1.60 \pm 0.10$; the two ranges have notably different values of $x_0$.  This group [Bermond et al, 1998] also studied Si, equilibrated at $900^\circ$C, near a (111) facet.  Many profiles were measured along a high-symmetry $\langle 111 \rangle$ zone of samples with various diameters of order a few $\mu$m, over the range $3^\circ \le \theta \le 17^\circ$.  The results are consistent with $\vartheta$ = 3/2, with an uncertainty estimated at 6\%.  Finally Gladi\'c et al.\ [2002] studied large (several mm.) spherical cuprous selenide (Cu$_{2-x}$Se) single crystals near a (111) facet. Study in this context of metal chalcogenide superionic conductors began some dozen years ago because, other than $^4$He, they are the only materials having sub-cm. size crystals that have an ECS form that can be grown on a practical time scale (viz.\ over several days) because their high ionic and electronic conductivity enable fast bulk atomic transport.  For $14.0^\circ \le \theta \le 17.1^\circ$ Gladi\'c et al.\ [2002] find $\vartheta = 1.499 \pm 0.003$. (They also report that farther from the facet $\vartheta \approx 2.5$, consistent with the Andreev mean field scenario.)

\subsection{Summary of Highlights of Novel Approach to Behavior Near Smooth Edges}

\begin{figure}[t]
\includegraphics[width=8 cm]{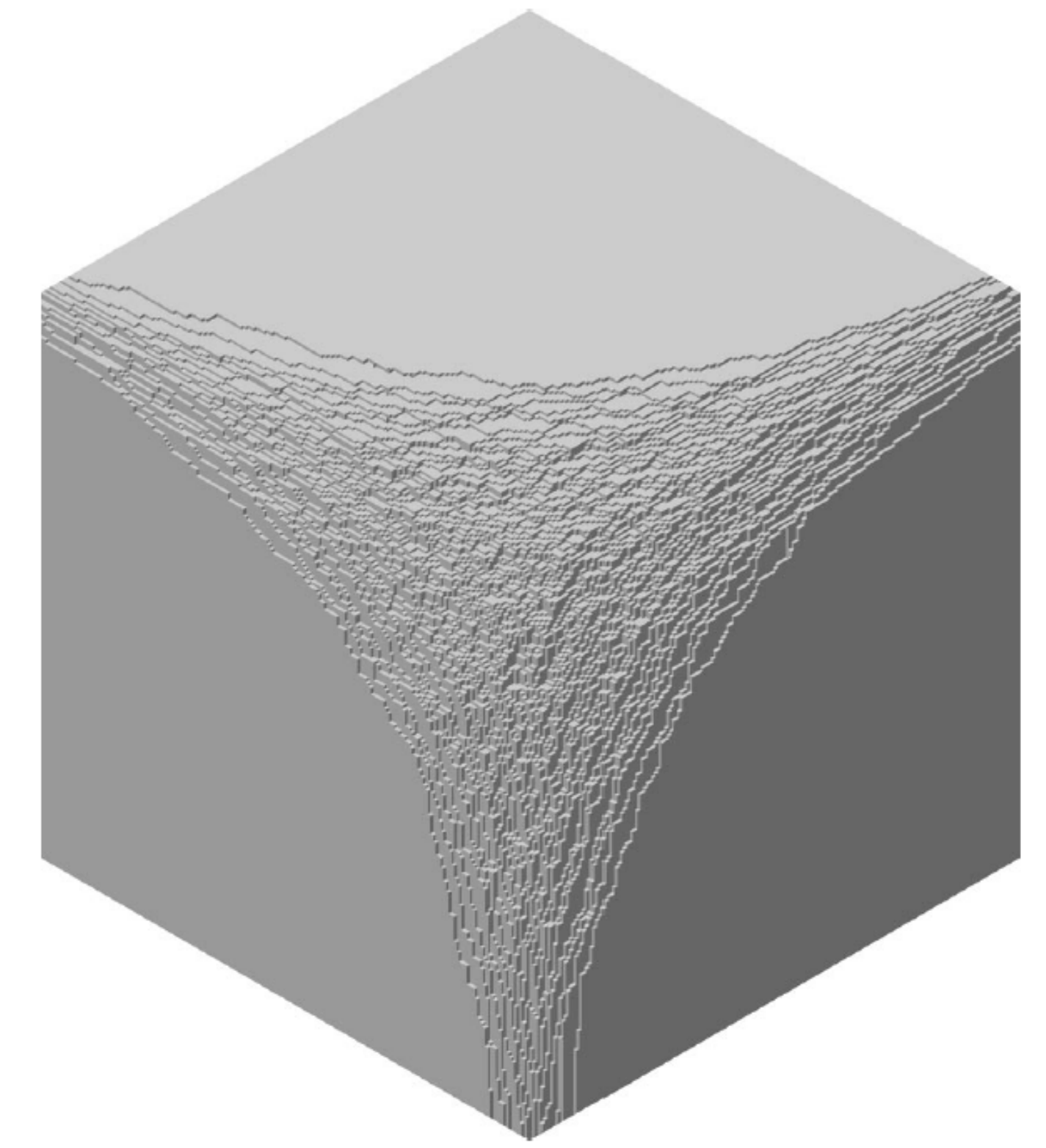}
\caption {Simple-cubic crystal corner viewed from the \{111\} direction. From Ferrari et al.\ [2004].}
\label{f:FPSCorner}
\end{figure}

Digressing somewhat, we note that Ferrari, Pr\"{a}hofer, and Spohn [2004] (FPS)
found novel static scaling behavior of the equilibrium fluctuations of an atomic ledge bordering a
crystalline facet surrounded by rough regions of the ECS in their examination of a 3D Ising corner (Fig.~\ref{f:FPSCorner}).
This boundary edge might be viewed as a ``shoreline" since it is the edge of an island-like region--the crystal facet--surrounded by a ``sea" of steps [Pimpinelli et al.\ 2005].

FPS assume that there are no interactions between steps other than entropic, and accordingly map the step configurations can be mapped to the world lines of free spinless fermions, as in treatments of vicinal surfaces [Jayaprakash et al.\ 1984].  However, there is the key new feature that the step number operator is weighted by the step number, along with a   Lagrange multiplier $\lambda^{-1}$ associated with volume conservation of the crystallite.  The asymmetry of this term is leads to the novel behavior they find.  They then derive an exact result for the step density and find that, near the shoreline,
\begin{equation}
\lim_{\lambda \rightarrow \infty} \lambda^{1/3} \rho_\lambda (\lambda^{1/3}x) = -x (\mathrm{Ai}(x))^2 +(\mathrm{Ai}^\prime(x))^2,
\end{equation}
\noindent where $\rho_\lambda$ is the step density (for the particular value of $\lambda$).

\begin{figure}[b]
\includegraphics[width=8.5 cm]{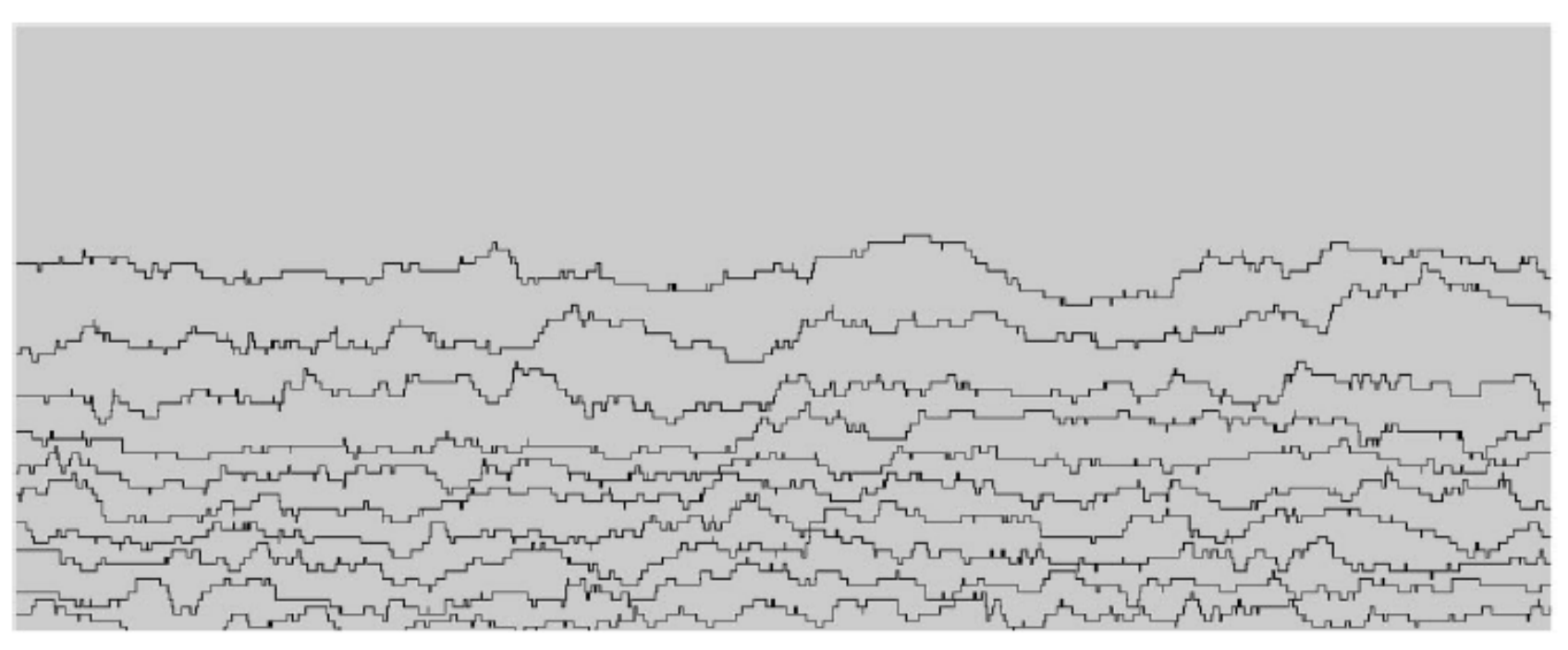}
\caption{Magnified detail of the steps near the facet edge, from Ferrari and Spohn [2003];  b) Snapshot of computed configurations of the top steps (those near a facet at the flattened side portion of a cylinder) for a terrace-step-kink (TSK) model with volume constraint. From Ferrari et al.\ [2004].
}
\label{f:FPSSteps}
\end{figure}

The presence of the Airy function Ai results from the asymmetric potential implicit in $\mathcal{H}_F$ and preordains exponents involving 1/3.  The variance of the wandering of the shoreline, the top fermionic world line in Fig.~\ref{f:FPSSteps} and denoted by $b$, is given by
\begin{equation}
\mathrm{Var}[b_\lambda(t) - b_\lambda(0)] \cong \lambda^{2/3} g(\lambda^{-2/3}t)
\label{e:var1}
\end{equation}
\noindent where $t$ is the fermionic ``time" along the step; $g(s) \sim 2|s|$ for small $s$ (diffusive meandering) and $\sim 1.6264 - 2/s^2$ for large $s$.  They then $1.202 \ldots$ is Apery's constant and $N$ is the number  of atoms in the crystal.  They find
\begin{equation}
\mathrm{Var}[b_\ell(\ell \tau + x) - b_\ell(\ell \tau)] \cong ({\cal A}\ell)^{2/3} g\left({\cal A}^{1/3}\ell^{-2/3}x\right),
\label{e:var2}
\end{equation}
where ${\cal A} = (1/2)b_\infty^{\prime\prime}$.  This leads to their central result that the width $w \sim \ell^{1/3}$, in contrast to the $\ell^{1/2}$ scaling of an  isolated step or the boundary of a single-layer island and to the $\ln \ell$ scaling of a step on a vicinal surface, i.e.\ in a step train.    Furthermore, the fluctuations are non-Gaussian.  They also show that near the shoreline, the deviation of the equilibrium crystal shape from the facet plane takes on the Pokrovsky-Talapov [1979, 1984] form  with $\vartheta$ = 3/2.

\begin{figure}[t]
\includegraphics[width=8 cm]{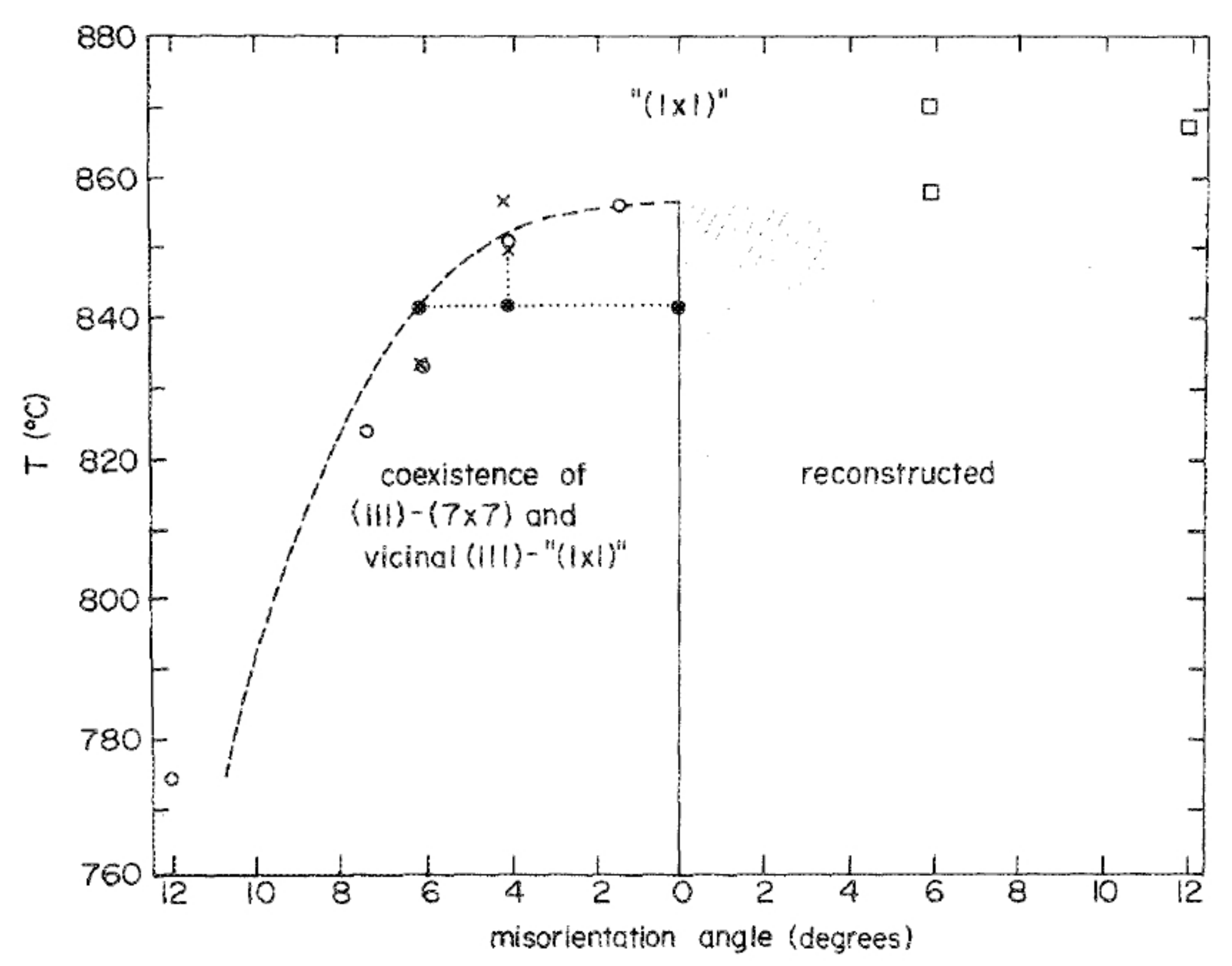}
\caption{Summary of experimental results
for vicinal Si (111) surface: $\bigcirc$ denotes
the temperature at which faceting
begins for surfaces misoriented towards
the (110] direction, $\times$ the faceting
temperatures for surfaces misoriented
towards the [ l 1 $\bar{2}$], and $\Box$ the temperatures
at which the step structure of surfaces
misoriented towards the [$\bar{1}\bar{1}2$] direction
change. The dashed line displays
a fit of the $\bar{1}$10] data to Eq.~(\ref{e:slope}). The dotted
lines show how a 4$^\circ$ sample phase-separates
into the states denoted by $\mathlarger{\bullet}$ as it is
 further cooled. From Bartelt et al.\ [1989].
}
\label{f:BarteltSiDiag}
\end{figure}

From this seminal work we could derive the dynamic exponents associated with this novel scaling and measure them with STM [Pimpinelli et al.\ 2005, Degawa et al.\ 2006, 2007], as reviewed in Einstein and Pimpinelli [2014].

\begin{figure}[t]
\includegraphics[width=8 cm]{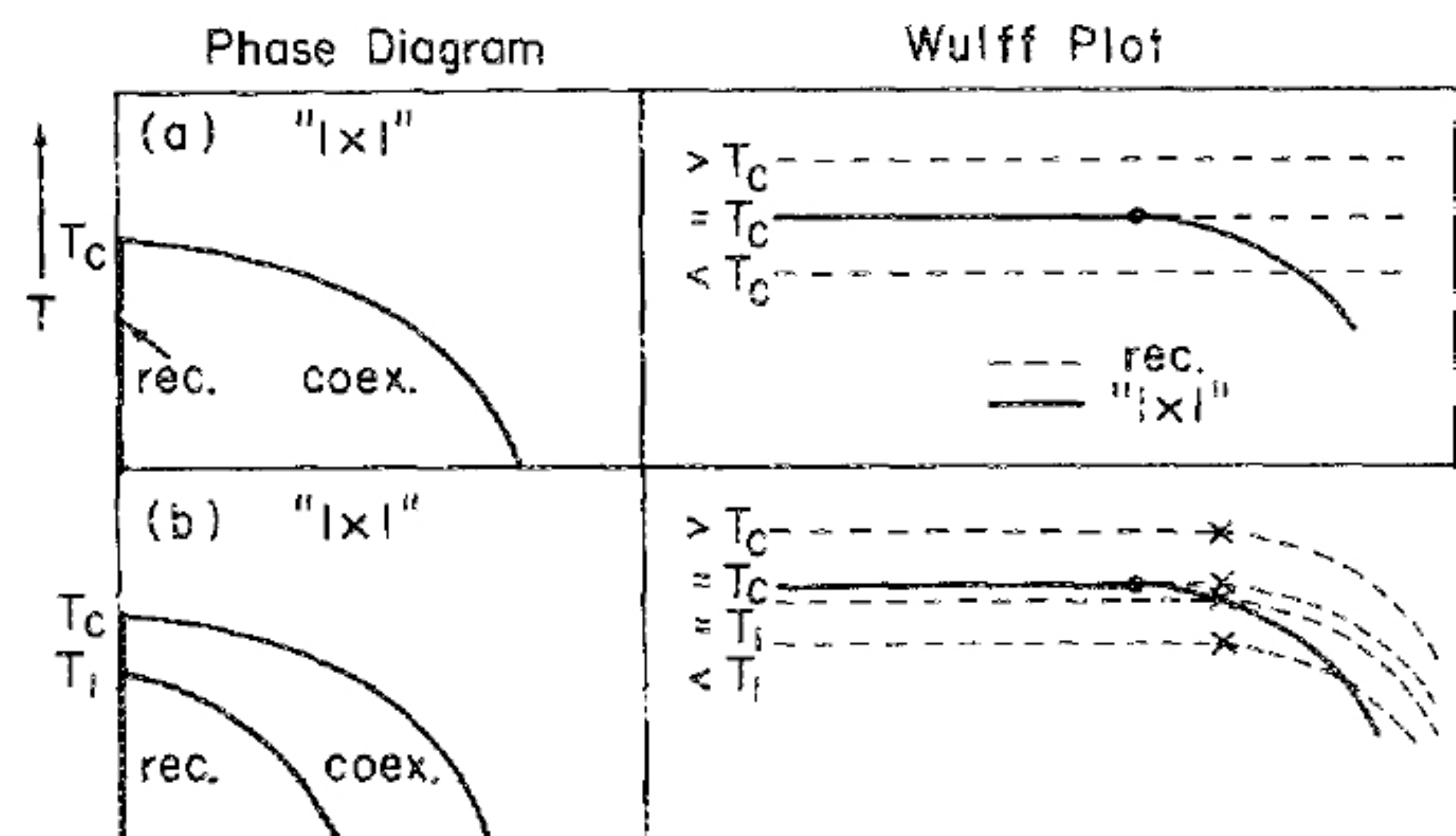}
\caption{Wulff plots illustrating the effect of a reconstructive transition on the
ECS, and corresponding temperature-[mis]orientation phase
diagrams. The solid curves represent the ECS with an unreconstructed
[``( 1 $\times$ 1)"] facet, while the dashed curves give the ECS with a reconstructed facet.
As temperature decreases, the free energy of the reconstructed facet, relative to that of the unreconstructed facet,
decreases. Below the transition temperature $T_c$ (called $T_7$ in the text), the two shapes intersect, giving
a ``net" ECS that is the inner envelope of the two. The
phase diagram shows regions where all orientations $\tan \theta$ (or $\mathbf{\hat{m}}$) are allowed for the
unreconstructed crystal [``(1 $\times$ 1)"], regions of phase separation (labeled
``coex."), and regions where the reconstruction (labeled ``rec.") is allowed for
ranges of orientation. The relative size of the reconstructed and unreconstructed
facets depends on the free energy to create a step on the reconstructed (111)
face, compared to its unreconstructed counterpart: (a) the behavior for
extremely large energy to create steps on the (7 $\times$ 7) terrace and (b) a smaller
such energy. Solid circles mark the sharp edge at the temperature at
which the crystal shapes cross. Crosses show the intersection of the facet
and the curved part (i.e., the smooth edge) of the crystal shape for the reconstructed phase.
From Bartelt et al.\ [1989].
}
\label{f:BarteltSiFE}
\end{figure}

\section{Sharp Edges and First-Order Transitions---Examples and Issues}
\subsection{Sharp Edges Induced by Facet Reconstruction}
Si near the (111) plane offers an easily understood entree into sharp edges  [Phaneuf and Williams 1987, Bartelt et al.\ 1989]. As Si is cooled from high temperatures, the (111) plane in the ``(1 $\times$ 1)" phase reconstructs into a (7 $\times$ 7) pattern [Schlier and Farnsworth 1959]  around 850$^\circ$C, to be denoted $T_7$ to distinguish it clearly from the other subscripted temperatures. (The notation ``(1 $\times$ 1)" is intended to convey the idea that this phase differs considerably from a perfect (111) cleavage plane but has no superlattice periodicity.)  For comparison, the melting temperature of Si is $\sim \! 1420^\circ$C, and the $T_R$ is estimated to be somewhat higher.   As shown in Fig.~\ref{f:BarteltSiDiag}, above $T_7$ surfaces of all orientations are allowed and are unreconstructed. At $T_7$  a surface in the (111) direction reconstructs but all other orientations are allow and are unreconstructed.  Below $T_7$, for surfaces misoriented toward [$\bar{1}\bar{1}$2] remain stable during cooling (although the step structure changes).  On the other hand, on surfaces misoriented toward [$\bar{1}$10] and [11$\bar{2}$] the temperature at which the (7 $\times$ 7) occurs  decreases with increasing misorientation angle $\mathbf{\hat{m}}$.  Furthermore, just as the (7 $\times$ 7) appears, the surface begins to separate into two phases, one a perfectly oriented (7 $\times$ 7) plane, $\mathbf{\hat{m}}=0$ and the second an unreconstructed phase with a misorientation greater than that at higher temperature.  As temperature further decreases, the misorientation of the unreconstructed phase increases.  Figure~\ref{f:BarteltSiDiag} depicts this scenario with solid circles and dotted lines for a 4$^\circ$ misoriented sample at 840$^\circ$C.  This behavior translates into the formation of a sharp edge on the ECS between a flat (7 $\times$ 7) line and a rounded ``(1 $\times$ 1)" curve.

To explain this behavior, one co-plots the ECS for the two phases, as in Fig.~\ref{f:BarteltSiFE} [Bartelt et al.\ 1989].  The free energy to create a step is greater in the (7 $\times$ 7) than in the ``(1 $\times$ 1)" phase.  In the top panels (a), the step energy for the (7 $\times$ 7) is taken as infinite (i.e.\  much larger than that of the
``(1 $\times$ 1)" phase) so its ECS never rounds.  At $T_7$ ($T_c$ on the figure) the free energies per area $f_0$ of the two facets are equal, call them $f_7$ and $f_1$, with associated energies $u_7$ and $u_1$ and entropies $s_7$ and $s_1$ for the (7 $\times$ 7) and ``(1 $\times$ 1)" phases, respectively, near $T_7$.  Then $T_7 = (u_1-u_7)/(s_1-s_7)$ and, assuming the internal energies and entropies are insensitive to temperature,
\begin{equation}\label{e:f117}
  f_1-f_7 = (T_7 - T)(s_1-s_7).
\end{equation}
Since $s_1 > s_7$ because the (7 $\times$ 7) phase is so highly ordered, we find that $f_1-f_7 > 0$ below $T_7$, as illustrated in Fig.~\ref{f:BarteltSiFE}.  Making connection to thermodynamics, we identify

\begin{equation}\label{e:fL}
 \frac{L}{T_7} = \left(s_1-s_7\right)_{T_7} = \left(\frac{\partial f_7}{\partial T}\right)_{T_7} - \left(\frac{\partial f_1}{\partial T}\right)_{T_7}
\end{equation}
\noindent where $L$ is the latent heat of the first-order reconstruction transition.

\begin{figure}[t]
\includegraphics[width=8 cm]{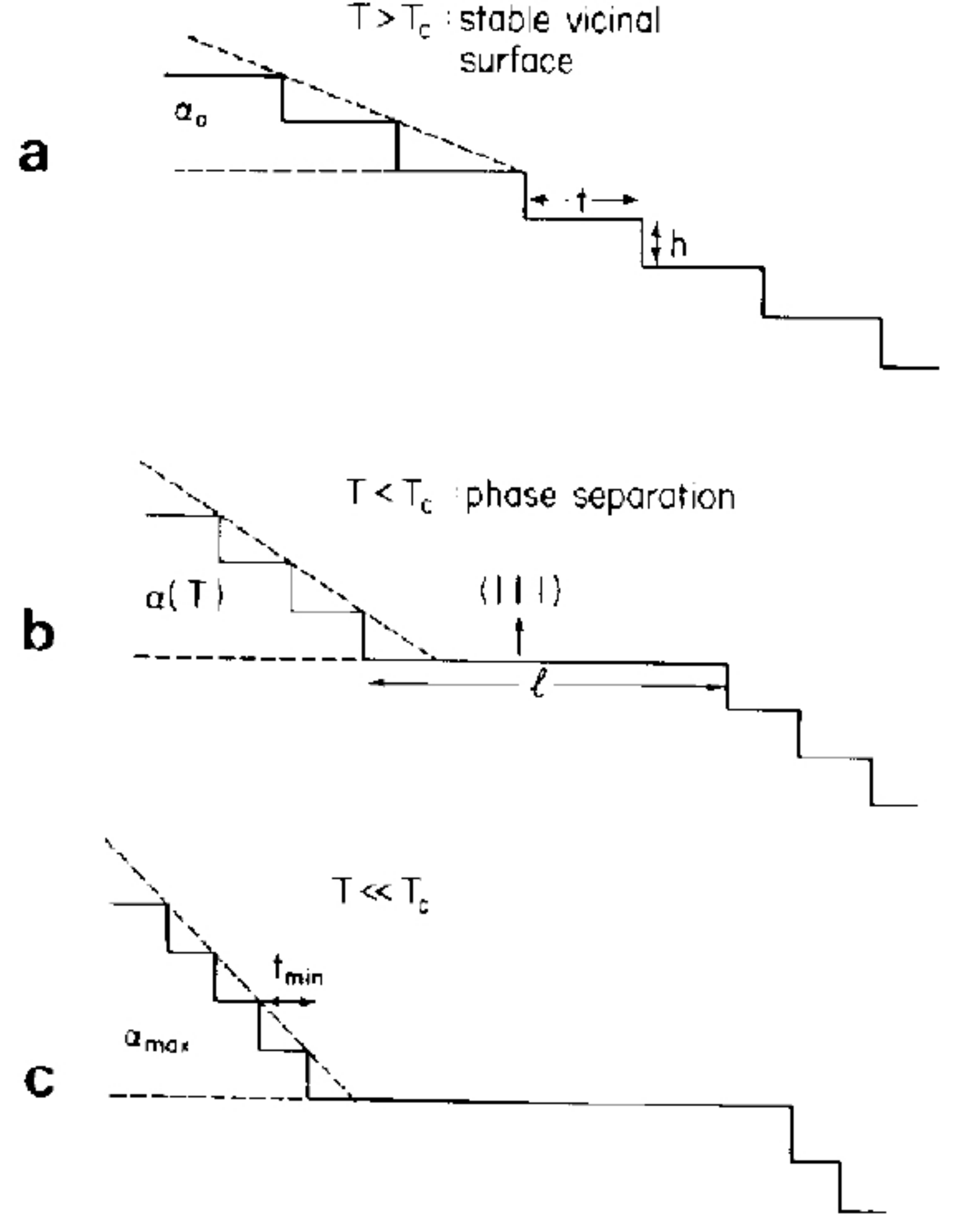}
\caption{Microscopic view of what happens to a misoriented surface in Fig.~\ref{f:BarteltSiFE} as temperature decreases.  (a) At high temperature, the Si(lll) vicinal surface
is a single, uniform phase. Initial terrace widths $t$ are
typically a few nm., as determined by the net angle of
miscut $\alpha_0$ (i.e., $\theta_0$), and the step-height $h$, which is one interplanar
spacing ($\sim 0.31$ nm). (b) Below the (7$\times$7) reconstruction temperature
~ ($\sim 850^\circ$C) the steps cluster to form a new
surface of misorientation angle $\alpha(T)$ (i.e., $\theta$). A facet of (111) orientation
with (7$\times$7) reconstruction forms simultaneously. The
width of the (111) facet, $\ell$, is larger than the experimentally
resolvable width of 500 \AA. (c) Well below the transition, the
step separation reaches a minimum distance, $t_{\rm min} \sim$ 1 nm. No
further narrowing occurs, perhaps because surface diffusion is too
slow $T \le 600^\circ$C.
From Williams and Bartelt [1989].
}
\label{f:EDWPhaseSep}
\end{figure}

Corresponding to the minimum of a free energy as discussed earlier, the ECS of the system will be the inner envelope of the dashed and solid traces: a flat (7 $\times$ 7) facet along the dashed line out to the point of intersection, the sharp edge, beyond which it is ``(1 $\times$ 1)" with continuously varying orientation.  If one tries to construct a surface with a smaller misorientation, it will phase separate into flat (7 $\times$ 7) regions and vicinal unreconstructed regions with the orientation at the curved (rough) side of the sharp edge.  Cf.\ Fig.~\ref{f:EDWPhaseSep}.

Using the leading term in Eq.~(\ref{e:Legendre14}) or (\ref{e:zx}) we can estimate the slope of the coexisting vicinal region and its dependence on temperature \footnote{There are some minor differences in prefactors from Bartelt et al.\ [1989]}: First we locate the sharp edge (recognizing $f_0$ as $f_1$ and $z_0$ as $z_1$ ) by noting
\begin{eqnarray}
  z_7 &=& z_1 - 2(\lambda/g)^{1/2}(x-x_0)^{3/2}  \\
(T_7 - T) \Delta s &\approx& \left(f_1-f_7\right)_T = \lambda^{3/2}g^{-1/2}(x-x_0)^{3/2}  \nonumber
\end{eqnarray}
Since the slope $m$ there is $-3\lambda(\lambda/g)^{1/2} (x-x_0)^{1/2}$, the temperature dependence of the slope is
\begin{equation}\label{e:slope}
m = -3 \left(\frac{L}{2g}\right)^{1/3} \left(1 - \frac{T}{T_7}\right)^{1/3}
\end{equation}

If the step free energy of the reconstructed phase were only modestly greater than that of the ``(1 $\times$ 1)", then, as shown in the second panel in Fig.~\ref{f:BarteltSiFE}, the previous high-$T$ behavior obtains only down to the  temperature $T_1$ at which the ``(1 $\times$ 1)" curve intersects the (7 $\times$ 7) curve at its [smooth] edge.  For $T < T_1$ the sharp edge associated with the interior of the curves is between a misoriented ``(1 $\times$ 1)" phase and a differently misoriented (7 $\times$ 7) phase, so that it is these two which coexist.  All orientations with smaller misorientation angles than this (7 $\times$ 7) plane are also allowed, so that the forbidden or coexistence regime has the depicted slivered crescent shape.  Some other, but physically improbable, scenarios are also discussed by  Bartelt et al.\ [1989].  Phaneuf and Williams [1987] show (their Fig.~3) the LEED-beam splitting for a surface misoriented by 6.4$^\circ$ is $\propto (T_7 - T)^{1/3}$ once the surface is cooled below the temperature (which is $< T_7$) when this orientation becomes unstable to phase separation; however, by changing the range of fitting, they could also obtain agreement with $(T_7 - T)^{1/2}$, i.e.\ $\vartheta$ = 2.  With high-resolution LEED Hwang et al.\ [1989] conclude that the exponent $\bar{\beta} \equiv (\vartheta -1)/\vartheta = 0.33\pm0.05$ (i.e., that $\vartheta$ = 3/2.  The result does depend somewhat on what thermal range is used in the fit, but they can decisively rule out the mean-field value $\vartheta$ = 2. Williams et al.\ [1993]  give a more general discussion of vicinal Si, with treatment of azimuthal in addition to polar misorientations.  In contrast, synchrotron x-ray scattering experiments by Noh et al.\ [1991, 1993] report the much larger $\vartheta = 2.3$ \raisebox{-0.5ex}{$\stackrel{\scriptstyle +0.8}{\scriptstyle -0.3}$}.  However, subsequent synchrotron x-ray scattering experiments by Held et al.\ [1995] obtain a decent fit of data with $\vartheta = 3/2$ and a best fit with $\vartheta = 1.75$  (i.e.,  $\bar{\beta} = 0.43\pm0.07$).  (They also report that above 1159K, the surface exists as a single, logarithmically rough phase.)  The origin of the curious value of $\vartheta$ in the Noh et al.\ experiments is not clear.  It would be possible to attribute the behavior to impurities, but there is no evidence to support this excuse, and indeed for the analogous behavior near the reconstructing (331) facet of Si (but perhaps a different sample), Noh et al.\ [1995] found $\vartheta = 1.47 \pm 0.1$.  It is worth noting that extracting information from x-ray scattering from vicinal surfaces requires great sophistication (cf.\ the extensive discussion in Dietrich and Haase [1995]) and attention to the size of the coherence length relative to the size of the scattering region [Ocko, 2014], as for other diffraction experiments.

Similar effects to reconstruction (viz.\ the change in $f_0$) could be caused by adsorption of impurities on the facet [Cahn 1982].  Some examples are given in a review by Somorjai and Van Hove [1989].  In small crystals of dilute Pb-Bi-Ni alloys, co-segregation of Bi-Ni to the surface has a similar effect of reversibly) changing the crystal structure to form \{112\} and \{110\} facets [Cheng and Wynblatt, 1994].  There is no attempt to scrutinize the ECS to extract an estimate of $\vartheta$.  Meltzman et al.\ [2011] considered the ECS of Ni on a sapphire support, noting that, unlike most fcc crystals, it exhibits a faceted shape even with few or no impurities, viz.\ with \{111\}, \{100\}, and \{110\} facets; \{135\} and \{138\} emerged at low oxygen pressure and additionally \{012\} and \{013\} at higher pressure.

\begin{figure}[t]
\includegraphics[width=8.5 cm]{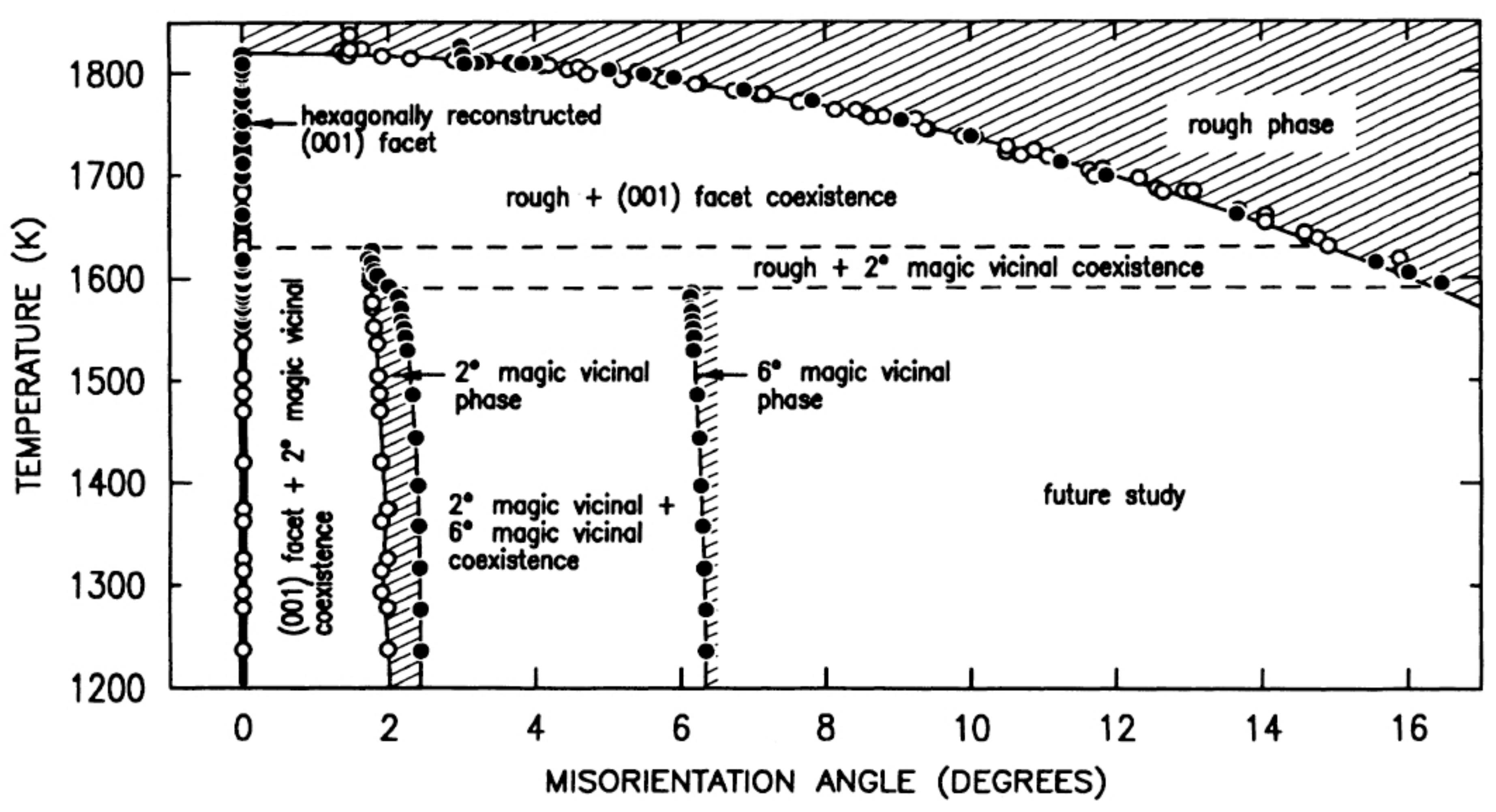}
\caption{Orientational phase diagram of vicinal Pt (001) misoriented toward the [110] direction. Single-phase regions are hatched, and two-phase coexistence regions are unhatched. Solid lines are boundaries between two phases. Dashed lines mark triple points. Open circles show misorientation angles measured for a sample miscut by 1.4$^\circ$ towards the [110] direction, while solid circles show tilt angles measured for a sample miscut by 3.0$^\circ$.
 From Yoon et al.\ [1994].
}
\label{f:MochriePt1}
\end{figure}

The phase diagram of Pt(001), shown in Fig.~\ref{f:MochriePt1} and studied by Yoon et al.\ [1994] using synchrotron x-ray scattering, at first seems similar to that of Si near (111) [Song et al.\ 1994, 1995], albeit with more intricate magic phases with azimuthal rotations at lower temperatures, stabilized by near commensurability of the period of their reconstruction and the separation of their constituent steps.  In the temperature-misorientation (surface slope) phase diagram, shown in Fig.~\ref{f:MochriePt1}, the (001) facet undergoes a hexagonal reconstruction at $T_6$ = 1820 K (well below the bulk melting temperature of 2045K).  For samples misoriented from the (100) direction (which are stable at high temperature), there is coexistence between flat reconstructed Pt(001) and a rough phase more highly misoriented than it was at high temperature, with a misorientation that increases as temperature decreases.  However, they find $\bar{\beta} = 0.49 \pm 0.05$, or $\vartheta = 1.96$, consistent with mean field and inconsistent with $\bar{\beta} = 1/3$ or $\vartheta = 3/2$ of Pokrovsky-Talapov.  The source of this mean-field exponent is that in this case the (001) orientation is rough above $T_6$.  Hence, in Eq.~(\ref{e:fp-thB}) $B$ vanishes, leaving the expansion appropriate to rough orientations  . Proceeding as before, Eq.~(\ref{e:Legendre14} becomes
\begin{equation}\label{e:rough}
 f_p(p) = f_0 + D p^2 \; \Rightarrow \; \tilde{f}(\eta) = f_0 - \eta^2/4D,
\end{equation}
\noindent where the result for $\tilde{f}(\eta)$ is reached by proceeding as before to reach the modification of  Eq.~(\ref{e:Legendre14}).  Thus, there is no smooth edge take-off point (no shoreline) in the equivalent of Fig.~\ref{f:BarteltSiFE}, and one finds the reported exponent $\vartheta$ near 2.

The effect of reactive and nonreactive gases metal catalysts has long been of interest [Flytzani-Stephanopoulos and Schmidt, 1979]. Various groups investigated adsorbate-induced faceting. Walko and Robinson [2001] considered the oxygen-induced faceting of Cu(115) into O/Cu(104) facets, using Wulff constructions to explain their observations.  They found three temperature regimes with qualitatively different faceting processes.
Szczepkowicz et al.\ [2005] studied the formation of \{211\} facets by depositing oxygen and paladium on tungsten, both on (111) facets and on soherical crystals.  While the shape of the facets is different for flat and curved surfaces, the distance between parallel facet edges is comparable, although the area of a typical facet on a curved surface is an order of magnitude greater.  There is considerable information about facet sizes, width of the facet-size distribution, and surface rms roughness.

For 2D structures on surfaces, edge decoration can change the shape of the islands.  A well-documented example is Pt on Pt(111).  As little as $10^{-3}$ ML of CO produces a 60$^\circ$ rotation of the triangular islands by change the balance of the edge free energies of the two different kinds of steps forming the island periphery [Kalff et al.\ 1998].  Stasevich et al.\ [2009] showed how decoration of single-layer Ag islands on Ag(111) by a single-strand ``necklace" of C$_{60}$ dramatically changes the shape from hexagonal to circular.  With lattice-gas modeling combined with STM measurements, they could estimate the strength of C$_{60}$-Ag and C$_{60}$-C$_{60}$ attractions.  Generalizations to decoration on systems with other symmetries is also discussed.

\section{Gold--Prototype or Anomaly of Attractive Step-Step Interaction?}

Much as $^4$He and Pb are the prototypical materials with smooth edges, Au is perhaps the prime example of a surface with sharp edges, around (111) and (100) facets. (Cf., e.g., Wortis [1988]).  Care must be taken to insure that the surface is not contaminated by atoms (typically C) from the supporting substrate [Wang and Wynblatt 1988].  (See similar comments by Handwerker et al.\ [1988] for ceramics, which have a rich set of ECS possibilities.)  To describe these systems phenomenologically, the projected free energy expansion in Eq.~(\ref{e:fp-th}) requires a negative term to generate a region with negative curvature, as in Fig.~\ref{f:HV}, so that the two orientations joined by the Maxwell double-tangent construction correspond to the two sides of the sharp edge.  Thus, for sharp edges around facets, the more-left minimum must be in the high-symmetry facet direction.

In a mean field based approach, Wang and Wynblatt [1988] included a negative quadratic term, with questionable physical basis.  Emundts et al.\ [2001] instead took the step-step interaction to be attractive ($g < 0$) in Eq.~(\ref{e:fp-th}).  Then proceeding as above they find
\begin{equation}\label{e:gneg}
 x_0 = \frac{1}{\lambda}\left[ B - \frac{4}{27}|g|\left(\frac{g}{c}\right)^2\right], \quad
 p_{\rm c}= \frac{2|g|}{3c},
\end{equation}
\noindent where $p_c$ is the tangent of the the facet contact angle.  Note that both the shift in the facet edge from $B$ and the contact slope increase with $|g|/c$.  Emundts et al.\ [2001] obtain estimates of the key energy parameters in the expansion for the sharp edges of both the (111) and (100) facets.  They also investigate whether it is the lowering of the facet free energy $f_0$ that brings about the sharp edges, in the manner of the case of Si(111) discussed above.  After reporting the presence of standard step-step repulsions (leading to narrowing of the TWD) in experiments on flame-annealed gold, Shimoni et al.\ [2000] then attribute to some effective long-range attraction---with undetermined dependence on $\ell$---the (non-equilibrium) movement of single steps toward step bunches whose steps are oriented along the high-symmetry $\langle 110 \rangle$.

Is it possible to find a generic long-range attractive $A\ell^{-2}$ step-step interaction ($A<0$) for metals and elemental semiconductors (where there is no electrostatic attraction between oppositely charged atoms)?  Several theoretical attempts have only been able to find such attractions when there is significant alternation between ``even" and ``odd" layered steps.  Redfield and Zangwill consider whether surface relaxation can produce such an attraction, pointing out a flaw in an earlier analysis assuming a rigid relaxation by noting that for large step separations, the relaxation must return to its value for the terrace orientation.  Since atomic displacements fall off inversely with distance from a step, the contribution to the step interaction can at most go like $\ell^{-2}$ and tend to mitigate the combined entropic and elastic repulsion.  They argue that this nonlinear effect is likely to be small, at least for metals.  It is conceivable that on an elastically highly anisotropic surface, the elastic interaction might not be repulsive in special directions, though I am not aware of any concrete examples.

By observing that the elastic field mediating the interaction between steps is that of a dipole applied on a stepped rather than on a flat surface, Kukta et al.\ [2002] deduce a correction to the $\ell^{-2}$ behavior of the Marchenko-Parshin [1980] formula that scales as $\ell^{-3}\ln \ell$.  Using what was then a state-of-the-art semiempirical potential, the embedded atom method (EAM)[Daw et al.\ 1993], they find that this can lead to attractive interactions at intermediate values of $\ell$.  However, their ``roughness correction" term exists only when the two steps have unlike orientations (i.e., one up and one down, such as on opposite ends of a monolayer island or pit).  For the like-oriented steps of a vicinal surface or near a facet edge, the correction term vanishes.  The oft-cited paper then invokes 3-step interactions, which are said to have the same size as the correction term, as a way to achieve attractive interactions.  Although the authors discuss how this idea relates to the interaction between an isolated step and a step bunch, they do not provide the explicit form of the threefold interaction; their promise that it will be ``presented elsewhere" has not, to the best of my searching, ever been fulfilled.  Pr\'evot and Croset [2004] revisited elastic interactions between steps on vicinals and found that with a buried-dipole model (rather than the surface-dipole picture of Marchenko and Parshin), they could achieve ``remarkable agreement" with molecular dynamics simulations for vicinals to Cu and Pt (001) and (111), for which data is fit by $E_2^{MD}\ell^{-2} + E_3^{MD}\ell^{-3}$.  The tabulated values of $E_2^{MD}$ indeed agree well with their computed results for their improved elastic model, which includes the strong dependence of the interaction energy on the force direction.  While there is barely any discussion of $E_3$, plots of the interaction are always repulsive.  Hecquet [2008] finds that surface stress modifies the step-step interaction compared to the Marchenko-Parshin result, enhancing the prefactor of $\ell^{-2}$ nearly threefold for Au(001); again, there is no mention of attractive interactions over any range of step separations.

In pursuit of a strictly attractive $\ell^{-2}$ step interaction to explain the results of Shimoni et al.\ [2000], Wang et al.\ [2007] developed a model based on the SSH model [Su et al.\ 1979] of polyacetylene (the original model extended to include electron-electron interaction), focusing on the dimerized atom rows of the (2 $\times$ 1) reconstruction of Si(001).  The model produces an attractive correction term to the formula derived by Alerhand et al.\ [1988] for interactions between steps on Si(001), where there is ABAB alternation of (2 $\times$ 1 and 1 $\times$ 2) reconstructions on neighboring terraces joined at single-height steps.  For this type of surface, the correction has little significance, being dwarfed by the logarithmic repulsion.  It also does not occur for vicinals to high-symmetry facets of metals.  However, for surfaces such as Au(110) with its missing row morphology [Copel and Gustafsson, 1986] or adsorbed systems with atomic rows, the row can undergo a Peierls [1955] distortion that leads to an analogous dimerization and an $\ell^{-2}$ attraction.  There have been no tests of these unsung predictions by electronic structure computation.

Returning to gold, applications of the glue potential (a semiempirical potential rather similar to EAM), Ercolessi et al.\ [1987] were able to account for reconstructions of various gold facets, supporting that the sharp edges on the ECS are due to the model used for Si(111) rather than attractive step interactions.  Studies by this group found no real evidence for attractive step interactions  [Tosatti, 2014].

In an authoritative review a decade ago, Bonzel [2003]---the expert in the field who has devoted the most sustained interest in ECS experiments on elemental systems---concluded that it was not possible to decide whether the surface reconstruction model or attractive interactions was more likely to prove correct.  In my view, mindful of Ockham's razor, the former seems far more plausible, particularly if the assumed attractive interaction has the $\ell^{-2}$ form.

The phase diagram of surfaces vicinal to Si(113) presents an intriguing variant of that vicinal to Si(111).  There is again a coexistence regime between the (113) orientation and progressively more highly misoriented vicinals as temperature is reduced below a threshold temperature $T_t$, associated with a first-order transition.  However, for higher temperatures $T > T_t$ there is a continuous transition, in contrast to the behavior on (111) surfaces for $T > T_7$.  Thus, Song and Mochrie [1995] identify the point along (113) at which coexistence vanishes, i.e.\ $T_t$, as a tricritical point, the first such point seen in a misorientation phase diagram.  To explain this behavior Song and Mochrie invoke a mean-field Landau-theory argument in which the cubic term in $p$ is proportional to $(T-T_t)$, so negative for $T < T_t$, with a positive quartic term.  Of course, this produces the observed generic behavior, but the exponent $\bar{\beta}$ is measured as 0.42$\pm$ 0.10 rather than the mean-field value 1.  Furthermore, the shape of the phase diagram differs from the mean field prediction and the amplitude of the surface stiffness below $T_t$ is larger than above it, the opposite of what happens in mean field.  Thus, it is not clear in detail what the interactions actually are, let alone how an attractive interaction might arise physically.

\section{Well-Established Attractive Step-Step Interactions Other Than $\ell^{-2}$}

For neutral crystals, there are two ways to easily obtain interactions that are attractive for some values of $\ell$.  In neither case are the interactions monotonic long-range.  The first is short-range local effects due to chemical properties of proximate steps while the other is the indirect Friedel-like interaction.

\subsection{Atomic-Range Attractions}

At very small step separations the long-range the long-range $\ell^{-2}$ monotonic behavior is expected to break down and depend strongly on the local geometry and chemistry.  Interactions between atoms near step edges are typically direct and so stronger than interactions mediated by substrate elastic fields or indirect electronic effects (see below). We saw earlier that a $\ell^{-3}$ higher-order term arises at intermediate separations [Najafabadi and Srolovitz 1994], and further such terms should also appear with decreasing $\ell$. On TaC(910) [vicinal to (001) and miscut toward the [010] direction], Zuo et al.\ [2001] explained step bunching using a weak $\ell^{-3 [\pm 0.5]}$ attraction in addition to the $\ell^{-2}$ repulsion.  [The double-height steps are electrically neutral.]  Density-functional theory (DFT) studies were subsequently performed for this system by Shenoy and Ciobanu [2003].  Similarly, Yamamoto et al.\ [2010] used an attractive $\ell^{-3}$ dipole-quadrupole interaction to explain anomalous decay of multilayer holes on SrTiO$_3$(001).

More interesting than such generic effects are attractions that occur at very short step separations for special situations.  A good example is Ciobanu et al.\ [2003], who find an attraction at the shortest separation due to the cancellation of force monopoles of two adjacent steps on vicinal Si(113) at that value of $\ell$.

As alluded to above, most of our understanding of the role of $\ell^{-2}$ step interactions comes from the mapping of classical step configurations in 2D to the world lines of spinless fermions in 1D.  Unlike fermions, however, steps can touch (thereby forming double-height steps), just not cross.  Such behavior is even more likely for vicinal fcc or bcc (001) surfaces, where the shortest possible ``terrace", some fraction of a lateral nearest-neighbor spacing, amounts to touching fermions when successive layers of the crystal are described with simple-cubic rather than layer-by-layer laterally offset coordinates.  Sathiyanarayanan et al.\ [2009] investigated some systematics of step touching, adopting a model in which touching steps on a vicinal cost an energy $\epsilon_t$.  Note that $\epsilon_t = \infty$ recoups the standard fermion model.  For simplicity, the short study concentrates on the ``free fermion" case $\tilde{A} = 0$, i.e. $\varrho =2$.  (Cf.\ Eq.~(\ref{e:avrh}).)  Even for $\epsilon_t = 0$ there is an effective attraction, i.e.\ $\varrho <2$, since the possibility of touching broadens the TWD. This broadening is even more pronounced for $\epsilon_t < 0$. In other words, such short-range effects can appear, for a particular system, to contribute a long-range attraction.  Closer examination shows that this attraction is a finite-size effect that fades away for large values of $\langle \ell \rangle$.  In our limited study, we found that fits of simulated data to the GWD expression could be well described by the following finite-size scaling form, with the indicated three fitting parameters:
\begin{equation}\label{e:fss}
  \varrho_{\rm eff} = 2 - (0.9 \pm 0.1) \langle \ell \rangle^{-0.29 \pm 0.07} \exp[-( 3.3 \pm 0.2)\epsilon_t/k_BT].
\end{equation}
\noindent While Eq.~(\ref{e:fss}) suggests that making the step touching more attractive (decreasing $\epsilon_t$) could decrease $\beta_{\rm eff}$ without limit, instabilities begin to develop, as expected since L\"assig [1996] showed that for $\tilde{A} < -1/4$, i.e. $A < -(k_BT)^2/4\tilde{\beta}$, a vicinal surface becomes unstable (to collapse to step bunches).  Correspondingly, the lowest value tabulated in Sathiyanarayanan et al.\ [2009] is $\epsilon_t/k_BT = -0.2$.

To distinguish true long-range ($\ell^{-2}$) attractions on vicinal surfaces requires measurements of several different vicinalities (i.e. values of $\langle \ell \rangle$).  Likewise, in analyses of ECS data, consideration of crystallites of different sizes would seem necessary.  Wortis [1988] had noted the importance of size dependence in other contexts.

\begin{figure}[b]
\includegraphics[width=8 cm]{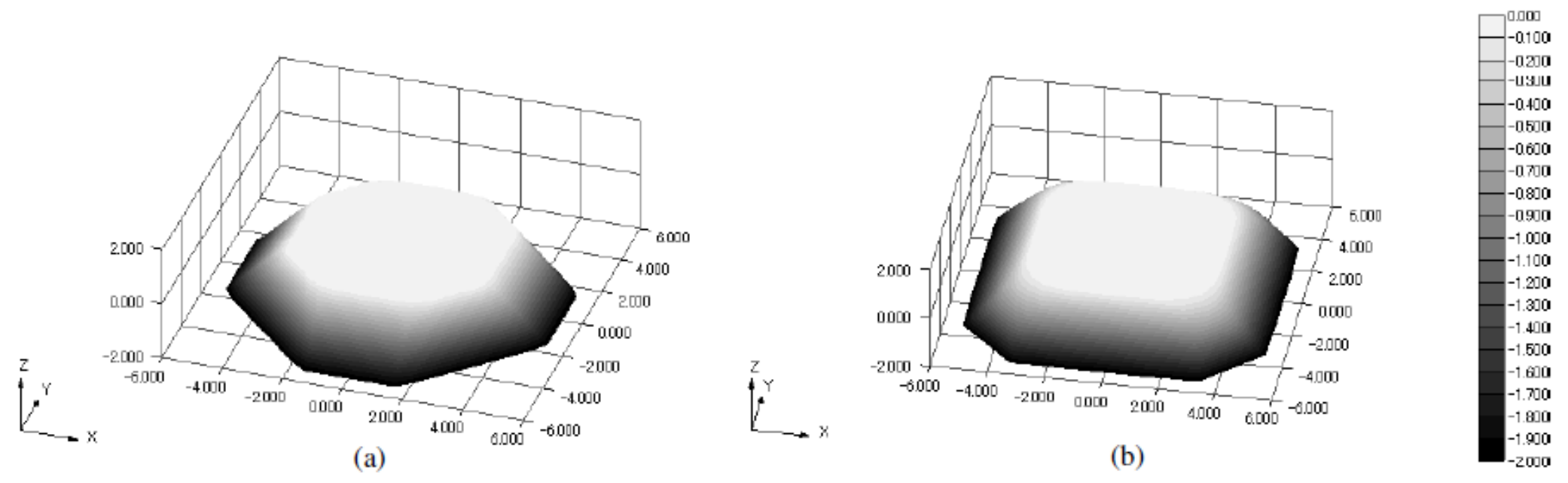}
\caption {Perspective views of essentially the ECS (actually the Andreev surface free energy divided by $k_BT$) around the (001) facet calculated by the transfer matrix method with the PWFRG algorithm at $k_BT/\epsilon_1 = 0.3$. (a) p-RSOS model ($\epsilon_{int}/\epsilon_1 = -0.5$).. (b) For comparison, the original unsticky RSOS model ($\epsilon_{int} = 0$) From Akutsu [2011] .
}
{\label{f:Akutsu1}}
\end{figure}

\begin{figure}[t]
\includegraphics[width=8 cm]{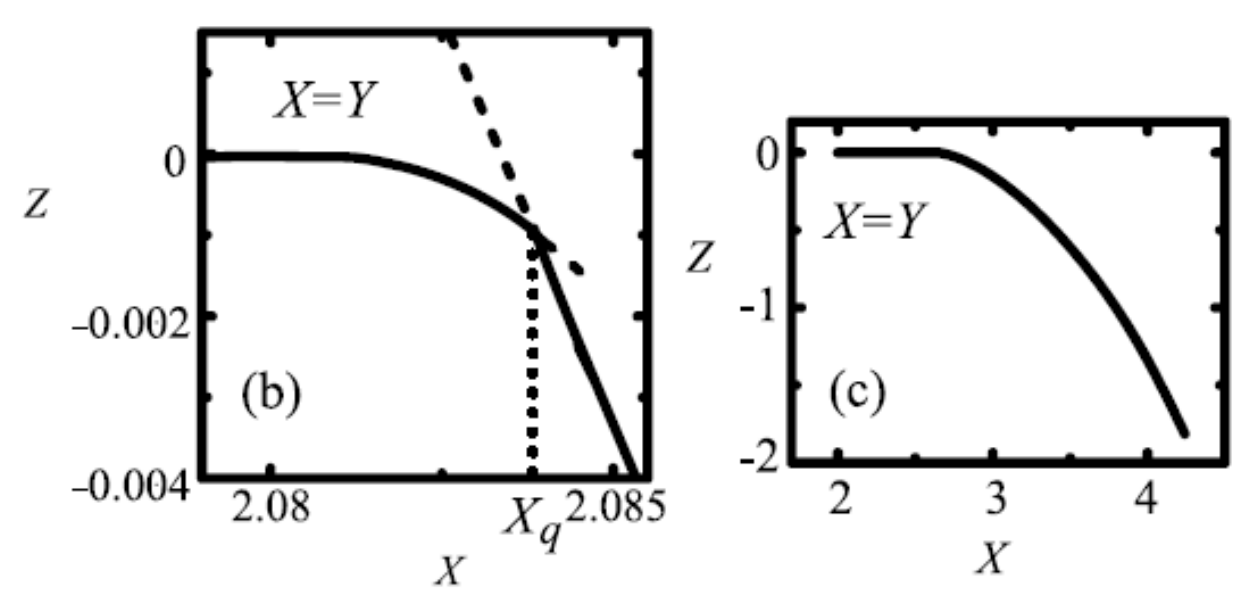}
\caption {Profiles in the diagonal direction of the surface in Fig.~\ref{f:Akutsu1}, still at $k_BT/\epsilon_1 = 0.3$.  Broken lines represent metastable lines.  (a) $k_BT/\epsilon_1 = 0.36$, $\epsilon_{int}/\epsilon_1 = -0.5$, on a very fine length scale. The edge of the (111) facet is denoted by $Xq$. (b) The original RSOS model ($\epsilon_{int} = 0$) on a much coarser scale.  On this scale (and on an intermediate scale not included here), the profile profiles are flat until the edge. On the intermediate scale, the region beyond $Xq$ is starts deviating rather smoothly for $k_BT/\epsilon_1 = 0.35$ but looks straight for $k_BT/\epsilon_1 = 0.36$ and 0.37. See text and source. From Akutsu [2011].
}
{\label{f:Akutsu2}}
\end{figure}

Along this theme, an instructive specific case is the ``sticky-step" or, more formally, the p-RSOS [restricted solid-on-solid with point-contact attractions between steps] model explored in detail by Akutsu [2011] using the product wavefunction renormalization group (PWFRG) method, calculating essentially the ECS (see Fig.~\ref{f:Akutsu1}) and related properties. Steps are zig-zag rather than straight as in the preceding Sathiyanarayanan model, so her ``stickiness" parameter $\epsilon_{int}$ is similar but not identical to $\epsilon_t$.  She finds that in some temperature regimes, non-universal non-Pokrovsky Talapov values of $\vartheta$ occur.  Specifically, let $T_{f,111}(\epsilon_{int}/\epsilon, \phi_0)$ and $T_{f,001}(\epsilon_{int}/\epsilon, \phi_0)$ be the highest temperature at which a first-order phase transition (sharp edge) occurs for the (111) and (001) facets, respectively, where $\phi_0$ indicates the position along the ECS.  Note $T_{f,111}(\epsilon_t/\epsilon, \phi_0) = (0.3610\pm 0.0005) \epsilon/k_B > T_{f,111}(\epsilon_{int}/\epsilon, \phi_0) = (0.3585\pm 0.0007)\epsilon/k_B$.  For $k_BT/\epsilon = 0.37$, so $T > T_{f,111}(-0.5, \pi/4)$, Akutsu recovers Pokrovsky-Talapov values for $\vartheta$ and $\vartheta_t$, but for $k_BT/\epsilon = 0.36$ (shown in Fig.~\ref{f:Akutsu2}), so $T_{f,111}(-0.5, \pi/4) > T > T_{f,001}(-0.5, \pi/4)$, the values are very different: $\vartheta = 1.98 \pm 0.03$ and $\vartheta_t = 3.96 \pm 0.08$, more like mean field.  For $\phi_0 = 0$ (tilting toward the $\langle 100 \rangle$\ direction) only standard Pokrovsky-Talapov exponents are found. 
Upon closer examination with Monte Carlo simulations, Akutsu finds large step bunches for $T < T_{f,100}$ but step droplets for $T_{f,001} < T < T_{f,111}$.  The details are beyond the scope of this review, but eventually Akutsu deduces an expansion of the projected free energy that includes either a quadratic term or a term after the linear term that has the form $|p|^\zeta$, with $\zeta >1$.

\subsection{Attractions at Periodic Ranges of Separation via Oscillatory Friedel-Type Interactions}

Oscillatory (in sign) interactions between steps, mediated by substrate conduction electrons, ipso facto lead to attractive interactions between steps.  As reviewed by Einstein [1996], such interactions have been know for many decades to account for the ordered patterns of adsorbates on metal surfaces [Einstein and Schrieffer 1973].  While at short range all electrons contribute, asymptotically the interaction is dominated by the electron[s] at the Fermi surface or, from another perspective, the non-analyticity in the response function at the nesting vector.  The interaction energy has the form
\begin{equation}\label{e:pair}
  E_{\rm pair}^{\rm asymp} \propto \ell^{-n} \cos (2 k_F \ell + \Phi)
\end{equation}
This, or its analogue for interacting local magnetic species, is called the RKKY [Ruderman and Kittel 1954,
Yosida 1957] interaction. (The community studying magnetism now labels as RKKY any interaction mediated by substrate electrons, not just the asymptotic limit written down in the RKKY papers.)  The phase factor $\Phi$ is the nonperturbative result is the scattering phase shifts at the two atoms that are interacting; it is absent in the perturbational approach to this problem used in the RKKY papers.  The exponent $n$ indicates the decay envelope.  For interacting bulk entities, $n$ = 3, the standard RKKY results.  On metal surfaces, the leading term in the propagator is canceled by the image charge, leading to $n$ = 5, with very rapid decay [Einstein and Schrieffer 1973, Einstein 1996].  Such effects are insignificant for adatom interactions but can be more potent when a whole step participates. Redfield and Zangwill [1992] show that a line of localized perturbations will generate an interaction with $n$  reduced by subtracting 1/2 and $\Phi$ augmented by $\pi/4$.  They used this result, with $n$ = 9/2, to account for Frohn et al.\ [1991] remarkable experimental results on vicinal Cu(001): from their observed bimodal TWD, Frohn et al.\ deduced that the step-step interaction is attractive for intermediate distances 3--5 atoms.  Indeed, it was their striking observation that led to several of the previously-discussed theory papers that claimed to find long-range step attractions.

When there are metallic surface states (i.e. surface states for which their 2D band dispersion relation crosses the Fermi energy $E_F$) of Shockley nature (lying in a 2D band gap containing $E_F$) , the indirect interaction has a much slower decay, with $n$ = 2 [Lau and Kohn 1978, Einstein 1996, Repp et al.\ 2000, Hyldgaard and Persson 2000, Knorr et al.\ 2002, Hyldgaard and Einstein 2005]  Furthermore, the Fermi wavevector typically is much smaller than that of bulk states, so the period of the oscillation in real space is much larger.  Perhaps the most familiar metallic surface on metals is that at the center of the surface Brillouin zone ($\bar{\Gamma}$) of the (111) surfaces of noble metals, which exist inside the necks of the Fermi surface, discussed in textbooks like Ashcroft and Mermin [1976].  This is the state produces the famous wave structure in Eigler's group's dramatic STM images [Crommie et al.\ 1993] of atoms on metal surfaces.  However, there is a less well known metallic surface state on Cu (001), discovered relatively late (compared to other surface states) by Kevan [1983], centered at the zone-edge center $\bar{X}$ rather than $\bar{\Gamma}$, that may provide a better explanation of Frohn et al.'s results in the Redfield-Zangwill framework.  For surface-state mediated interactions between steps, their formula indicates $n$ = 3/2, comparable to the entropic and elastic repulsions.

The effect of surface-state mediated interactions on TWDs was elucidated by Pai et al.\ [1994] in combined experimental and theoretical examination of vicinal Ag(110), which has a metallic surface state centered at $\bar{Y}$, the middle of the shorter edge of the rectangular surface Brillouin zone [Liu et al.\ 1984].  In essence, the surface state introduces a second length scale, the Fermi wavelength $\lambda_F$, in addition to $\langle \ell \rangle$, with the major consequence that the TWD is no longer a function of the single scaled dimensionless variable $s$ but depends also on $\langle \ell \rangle$.  With a suitable model potential Pai et al.\ [1994] could then account for the different TWDs at a few different misorientations (i.e., mean step spacings).  Indeed, to establish convincingly that this Friedel-like effect is significant, one must measure several different values of $\langle \ell \rangle$.  While this paper has been cited with regards to other modifications of TWDs (cf.\ e.g., Mugarza et al.\ [2006] and Li et al.\ [2010]), I have found no other investigations of Friedel-like effects on TWDs for several misorientations of the same substance.

Patrone and Einstein [2012] discuss other issues related to possible anisotropic surface state dispersion as well as showing the insensitivity to the point in the surface Brillouin zone about which the state is centered.

\section{Conclusions}

An aspect of ECS studies on which there has been substantial progress since the 1980's, but which has received little attention in this chapter, is comparing and reconciling the values of the characteristic energies (surface free energy per area, step free energy per length, and step-step repulsion strength) that are extracted from experimental measurements with ever-improving calculations (using density functional theory) of these energies.  Bonzel's review [2003],as well as Nowicki and Bonzel [2004], Yu et al. [2006], Bonzel et al.\ [2007],Barreteau et al.\ [2003]  contain extensive coverage of this issue for the soft metals to which his group has devoted exhaustive attention.  Jeong and Williams [1999] review most results for silicon. Such efforts to find absolute energies has also taken place in studies of island shapes, e.g.\ of TiN (001) [Kodambaka et al.\ 2002] and (111)  [Kodambaka et al.\ 2003].

There are several significant advances in generic understanding of ECS since the 1980's.  The Pokrovsky-Talapov ($\vartheta$ = 3/2) critical phenomena near the edge of the smoothly-curved region near a facet has proved to be far more robust and general than originally realized, while novel behavior is predicted in a very special direction.  Even though invoked in many accounts of sharp edges, long-range attractive $\ell^{-2}$ do not have an apparent physical basis, except perhaps in idiosyncratic cases.  The likely cause is a reconstruction or adsorption that changes the surface free energy of the facet orientation.  On the other hand, hill and valley structures are widely seen, and the possibility of azimuthal in addition to polar misorientation can lead to astonishingly rich phase diagrams.  Of course, non equilibrium considerations open up a whole new universe of behavior.  Furthermore, at the nanoscale cluster shape is very sensitive to the particulars of a system, with the addition or removal of a single atom leading to a substantial change in shape, rather like biological systems, in contrast to the macroscale phenomena that have been treated in this chapter.

\section*{Acknowledgements}
My research related to this subject was long supported by the UMD-NSF MRSEC; it is now supported partially by NSF-CHE 13-05892.  Much of this paper is based on extensive collaboration with the experimental surface physics group at U. of Maryland, led by Ellen D.\ Williams until 2010, with ongoing guidance by Janice Reutt-Robey and William G. Cullen, as well as with Margret Giesen and Harald Ibach, and Hans P.\ Bonzel, at FZ-J\"ulich, partially sponsored by a Humboldt Senior U.S. Scientist Award, and, over the last decade, with Alberto Pimpinelli.  I also benefited from theory interactions at Maryland with John D. Weeks and Dionisios Margetis, postdocs Olivier Pierre-Louis, Howard L. Richards, Ferenc Szalma, and Kwangmoo Kim, and graduate students Norman C. Bartelt, Raymond C. Nelson, Sanjay V. Khare, Hailu Gebremarian, Timothy J. Stasevich, Rajesh Sathiyanarayan, Paul N. Patrone, and Josue R. Morales-Cifuentes.

\leftskip 0.15in
\parindent -0.15in
\section*{Bibliography}
\frenchspacing

D.\,B.\ Abraham and P.\ Reed, Phase Separation in the Two-Dimensional Ising Ferromagnet, Phys.\ Rev.\ Lett.\ 33 (1974) 377.

D.\,B.\ Abraham and P.\ Reed, Interface profile of the Ising ferromagnet in two dimensions, Commun.\ Math.\ Phys.\ 49 (1976) 35.

Y.\ Akutsu, N.\ Akutsu, and T.\ Yamamoto, Universal Jump of Gaussian Curvature at the Facet Edge of a Crystal, Phys.\ Rev.\ Lett.\ 61 (1988) 424.

Y.\ Akutsu, N.\ Akutsu, and T.\ Yamamoto, Universality of the Tangential Shape Exponent at the Facet Edge of a Crystal, arXiv:cond-mat/9803189 (1998).

Y.\ Akutsu and N.\ Akutsu, The Equilibrium Facet Shape of the Staggered Body-Centered-Cubic Solid-on-Solid Model, Prog.\ Theor.\ Phys.\ 116 (2006) 983.

N.\ Akutsu, Non-universal equilibrium crystal shape results from sticky steps, J.\ Phys.: Condens.\ Matter 23 (2011) 485004.

O.\,L.\ Alerhand, D.\ Vanderbilt, R.\,D.\ Meade, and J.\,D.\ Joannopoulos, Spontaneous Formation of Stress Domains on Crystal Surfaces, Phys.\ Rev.\ Lett.\ 61 (1988) 1973.

F.\ Almgren and J.\,E.\ Taylor, Optimal Geometry in Equilibrium and Growth, Fractals 3 (1996) 713.

A.\,F.\ Andreev, Faceting phase transitions of crystals, Sov.\ Phys.-JETP 53 (1982) 1063 [Zh.\ Eksp.\ Teor.\ Phys 80 (1981) 2042.

N.\,W.\ Ashcroft and N.\,D.\ Mermin, Solid State Physics (Cengage Learning, 1976).

J.\,E.\ Avron, H.\ van Beijeren, L.\,S.\ Schulman, and R.\,K.\,P.\ Zia, Roughening transition, surface tension and equilibrium droplet shapes in a two-dimensional Ising system, J.\ Phys.\ A: Math.\ Gen.\ 15 (1982) L81.

S.\ Balibar and B.\ Castaing, J.\ Phys.\ Lett.\ 41 (1980) L329.

S.\ Balibar, H.\ Alles, and A.\ Ya.\ Parshin, The surface of helium crystals, Rev.\ Mod.\ Phys.\ 77 (2005) 317.

C.\ Barreteau, F.\ Raouafi, M.\,C.\ Desjonqu\`eres, and D.\ Spanjaard, J.\ Phys.: Condens.\ Matter 15 (2003) 3171.

N.\,C.\ Bartelt, E.\,D.\ Williams, R.\,J.\ Phaneuf, Y.\ Yang, and S.\ Das Sarma, Orientational stability of silicon surfaces, J.\ Vac.\ Sci.\ Technol.\ A 7 (1989) 1898.

N.\,C.\ Bartelt, T.\,L.\ Einstein, and E.\,D.\ Williams, The Influence of Step-Step Interactions on Step Wandering, Surface Sci.\ 240 (1990) L591.

A.\,A.\ Baski, S.\,C.\ Erwin, and L.\,J.\ Whitman, The structure of silicon surfaces from (001) to (111), Surface Sci.\ 392 (1997) 69.

J.\,M.\ Bermond, J.\,J.\ M\'etois, J.\,C.\ Heyraud, and F.\ Floret, Shape universality of equilibrated silicon crystals, Surface Science 416 (1998) 430.

F.\ Bonczek, T.\ Engel, and E.\ Bauer, Surface Sci.\ 97 (1980) 595.

H.\,P.\ Bonzel and E.\ Preuss, Morphology of periodic surface profiles below the roughening temperature: aspects of continuum theory, Surface Sci.\ 336 (1995) 209.

H.\,P.\ Bonzel, 3D equilibrium crystal shapes in the new light of STM and AFM, Physics Reports 385 (2003) 1.

H.\,P.\ Bonzel, D.\,K.\ Yu, and M.\ Scheffler, The three-dimensional equilibrium crystal shape of Pb: Recent results of theory and experiment, Appl.\ Phys.\ A 87 (2007) 391.

H.\,P.\ Bonzel and M.\ Nowicki, Absolute surface free energies of perfect low-index orientations of metals and semiconductors, Phys.\ Rev.\ B 70 (2004) 245430.

W.\,K.\ Burton, N.\ Cabrera, and F.\,C.\ Frank, The growth of crystals and the equilibrium structure of their surfaces, Phil.\ Trans.\ Roy.\ Soc.\ A 243 (1951) 299.

N.\ Cabrera and N.\ Garc\'{\i}a, Roughening transition in the interface between superfluid and solid $^4$He, Phys.\ Rev.\ B 25 (1982) 6057.

N.\ Cabrera, The equilibrium of crystal surfaces, Surface Sci.\ 2 (1964) 320.

J.\,W.\ Cahn, and W.\,C.\ Carter, Crystal Shapes and Phase Equilibria: A Common Mathematical Basis, Metallurgical and Materials Transactions A-Physical Metallurgy and Materials Science 27 (1996) 1431.\ [arXiv:cond-mat/0703564v1] .

J.\,W.\ Cahn, Transitions and Phase-Equilibria Among Grain-Boundary Structures, J.\ de Physique, 43(C6) (1982) 199, Proceedings of Conference on the Structure of Grain Boundaries, Caen, France.

H.\,B.\ Callen, Thermodynamics and an Introduction to Themostatistics, 2nd ed.\ (Wiley, New York, 1985).

F.\ Calogero, Solution of a Three-Body Problem in One Dimension, J.\ Math.\ Phys.\ 10 (1969) 2191, Ground State of a One-Dimensional \textit{N}-Body System, J.\ Math.\ Phys.\ 10 (1969) 2197.

E.\ Carlon and H.\ van Beijeren, Equilibrium shapes and faceting for ionic crystals of body-centered-cubic type, Phys.\ Rev.\ E 62 (2000) 7646.

Y.\ Carmi, S.\,G.\ Lipson, and E.\ Polturak, Critical behavior of vicinal surfaces of $^4$He, Phys.\ Rev.\ B 36 (1987) 1894.

R.\ Cerf and J.\ Picard, The Wulff Crystal in Ising and Percolation Models (Springer, Berlin, 2006).

W.-C.\ Cheng and P.\ Wynblatt, Coupled compositional and roughening phase transitions at the surface of a Pb-Bi-Ni alloy, Surface Sci.\ 302 (1994) 185.

A.\,A.\ Chernov, Layer-Spiral Growth of Crystals, Sov.\ Phys.-Uspekhi 4 (1961) 116.

C.\,V.\ Ciobanu, D.\,T.\ Tambe, V.\,B.\ Shenoy, C.\,Z.\ Wang, and K.\,M.\ Ho, Atomic-scale perspective on the origin of attractive step interactions on Si(113), Phys.\ Rev.\ B 68 (2003) 201302R.

M.\ Copel and T.\ Gustafsson, Structure of Au(110) Determined with Medium-Energy-Ion scattering, Phys.\ Rev.\ Lett.\ 57 (1986) 723.

M.\,F.\ Crommie, C.\,P.\ Lutz, and D.\,M.\ Eigler.\ Imaging standing waves in a two-dimensional electron gas, Nature 363 (1993) 524; Confinement of electrons to quantum corrals on a metal surface, Science 262 (1993) 218.

P.\ Curie, Bull.\ Soc.\ Min.\ de France 8 (1885) 145.

B.\ Dacorogna and C.\,E.\ Pfister, Wulff theorem and best constant in Sobolev inequality, J.\ Math.\ Pures.\ Appl.\ 71 (1992) 97.

S.\,R.\ Dahmen, B.\ Wehefritz, and G.\ Albertini, A novel exponent in the Equilibrium Shape of Crystals, arXiv:cond-mat/9802152 (1998).

M.\,S.Daw, S.\,M.Foiles, and M.\,I.Baskes, The Embedded-Atom Method - a Review of Theory and Applications, Mater.\ Sci.\ Rep.\ 9 (1993) 251.

J.\ De Coninck, F.\ Dunlop, and V.\ Rivasseau, On the microscopic validity of the Wulff construction and of the generalized Young equation, Commun.\ Math.\ Phys.\ 121 (1989) 401.

P.\,G.\ de Gennes, Soluble Model for Fibrous Structures with Steric Constraints, J.\ Chem.\ Phys.\ 48 (1968) 2257.

M.\ Degawa, T.\,J.\ Stasevich, W.\,G.\ Cullen, A.\ Pimpinelli, T.\,L.\ Einstein, and E.\,D.\ Williams, Distinctive Fluctuations in a Confined Geometry, Phys.\ Rev.\ Lett.\ 97 (2006) 080601.

M.\ Degawa, T.\,J.\ Stasevich, A.\ Pimpinelli, T.\,L.\ Einstein, and E.\,D.Williams, Facet-edge Fluctuations with Periphery Diffusion Kinetics,  Surface Sci.\ 601 (2007) 3979 [Proc.\ ECOSS 2006].

For a review of fermionic methods, see M.\ den Nijs, ``The Domain Wall Theory of Two-dimensional Commensurate-Incommensurate Phase Transitions," in: Phase Transitions and Critical Phenomena, Vol.\ 12, edited by C.\ Domb and J.\,L.\ Lebowitz (Academic, London, 1989) pp.\ 219--333.

S.\ Dieluweit, H.\ Ibach, M.\ Giesen, and T.\,L.\ Einstein, Orientation Dependence of Step Stiffness: Failure of Solid-on-Solid and Ising Models to Describe Experimental Data, Phys.\ Rev.\ B 67 (2003) 121410(R).

S.\ Dietrich and A.\ Haase, Scattering of X-rays and neutrons at interfaces, Physics Reports 260 (1995) 1.

A.\ Dinghas, \"Uber einen Gcometrischen Satz von Wulff f\"ur die Gleichgewichtsform von Kristallen, Z.\ Kristallog.\ 105 (1944) 304.

R.\,L.\ Dobrushin, R.\ Koteck\'y, and S.\,B.\ Shlosman, A Microscopic Justification of the Wulff Construction, J.\ Stat.\ Phys.\ 12 (1993) 1.

R.\,L.\ Dobrushin, R.\ Koteck\'y, and S.\ Shlosman, The Wulff Construction: A Global Shape from Local Interactions (AMS, Providence, Rhode Island, 1992); preprint: http://www.cpt.univ-mrs.fr/dobrushin/DKS-book.pdf.

P.\,M.\ Duxbury and T.\,J.\ Pence (eds.), Dynamics of Crystal Surfaces and Interfaces (Plenum, New York, 1997).

F.\,J.\ Dyson, Statistical Theory of the Energy Levels of Complex Systems.\ III, J.\ Math.\ Phys.\ 3 (1962) 166.

F.\,J.\ Dyson, Correlations between eigenvalues of a random matrix, Commun.\ Math.\ Phys.\ 19 (1970) 235.

T.\,L.\ Einstein and J.\,R.\ Schrieffer, Indirect Interaction Between Adatom Pairs on a Tight-Binding Solid, Phys.\ Rev.\ B7 (1973) 3629.

T.\,L.\ Einstein, ``Interactions Between Adsorbate Particles," in: Physical Structure of Solid Surfaces, edited by W.\,N.\ Unertl (Elsevier, Amsterdam, 1996), Handbook of Surface Science, vol.\ 1, S.\ Holloway and N.\,V.\ Richardson, series eds., pp.\ 577–-650.

T.\,L.\ Einstein, H.\,L.Richards, S.\,D.Cohen, and O.\ Pierre-Louis, Terrace-Width Distributions and Step-Step Repulsions on Vicinal Surfaces: Symmetries, Scaling, Simplifications, Subtleties, and Schr\"odinger, Surface Sci.\ 493 (2001) 460.

T.\,L.\ Einstein, Using the Wigner-Ibach Surmise to Analyze Terrace-Width Distributions: History, User's Guide, and Advances, Appl.\ Phys.\ A 87 (2007) 375 [cond-mat/0612311].

T.\,L.\ Einstein and Alberto Pimpinelli, Dynamical Scaling Implications of Ferrari, Prähofer, and Spohn's Remarkable Spatial Scaling Results for Facet-Edge Fluctuations, J.\ Statistical Phys.\ 155 (2014) 1178; longer version posted at arXiv 1312.4910v1.

A.\ Emundts, H.\,P.\ Bonzel, P.\ Wynblatt, K.\ Thürmer, J.\ Reutt-Robey, and E.\,D.\ Williams, Continuous and discontinuous transitions on 3D equilibrium crystal shapes: a new look at Pb and Au, Surface Sci.\ 481 (2001) 13.

F.\ Ercolessi, A.\ Bartolini, M.\ Garofalo, M.\ Parrinello, and E.\ Tosatti, Au Surface Reconstructions in the Glue Model, Surface Sci 189/190 (1987) 636.

P.\,L.\ Ferrari and H.\ Spohn, Step Fluctuations for a Faceted Crystal, J.\ Stat.\ Phys.\ 113 (2003) 1.

P.\,L.\ Ferrari, M.\ Pr\"{a}hofer, and H.\ Spohn, Fluctuations of an atomic ledge bordering a crystalline facet, Phys.\ Rev.\ E 69 (2004) 035102.

D.\,S.\ Fisher and J.\,D.\ Weeks, Shape of Crystals at Low Temperatures: Absence of Quantum Roughening, Phys.\ Rev.\ Lett.\ 50 (1983) 1077.

M.\,P.\,A.\ Fisher, D.\,S.\ Fisher, and J.\,D.\ Weeks, Agreement of Capillary-Wave Theory with Exact Results for the Interface Profile of the Two-dimensional Ising Model, Phys.\ Rev.\ Lett.\ 48 (1982) 368.

M.\ Flytzani-Stephanopoulos and L.\,D.\ Schmidt, Morphology and Etching Processes on Macroscopic Metal Catalysts, Prog.\ Surf.\ Sci.\ 9 (1979) 83.

I.\ Fonseca and S.\ M\"uller, An uniqueness proof for the Wulff problem, Proc.\ Royal Soc.\ Edinburgh 119 (1991) 125.

I.\ Fonseca, The Wulff theorem revisited, Proc.\ R.\ Soc.\ London A 432 (1991) 125.

F.\,C.\ Frank, Metal Surfaces, (American Society for Metals, Metals Park, Ohio 1962/3) chap.\ 1, pp.\ 1--15.

J.\ Frohn, M.\ Giesen, M.\ Poensgen, J.\,F.\ Wolf, and H.\ Ibach, Attractive interaction between steps, Phys.\ Rev.\ Lett.\ 67 (1991) 3543.

G.\ Gallavotti, The phase separation line in the two-dimensional Ising model, Commun.\ Math.\ Phys.\ 27 (1972) 103.

N.\ Garc\'{\i}a, J.\,J.\ S\'aenz, and N.\ Cabrera, Cusp Points in Surface Free Energy: Faceting and First Order Phase Transitions, Physica 124B (1984) 251.

Hailu Gebremariam, S.\,D.\ Cohen, H.\,L.\ Richards, and T.\,L.\ Einstein, Analysis of Terrace Width Distributions Using the Generalized Wigner Surmise: Calibration Using Monte Carlo and Transfer-Matrix Calculations, Phys.\ Rev.\ B 69 (2004) 125404.

J.\,W.\ Gibbs, On the Equilibrium of Heterogeneous Substances, Transactions of the Connecticut Academy of Arts and Sciences, 3, 108–248, 343–524, (1874–1878).\ Reproduced in both The Scientific Papers (1906), pp.\ 55-–353 and The Collected Works of J.\ Willard Gibbs, edited by W.\,R.\ Longley and R.\,G.\ Van Name (Yale University Press, New Haven, 1957 [1928]) pp.\ 55-–353.

J.\,W.\ Gibbs, Trans Connecticut Acad.\ 3, 108-248 (1877) 343-524; see Collected Works 1928 (previous entry).

M.\ Giesen, Step and island dynamics at solid/vacuum and solid/liquid interfaces, Prog.\ Surface Sci.\ 68 (2001) 1.

J.\ Gladi\'c, Z.\ Vu\v{c}i\'c, and D.\ Lovri\'c, Critical behaviour of the curved region near 111-facet edge of equilibrium shape cuprous selenide large single crystals, J.\ Crystal Growth 242 (2002) 517–532.

E.\,E.\ Gruber and W.\,W.\ Mullins, J.\ Phys.\ Chem.\ Solids 28 (1967) 875.

T.\ Guhr, A.\ M\"uller-Groeling, and H.\,A.\ Weidenm\"uller, Random-matrix theories in quantum physics: common concepts, Phys.\ Rept 299 (1998) 189.

F.\ Haake, Quantum Signatures of Chaos, 2nd ed.\ (Springer, Berlin, 1991).

F.\,D.\,M.\ Haldane and J.\ Villain, J.\ Phys.\ (Paris) 42 (1981) 1673.

C.\,A.\ Handwerker, M.\,D.\ Vaudin, and J.\,E.\ Blendell, Equilibrium Crystal Shapes and Surface Phase Diagrams at Surfaces in Ceramics, J.\ Phys.\ Colloques 49 (1988) C5-367.

P.\ Hecquet, Surface stress modifies step–step interaction energy with respect to the Marchenko–Parshin model, Surface Sci.\ 602 (2008) 819.

G.\,A.Held, D.\,M.Goodstein, and J.\,D.Brock, Phase separation and step roughening of vicinal Si(ill): An x-ray-scattering study, Phys.\ Rev.\ B 51 (1995) 7269.

C.\ Herring, Some Theorems on the Free Energies of Crystal Surfaces, Phys.\ Rev.\ 82 (1951) 87.

C.\ Herring, The use of classical macroscopic concepts in surface energy problems, in: Structure and Properties of Solid Surfaces, edited by R.\ Gomer and C.\,S.Smith (University of Chicago Press, Chicago, 1953), Chapter 1, pp.\ 5--81.

H.\ Hilton, Mathematical Crystallography (Oxford, 1903).

D.\,W.\ Hoffman and J.\,W.\ Cahn, A Vector Thermodynamics for Anisotropic Surfaces: I.\ Fundamentals and Application to Plane Surface Junctions, Surface Sci.\ 31 (1972) 368; J.\,W.\ Cahn and D.\,W.\ Hoffman, Vector Thermodynamics for Anisotropic Surfaces: II.\ Curved and Faceted Surfaces, Acta Met.\ 22 (1974) 1205.

R.\,Q.\ Hwang, E.\,D.\ Williams, and R.\,L.\ Park, High-resolution low-energy electron-diffraction study of the phase diagram of vicinal Si(111) surfaces, Phys.\ Rev.\ B 40 (1989) 11716.

P.\ Hyldgaard and T.\,L.\ Einstein, Interactions Mediated by Surface States: From Pairs and Trios to Adchains and Ordered Overlayers, J.\ Crystal Growth 275 (2005) e1637 [cond-mat/0408645].

P.\ Hyldgaard and M.\ Persson, Long-ranged adsorbate–adsorbate interactions mediated by a surface-state band, J.\ Phys.: Condes.\ Matter 12 (2000) L13.

H.\ Ibach and W.\ Schmickler, Step Line Tension on a Metal Electrode, Phys.\ Rev.\ Lett.\ 91 (2003) 016106.

T.\ Ihle, C.\ Misbah, and O.\ Pierre-Louis, Equilibrium step dynamics on vicinal surfaces revisited, Phys.\ Rev.\ B 58 (1998) 2289.

K.\,A.\ Jackson, ``Theory of Melt Growth," in: Crystal Growth and Characterization, edited by R.\ Ueda and J.\,B.\ Mullin (North-Holland, Amsterdam, 1975) [Proc.\ ISSCG-2 Spring School, Lake Kawaguchi, Japan (1974)], pp.\ 21--32.

C.\ Jayaprakash and W.\,F.\ Saam, Thermal evolution of crystal shapes: The fcc crystal, Phys.\ Rev.\ B 30 (1984) 3916.

C.\ Jayaprakash, C.\ Rottman, and W.\,F.\ Saam, Simple model for crystal shapes: Step-step interactions and facet edges, Phys.\ Rev.\ B 30 (1984) 6549; in the Hamiltonian in their Eq.~3, the factor $t$/2 should have been $t$.\ See Williams et al.~[1994].

C.\ Jayaprakash, W.\,F.\ Saam, and S.\ Teitel, Roughening and Facet Formation in Crystals, Phys.\ Rev.\ Lett.\ 50 (1983) 2017.

H.-C.\ Jeong and E.\,D.\ Williams, Steps on Surfaces: Experiment and Theory, Surface Sci.\ Rept.\ 34 (1999) 171.

H.-C.\ Jeong and J.\,D.\ Weeks, Effects of step–step interactions on the fluctuations of an individual step on a vicinal surface and its wavelength dependence, Surface Sci.\ 432 (1999) 101, and references therein.

B.\ Jo\'os, T.\,L.\ Einstein, and N.\,C.Bartelt, Distribution of Terrace Widths on a Vicinal Surface in the One-Dimensional Free-Fermion Model, Phys.\ Rev.\ B 43 (1991) 8153.

M.\ Kalff, G.\ Comsa, and T.\ Michely, How Sensitive is Epitaxial Growth to Adsorbates?, Phys.\ Rev.\ Lett.\ 81 (1998) 1255.

M.\ Kardar and D.\,R.\ Nelson, Commensurate-Incommensurate Transitions with Quenched Random Impurities, Phys.\ Rev.\ Lett.\ 55 (1985) 1157.

K.\,O.\ Keshishev, A.\,Ya.\ Parshin, and A.\,V.\ Babkin, Crystallization Waves in He-4, Sov.\ Phys.\ JETP 53 (1981) 362.

S.\,D.\ Kevan, Observation of a new surface state on Cu(001), Phys.\ Rev.\ B 28 (1983) 2268(R).

N.\ Knorr, H.\ Brune, M.\ Epple, A.\ Hirstein, M.\,A.\ Schneider, and K.\ Kern, Long-range adsorbate interactions mediated by a two-dimensional electron gas, Phys.\ Rev.\ B 65 (2002) 115420.

S.\ Kodambaka, S.\,V.\ Khare, V.\ Petrova, A.\ Vailionis, I.\,Petrov, and J.\,E.\ Greene, Absolute orientation-dependent TiN(001) step energies from two-dimensional equilibrium island shape and coarsening measurements on epitaxial TiN(001) layers, Surface Sci.\ 513 (2002) 468.

S.\ Kodambaka, S.\,V.\ Khare, V.\ Petrova, D.\,D.\ Johnson, I.\ Petrov, and J.\,E.\ Greene,.\ Absolute orientation-dependent anisotropic TiN(111) island step energies and stiffnesses from shape fluctuation analyses, Phys.\ Rev.\ B 67 (2003) 035409.

S.\ Kodambaka, S.\,V.\ Khare, I.\ Petrov, J.\,E.\ Greene, Two-dimensional island dynamics: Role of step energy anisotropy, Surface Sci.\ Reports 60 (2006) 55.

W.\ Kossel, Extenoling the Law of Bravais, Nachr.\ Ges.\ Wiss.\ G\"ottingen (1927) 143.

W.\ Kossel, Zur Energetik von Oberfl\"achenvorg\"angen, Annal.\ Physik 21 (1934) 457.

J.\,M.\ Kosterlitz and D.\,J.\ Thouless, Ordering, metastability and phase transitions in two-dimensional systems, J.\ Phys.\ C6 (1973) 1181.

J.\,M.\ Kosterlitz, The critical properties of the two-dimensional xy model, J.\ Phys.\ C 7 (1974) 1046.

R.\,V.\ Kukta, A.\ Peralta, and D.\ Kouris, Elastic Interaction of Surface Steps: Effect of Atomic-Scale Roughness, Phys.\ Rev.\ Lett.\ 88 (2002) 186102.

M.\ L\"assig, Vicinal Surfaces and the Calogero-Sutherland Model, Phys.\ Rev.\ Lett 77 (1996) 526.

L.\,D.\ Landau and E.\,M.\ Lifshitz, Statistical Physics, Part 1, 3rd edition revised and enlarged by E.\,M.\ Lifshitz and L.\,P.\ Pitaevskii (Pergamon Press, Oxford, 1980), \S155.

K.\,H.\ Lau and W.\ Kohn, Indirect Long-Range Oscillatory Interaction Between Adsorbed Atoms, Surface Sci.\ 75 (1978) 69; T.\,L.\ Einstein, Comment on K.\,H.Lau and W.\ Kohn: ``Oscillatory Indirect Interaction between Adsorbed Atoms"-Non-Asymptotic Behavior in Tight-Binding Models at Realistic Parameters, Surface Sci.\ 75 (1978) L161.

H.\,J.\ Leamy, G.\,H.\ Gilmer, and K.\,A.\ Jackson, ``Statistical Thermodynamics of Clean Surfaces," in: Surface Physics of Materials, edited by J.\,M.\ Blakely, vol.\ 1 (Academic Press New York, 1975), chap.\ 3, pp.\ 121--188.

F.\ Li, F.\ Allegretti, S.\ Surnev, and F.\,P.\ Netzer, Atomic engineering of oxide nanostructure superlattices, Surface Sci.\ 604 (2010) L43.

E.\,H.\ Lieb,  Exact Solution of the Two-Dimensional Slater KDP Model of a Ferroelectric, Phys.\ Rev.\ Lett.\ 19 (1967) 108.

E.\,H.\ Lieb and F.\,Y.\ Wu, in: Phase Transitions and Critical Phenomena, edited by C.\ Domb and M.\,S.\ Green (Academic Press, London, 1972) vol.\ 1, p.\ 331.

H.\ Liebmann, Der Curie-Wulff'sche Satz \"uber Combinationsformen von Krystallen, Z.\ Kristallog.\ 53 (1914) 171.

S.\,H.\ Liu, C.\ Hinnen, C.\,N.\ van Huong, N.\,R.\ de Tacconi, and K.-M.\ Ho, Surface State Effects on the Electroreflectance Spectroscopy of Au Single Crystal Surfaces, J.\ Electroanal.\ Chem.\ 176 (1984) 325.

F.\ Liu, ``Modeling and Simulation of Strain-mediated Nanostructure Formation on Surface," in: Handbook of Theoretical and Computational Nanotechnology, vol.\ 4, edited by Michael Rieth and Wolfram Schommers, American Scientific Publishers (2006), pp.\ 577--625.

I.\ Lyuksyutov, A.\,G.\ Naumovets, and V.\ Pokrovsky, Two-Dimensional Crystals (Academic Press, San Diego, 1992).

T.\,E.\ Madey, C.-H.\ Nien, K.\ Pelhos, J.\,J.\ Kolodziej, I.\,M.\ Abdelrehim, and H.-S.\ Tao, Faceting induced by ultrathin metal films: structure, electronic, Surface Sci.\ 438 (1999) 191.

T.\,E.\ Madey, J.\ Guan, C.-H.\ Nien, H.-S.\ Tao, C.-Z.\ Dong, and R.\,A.\ Campbell, Surf.\ Rev.\ Lett.\ 3 (1996) 1315.

V.\,I.\ Marchenko and A.\,Ya.\ Parshin, Elastic properties of crystal surfaces, Sov.\ Phys.\ JETP 52 (1980) 129 [Zh.\ Eksp.\ Teor.\ Fiz.\ 79 (1980) 257.]

J.\,J.\ M\'etois and J.\,C.\ Heyraud, Analysis of the critical behaviour of curved regions in equilibrium shapes of In crystals, Surface Sci.\ 180 (1987) 647.

M.\,L.\ Mehta, Random Matrices, 3rd ed.\ (Academic, New York, 2004).

H.\ Meltzman, D.\ Chatain, D.\ Avizemer, T.\,M.\ Besmann, and W.\,D.\ Kaplan, The equilibrium crystal shape of nickel, Acta Materialia 59 (2011) 3473.

T.\ Michely and J.\ Krug, Islands, Mounds, and Atoms: Patterns and Process in Crystal Growth Far from Equilibrium, Springer, Berlin, 2004, chap.\ 3.

S.\ Miracle-Sole, Facet Shapes in a Wulff Crystal, in: Mathematical Results in Statistical Mechanics, edited  by S.\ Miracle-Sole, J.\ Ruiz, and V.\ Zagrebnov (World Scientific, Singapore, 1999) pp.\ 83--101 [arXiv: 1206.3736v1].

S.\ Miracle-Sole and J.\ Ruiz, On the Wulff construction as a problem of equivalence of statistical ensembles, in: On Three Levels: Micro, Meso and Macroscopic Approaches in Physics, edited by M.\ Fannes and A.\ Verbeure (Plenum Press, New York, 1994), pp.\ 295--302 [arXiv:1206.3739v1].

S.\ Miracle-Sole, Wulff shape of equilibrium crystals (2013) [arXiv: 1307.5180v1].

A.\ Mugarza, F.\ Schiller, J.\ Kuntze, J.\ Cord\'on, M.\ Ruiz-Os\'es, and J.\,E.\ Ortega, Modelling nanostructures with vicinal surfaces, J.\ Phys.: Condens.\ Matter 18 (2006) S27.

B.\ M\"uller, L.\ Nedelmann, B.\ Fischer, H.\ Brune, J.\,V.\ Barth, and K.\ Kern, Island Shape Transition in Heteroepitaxial Metal Growth on Square Lattices, Phys.\ Rev.\ Lett.\ 80 (1998) 2642.

W.\,W.\ Mullins, ``Solid Surface Morphologies Governed by Capillarity," in: Metal Surfaces: Structure, Energetics, and Kinetics, ed.\ by W.\,D.\ Robertson and N.\,A.\ Gjostein (American Society for Metals, Metals Park (OH), 1962/3), chap.\ 2, pp.\ 17–-62.

R.\ Najafabadi and D.\,J.\ Srolovitz, Elastic step interactions on vicinal surfaces of fcc metals, Surface Sci.\ 317 (1994) 221.

R.\,C.\ Nelson, T.\,L.\ Einstein, S.\,V.\ Khare, and P.\,J.\ Rous, Energies of Steps, Kinks, and Defects on Ag{100} and {111} Using Embedded Atom Method, and Some Consequences, Surface Sci.\ 295 (1993) 462.

T.\ Nishino and K.\ Okunishi, J.\ Phys.\ Soc.\ Jpn.\ 64 (1995) 4084.

D.\,Y.\ Noh, K.\,I.\ Blum, M.\,J.\ Ramstad, and R.\,J.\ Birgeneau, Long-range coherence and macroscopic phase separation of steps on vicinal Si(111), Phys.\ Rev.\ B 44 (1991) 10969.

D.\,Y.\ Noh, K.\,I.\ Blum, M.\,J.\ Ramstad, and R.\,J.\ Birgeneau, Faceting, roughness, and step disordering of vicinal Si(111) surfaces: An x-ray-scattering study, Phys.\ Rev.\ B 48 (1993) 1612.

D.\,Y.\ Noh, K.\,S.\ Liang, Y.\ Hwu, and S.\ Chandavarkar, Surface Sci.\ 326 (1995) L455.

I.\,M.\ Nolden and H.\ van Beijeren, Equilibrium shape of bcc crystals: Thermal evolution of the facets, Phys.\ Rev.\ B 49 (1994) 17224.

M.\ Nowicki, C.\ Bombis, A.\ Emundts, H.\,P.\ Bonzel, and P.\ Wynblatt, Universal exponents and step-step interactions on vicinal Pb(111) surfaces, Eur.\ Phys.\ Lett.\ 59 (2002a) 239.

M.\ Nowicki, C.\ Bombis, A.\ Emundts, H.\,P.\ Bonzel, and P.\ Wynblatt, Step–step interactions and universal exponents studied via three-dimensional equilibrium crystal shapes, New J.\ Phys.\ 4 (2002b) 60.

M.\ Nowicki, C.\ Bombis, A.\ Emundts, and H.\,P.\ Bonzel, Absolute step and kink formation energies of Pb derived from step roughening of two-dimensional islands and facets, Phys.\ Rev.\ B 67 (2003) 075405.

P.\ Nozi\`eres, Shape and Growth of Crystals, in: Solids Far From Equilibrium, edited by C.\ Godr\`eche, Cambridge U.\ Press, Cambridge, 1992.

B.\ Ocko, private discussion, 2014.

K.\ Okunishi, Y.\ Hieida and Y.\ Akutsu, $\delta$-function Bose-gas picture of S=1 antiferromagnetic quantum spin chains near critical fields, Phys.\ Rev.\ B 59 (1999) 6806; Middle-field cusp singularities in the magnetization process of one-dimensional quantum antiferromagnets, ibid.\ 60 (1999) R6953.

B.\ Olshanetsky and V.\ Mashanov, LEED Studies of Clean High Miller Index Surfaces of Silicon, Surface Sci.\ 111 (1981) 414.

W.\,W.\ Pai, J.\,S.\ Ozcomert, N.\,C.\ Bartelt, T.\,L.\ Einstein, and J.\,E.\ Reutt-Robey, Terrace-Width Distributions on Vicinal Ag(110): Evidence of Oscillatory Interactions, Surface Sci.\ 307-309 (1994) 747.

P.\,N.\ Patrone and T.\,L.\ Einstein, Anisotropic Surface State Mediated RKKY Interaction Between Adatoms, Phys.\ Rev.\ B 85 (2012) 045429.

R.\,F.\ Peierls, Quantum Theory of Solids (Clarendon, Oxford, 1955), p.\ 108.\ (Also H.\ Fr\"ohlich, Proc.\ Roy.\ Soc.\ A 223 (1954) 296.

D.\ Peng, S.\ Osher, B.\ Merriman, and H.-K.\ Zhao, ``The geometry of Wulff crystal shapes and its relation with Riemann problems," in: Nonlinear Partial Differential Equations, Contemp.\ Math.\ 238, AMS, Providence, RI, 1999, pp.\ 251-–303.

C.\,E.\ Pfister, Large deviations and phase separation in the two-dimensional Ising model, Helv.\ Phys.\ Acta 64 (1991) 953.

R.\,J.\ Phaneuf and E.\,D.\ Williams, Surface Phase Separation of Vicinal Si (111), Phys.\ Rev.\ Lett.\ 58 (1987) 2563.

R.\,J.\ Phaneuf, N.\,C.\ Bartelt, E.\,D.\ Williams, W.\ Swiech, and E.\ Bauer, Crossover from Metastable to Unstable Facet Growth on Si(1ll), Phys.\ Rev.\ Lett.\ 71 (1993) 2284.

A.\ Pimpinelli and J.\ Villain, Physics of Crystal Growth (Cambridge University Press, Cambridge, England, 1998).

A.\ Pimpinelli, M.\ Degawa, T.\,L.\ Einstein, and E.\,D.\ Williams, A Facet Is Not an Island: Step-Step Interactions and the Fluctuations of the Boundary of a Crystal Facet, Surface Sci.\ 598 (2005) L355.

V.\,L.\ Pokrovsky and A.\,L.\ Talapov, Ground State, Spectrum, and Phase Diagram of Two-Dimensional Incommensurate Crystals, Phys.\ Rev.\ Lett.\ 42 (1979) 65; Sov.\ Phys.-JETP 51 (1980) 134.

V.\,L.\ Pokrovsky and A.\,L.\ Talapov, Theory of Incommensurate Crystals, Soviet Scientific Reviews Supplement Series Physics vol.\ 1, Harwood Academic Publishers, Chur, 1984, and references therein.

G.\ Pr\'evot and B.\ Croset, Revisiting Elastic Interactions between Steps on Vicinal Surfaces: The Buried Dipole Model, Phys.\ Rev.\ Lett.\ 92 (2004) 256104.

A.\,C.\ Redfield and A.\ Zangwill, Attractive interactions between steps, Phys.\ Rev.\ B 46 (1992) 4289.

J.\ Repp, F.\ Moresco, G.\ Meyer, K.-H.\ Rieder, P.\ Hyldgaard, and M.\ Persson, Substrate Mediated Long-Range Oscillatory Interaction between Adatoms: Cu /Cu(111), Phys.\ Rev.\ Lett.\ 85 (2000) 2981.

C.\ Rottman and M.\ Wortis, Statistical mechanics of equilibrium crystal shapes: Interfacial phase diagrams and phase transitions, Phys.\ Rept.\ 103 (1984a) 59.

C.\ Rottman and M.\ Wortis, Equilibrium crystal shapes for lattice models with nearest- and next-nearest-neighbor interactions, Phys.\ Rev.\ B 29 (1984b) 328.

C.\ Rottman, M.\ Wortis, J.\,C.\ Heyraud, and J.\,J.\ M\'etois, Equilibrium Shapes of Small Lead Crystals: Observation of Pokrovsky-Talapov Critical Behavior, Phys.\ Rev.\ Lett.\ 52 (1984) 1009.

S.\ Rousset, J.\,M.\ Berroir, V.\ Repain, Y.\ Garreau, V.\,H.\ Etgens, J.\ Lecoeur, and R.\ Pinchaux, Thermal faceting behavior of Au(4,5,5), Surface Sci.\ 443 (1999) 265.

M.\,A.\ Ruderman and C.\ Kittel, Indirect Exchange Coupling of Nuclear Magnetic Moments by Conduction Electrons, Phys.\ Rev.\ 96 (1954) 99.

J.\,J.\ Sa\'enz and N.\ Garc\'{\i}a, Classical Critical Behaviour in Crystal Surfaces Near Smooth and Sharp Edges, Surface Sci.\ 155 (1985) 24.

R.\ Sathiyanarayanan, A.\ BH.\ Hamouda, and T.\,L.\ Einstein, Terrace-width Distributions of Touching Steps: Modification of the Fermion, with Implications for Measuring Step-step Interactions, Phys.\ Rev.\ B 80 (2009) 153415.

R.\ Sato and Y.\ Akutsu, Curvature Jump at the Facet Edge of a Crystal for Arbitrary Surface Orientation, J.\ Phys.\ Soc.\ Jpn.\ 64 (1995) 3593.

R.\,E.\ Schlier and H.\,E.\ Farnsworth, Structure and Adsorption Characteristics of Clean Surfaces of Germanium and Silicon, J.\ Chem. Phys.\ 30 (1959) 917.

H.\,J.\ Schulz, B.\,I.\ Halperin, and C.\,L.\ Henley, Dislocation interaction in an adsorbate solid near the commensurate-incommensurate transition, Phys.\ Rev.\ B 26 (1982) 3797.

R.\,F.\ Sekerka, Theory of Crystal Growth Morphology, in: Crystal Growth - From Fundamentals to Technology, edited by G.\ M\"uller, J.-J.\ M\'etois, and P.\ Rudolph (Elsevier, Amsterdam, 2004), pp.\ 55--93.

V.\,B.\ Shenoy and C.\,V.\ Ciobanu, Ab initio density functional studies of stepped TaC surfaces, Phys.\ Rev.\ B 67 (2003) 081402.

N.\ Shimoni, S.\ Ayal, and O.\ Millo, Step dynamics and terrace-width distribution on flame-annealed gold films: The effect of step-step interaction, Phys.\ Rev.\ B 62 (2000) 13147.

G.\,A.\ Somorjai and M.\,A.\ Van Hove, Adsorbate-Induced Restructuring of Surfaces, Prog.\ Surf.\ Sci.\ 30 (1989) 201.

S.\ Song and S.\,G.\,J.\ Mochrie, Tricriticality in the orientational phase diagram of stepped Si(113) surfaces, Phys.\ Rev.\ Lett.\ 73 (1994) 995.

S.\ Song and S.\,G.\,J.\ Mochrie, Attractive step-step interactions, tricriticality, and faceting in the orientational phase diagram of silicon surfaces between [113] and [114], Phys.\ Rev.\ B 51 (1995) 10068.

S.\ Song, M.\ Yoon, and S.\,G.\,J.\ Mochrie, Faceting, tricriticality, and attractive interactions between steps in the orientational phase diagram of silicon surfaces between [113] and [55 12], Surface Sci.\ 334 (1995) 153.

T.\,J.\ Stasevich and T.\,L.\ Einstein, Analytic Formulas for the Orientation Dependence of Step Stiffness and Line Tension: Key Ingredients for Numerical Modeling, [SIAM-]Multiscale Model.\ Simul.\ 6 (2007) 90.

T.\,J.\ Stasevich, C.\ Tao, W.\,G.\ Cullen, E.\,D.\ Williams, and T.\,L.\ Einstein, Impurity Decoration for Crystal Shape Control: C$_{\rm 60}$ on Ag(111), Phys.\ Rev.\ Lett.\ 102 (2009) 085501.

T.\,J.\ Stasevich, Hailu Gebremariam, T.\,L.\ Einstein, M.\ Giesen, C.\ Steimer, and H.\ Ibach, Low-Temperature Orientation Dependence of Step Stiffness on \{111\} Surfaces, Phys.\ Rev.\ B 71 (2005) 245414.

T.\,J.\ Stasevich, Ph.\,D.\ dissertation, Univ. of Maryland, 2006 (unpublished) [\texttt{http://drum.lib.umd.edu/} \texttt{bitstream/1903/4071/1/umi-umd-3818.pdf}].

T.\,J.\ Stasevich, T.\,L.\ Einstein, R.\,K.\,P.Zia, M.\ Giesen, H.\ Ibach, and F.\ Szalma, The Effects of Next-Nearest-Neighbor Interactions on the Orientation Dependence of Step Stiffness: Reconciling Theory with Experiment for Cu(001), Phys.\ Rev.\ B 70 (2004) 245404.

T.\,J.\ Stasevich, T.\,L.\ Einstein, and S.\ Stolbov, Extended Lattice Gas Interactions of Cu on Cu(111) and Cu(001): Ab-Initio Evaluation and Implications, Phys.\ Rev.\ B 73 (2006) 115426.

J.\ Stewart, O.\ Pohland, and J.\,M.\ Gibson, Elastic-displacement field of an isolated surface step, Phys.\ Rev.\ B 49 (1994) 13848.

I.\,N.\ Stranski, Zur Theorie des Kristallwachstums, Z.\ Phys.\ Chem.\ (Leipzig) 136 (1928) 259.

W.\,P.\ Su, J.\,R.\ Schrieffer, and A.\,J.\ Heeger, Solitons in Polyacetylene, Phys.\ Rev.\ Lett.\ 42, 1698 (1979); Soliton excitations in polyacetylene, Phys.\ Rev.\ B 22 (1980) 2099.

S.\ Surnev, K.\ Arenhold, P.\ Coenen, B.\ Voigtl\"ander, and H.\,P.\ Bonzel, Scanning tunneling microscopy of equilibrium crystal shapes, J.\ Vac.\ Sci.\ Technol.\ A 16 (1998) 1059.

B.\ Sutherland, C.\,N.\ Yang, and C.\,P.\ Yang, Exact Solution of a Model of Two-Dimensional Ferroelectrics in an Arbitrary External Electric Field, Phys.\ Rev.\ Lett.\ 19 (1967) 588.

B.\ Sutherland, Quantum Many-Body Problem in One Dimension: Ground State, J.\ Math.\ Phys.\ 12 (1971) 246; Exact Results for a Quantum Many-Body Problem in One Dimension, Phys.\ Rev.\ A 4 (1971) 2019.

A.\ Szczepkowicz, A.\ Ciszewski, R.\ Bryl, C.\ Oleksy, C.-H.\ Nien, Q.\ Wu, and T.\,E.\ Madey, A comparison of adsorbate-induced faceting on flat and curved crystal surfaces, Surface Sci.\ 599 (2005) 55.

K.\ Th\"{u}rmer, J.\,E.\ Reutt-Robey, and E.\,D.\ Williams, Nucleation limited crystal shape transformations, Surface Sci.\ 537 (2003) 123.

E.\ Tosatti, private communication, March 2014.

S.\,B.\ van Albada, M.\,J.\ Rost, and J.\,W.\,M.\ Frenken, Asymmetric and symmetric Wulff constructions of island shapes on a missing-row reconstructed surface, Phys.\ Rev.\ B 65 (2002) 205421.

H.\ van Beijeren, Exactly Solvable Model for the Roughening Transition of a Crystal Surface Phys.\ Rev.\ Lett.\ 38 (1977) 993.

H.\ van Beijeren and I Nolden, ``The Roughening Transition," in: Structure and Dynamics of Surfaces II: Phenomena, Models, and Methods, edited by W.\ Schommers and P.\ von Blanckenhagen, Topics in Current Physics, vol.\ 43 (Springer, Berlin, 1987), pp.\ 259--300.

J.\ Villain, ``Two-Dimensional Solids and Their Interactions with Substrates" in: Ordering in Strongly Fluctuating Condensed Matter Systems, edited by T.\ Riste (Plenum, New York, 1980), pp.\ 221--260.

M.\ von Laue, Der Wulffsche Satz f\"ur die Gleichgewichtsform von Kristallen, Z.\ Kristallogr., Mineral.\ Petrogr.\ 105 (1943) 124.

D.\,A.\ Walko and I.\,K.\ Robinson, Energetics of oxygen-induced faceting on Cu(115), Phys.\ Rev.\ B 64 (2001) 045412.

Z.\ Wang and P.\ Wynblatt, The equilibrium form of pure gold crystals, Surface Sci.\ 398 (1998) 259.

G.\ Wang, J.\,F.\ Webb, and J.\ Zi, The strictly attractive 1/$\ell^2$ interaction between steps of crystal surfaces, Surface Sci.\ 601 (2007) 1944.

X.-S.\ Wang, J.\,L.\ Goldberg, N.\,C.\ Bartelt, T.\,L.\ Einstein, and E.\,D.\ Williams, Terrace Width Distributions on Vicinal Si(111), Phys.\ Rev.\ Lett.\ 65 (1990) 2430.

J.\,D.\ Weeks, ``The Roughening Transition" in: Ordering in Strongly Fluctuating Condensed Matter Systems, edited by T.\ Riste (Plenum, New York, 1980), pp.\ 293--317.

J.\,D.\ Weeks, private discussions (2014).

A.\,A.\ Wheeler, Cahn-Hoffman $\xi$-Vector and Its Relation to Diffuse Interface Models of Phase Transitions, J.\ Stat.\ Phys.\ 95 (1999) 1245.

E.\,D.Williams, R.\,J.\ Phaneuf, Jian Wei, N.\,C.\ Bartelt, and T.\,L.\ Einstein, Thermodynamics and statistical mechanics of the faceting of stepped Si(111), Surface Sci.\ 294 (1993) 219; erratum 310 (1994) 451.

E.\,D.\ Williams and N.\,C.\ Bartelt,``Thermodynamics and Statistical Mechanics of Surfaces" in: Physical Structure of Solid Surfaces, edited by W.\,N.\ Unertl (Elsevier, Amsterdam, 1996), Handbook of Surface Science, vol.\ 1, S.\ Holloway and N.\,V.\ Richardson, series eds., pp.\ 51--99.

E.\,D.\ Williams and N.\,C.\ Bartelt, Surface Faceting and the Equilibrium Crystal Shape, Ultramicroscopy 31 (1989) 36.

P.\,E.\ Wolf, F.\ Gallet, S.\ Balibar, E.\ Rolley, and P.\ Nozi\`eres, Crystal growth and crystal curvature near roughening transitions in hcp $^4$He, J.\ Phys.\ (France) 46 (1985) 1987.

P.\,E.\ Wolf, S.\ Balibar, and F.\ Gallet, Experimental Observation of a Third Roughening Transition on hcp $^4$He Crystals, Phys.\ Rev.\ Lett.\ 51 (1983) 1366.

M.\ Wortis, ``Equilibrium Crystal Shapes and Interfacial Phase Transitions," in: Chemistry and Physics of Solid Surfaces, VII, edited by R.\ Vanselow and R.\ Howe (Springer-Verlag, Berlin, 1988), pp.\ 367--405.

G.\ Wulff, Zur Frage der Geschwindigkeit des Wachstums und der Aufl\"osung der Krystallflachen, Z.\ Krystallographie und Mineralogie 34 (1901) 449.

M.\ Yamamoto, K.\ Sudoh, H.\ Iwasaki, and E.\,D.\ Williams, Anomalous decay of multilayer holes on SrTiO$_3$(001), Phys.\ Rev.\ B 82 (2010) 115436.

C.\,P.\ Yang, Exact Solution of a Model of Two-Dimensional Ferroelectrics in an Arbitrary External Electric Field, Phys.\ Rev.\ Lett.\ 19 (1967) 586.

M.\ Yoon, S.\,G.\,J.\ Mochrie, D.\,M.\ Zehner, G.\,M.\ Watson, and D.\ Gibbs, Faceting and the orientational phase diagram of stepped Pt(001) surfaces, Phys.\ Rev.\ B 49 (1994) 16702.

K.\ Yosida, Magnetic Properties of Cu-Mn Alloys, Phys.\ Rev.\ 106 (1957) 893.

D.\,K.\ Yu, H.\,P.\ Bonzel, M.\ Scheffler, The stability of vicinal surfaces and the equilibrium crystal shape of Pb by first principles theory, New J.\ Phys.\ 8 (2006) 65.

R.\,K.\,P.\ Zia, ``Anisotropic surface tension and equilibrium crystal shapes," in: Progress in Statistical Mechanics, edited by C.\,K.\ Hu (World Scientific, Singapore, 1988), pp.\ 303--357.

R.\,K.\,P.\ Zia, Exact equilibrium shapes of Ising crystals on triangular/honeycomb lattices, J.\ Stat.\ Phys.\ 45 (1986) 801.

J.-K.\ Zuo, T.\ Zhang, J.\,F.\ Wendelken, and D.\,M.\ Zehner, Step bunching on TaC(910) due to attractive step-step interactions, Phys.\ Rev.\ B 63 (2001) 033404.

\end{document}